\documentclass[showpacs,
aps,superscriptaddress, 10pt,
prd,notitlepage,showkeys,
nofootinbib, floatfix]{revtex4-2}
\usepackage{bm}
\usepackage{amsfonts}
\usepackage{latexsym}
\usepackage[normalem]{ulem}
\usepackage{graphicx}
\usepackage{amsmath}
\usepackage{palatino}
\usepackage{mathpazo}
\usepackage{textcomp}
\usepackage{float}
\usepackage{booktabs}
\usepackage{dcolumn}
\usepackage{ragged2e}
\usepackage{hyperref}
\hypersetup{colorlinks,citecolor=red}
\hypersetup{colorlinks=true,linkcolor=blue,filecolor=magenta,urlcolor=blue}
\usepackage{amsmath}
\usepackage{subfigure}
\usepackage[]{natbib}
\usepackage[table,xcdraw]{xcolor}
\DeclareUnicodeCharacter{2212}{-}
\usepackage{orcidlink}
\usepackage{epsfig}
\usepackage{caption}
\usepackage[toc]{appendix}
\usepackage{commath}
\usepackage{cancel}
\usepackage{csquotes}
\usepackage{placeins}
\usepackage{multirow}
\usepackage{parskip}
\usepackage{threeparttable}
\usepackage{tcolorbox}
\usepackage{pifont}

\begin{document}
\raggedbottom
\newcommand{\BITS}{Department of Physics, Birla Institute of Technology and Science - Pilani, K. K. Birla Goa Campus, NH-17B, Zuarinagar, Sancoale, Goa- 403726, India}

\newcommand{\Tezpur}{Department of Physics, Tezpur University, Napaam, Tezpur, 784028, Assam, India}

\newcommand{\IUCAA}{Inter-University Centre for Astronomy and Astrophysics, PostBag 4, Ganeshkhind, Pune-411007, Maharashtra, India}

\title{Implications of Fermionic Dark Matter Interactions on Anisotropic Neutron Stars}

\author{Premachand Mahapatra~\orcidlink{0000-0002-3762-8147}}\email{p20210039@goa.bits-pilani.ac.in}
\affiliation{\BITS}

\author{Chiranjeeb Singha~\orcidlink{0000-0003-0441-318X}}\email{chiranjeeb.singha@iucaa.in}
\affiliation{\IUCAA}

\author{Ayush Hazarika~\orcidlink{0009-0004-5255-0730}}\email{ayush.hazarika4work@gmail.com}
\affiliation{\Tezpur}

\author{Prasanta Kumar Das~\orcidlink{0000-0002-2520-7126}}\email{pdas@goa.bits-pilani.ac.in}
\affiliation{\BITS}

\date{\today}

\begin{abstract}

The presence of Dark matter (DM) within a neutron star (NS) can substantially influence the macroscopic properties. It is commonly assumed that the pressure inside an NS is isotropic, but in reality, pressure is locally anisotropic. This study explores the properties of anisotropic NS with a subfraction of DM (isotropic) trapped inside. Implementing a two-fluid formalism with three Equations of State (EOS): AP3 (a realistic nucleon-nucleon interaction model), BSk22 (modeling atomic nuclei and neutron-matter), and MPA1 (considering relativistic effects in nuclear interactions). The properties of NS, such as mass ($M$), radius ($R$), and dimensionless tidal deformability ($\Lambda$), for various DM-anisotropic configurations, have been rigorously tested against observational constraints. These constraints include data from the binary NS merger GW170817, NICER x-ray measurements, and pulsar mass-radius observations. We observe that with increasing DM subfraction, higher anisotropies could also satisfy the observational constraints. Furthermore, increasing the coupling ($g$) between DM and its mediator leads to the formation of a core-halo structure, with a DM halo surrounding the baryonic matter (BM). Specifically, for coupling values of $g = 10^{-4}$, $10^{-3.7}$, and $10^{-3.5}$, we observe that the maximum radius ($R_{max}$) decreases with increasing anisotropy, which contrasts with the behavior at $g = 10^{-5}$ and in scenarios with no DM. Our analysis indicates that binary pulsar systems could potentially constrain the extent of admixed anisotropic NS or, more optimistically, provide evidence for the existence of DM-admixed anisotropic NS.

\end{abstract}

\maketitle

\section{Introduction}

Neutron stars (NSs) are among the densest and most compact celestial objects known to science. Understanding the internal composition of such compact objects is based on our knowledge of high-density matter \cite{lattimer2004physics, lattimer2014neutron}. Precision in determining neutron star (NS) properties at densities exceeding those found in nuclei faces two main challenges: limited understanding of nuclear interactions \cite{burgio2021neutron, steiner2010equation, lattimer2021neutron} and gravitational interactions \cite{lasky2015gravitational, belvedere2012neutron, sieniawska2019continuous}. The repulsive nuclear equation of state (EOS), which describes the relationship between matter pressure and energy density, and the opposing attractive strong-field gravitational physics are interconnected through the Tolman-Oppenheimer-Volkoff (TOV) equations \cite{lattimer2004physics, lattimer2014neutron, lattimer2001neutron}. Uncertainties in these equations can affect predictions related to the structure and characteristics of NSs.
\par In recent years, our understanding of NSs has significantly evolved due to multimessenger observations \cite{meszaros2019multi}, particularly from binary neutron star mergers \cite{Dietrich:2020eud}. These observations have enabled a more detailed examination of various parameters influencing NS global properties, such as radius, tidal deformability, maximum mass, and other characteristics closely tied to the equation of state (EOS) governing NS behavior. There has been an abundance of observational data on NSs from both gravitational wave detectors, such as Advanced LIGO \cite{aasi2015advanced} and Advanced Virgo \cite{acernese2014advanced}, and electromagnetic radiation, exemplified by the NICER mission \cite{gendreau2016neutron}. These astrophysical observations have led to numerous efforts to constrain the EOS of NSs. These endeavors include using NICER's measurements of mass and radius \cite{miller2019psr, vinciguerra2024updated}, assessing tidal deformability through gravitational waves, and applying joint constraints \cite{miller2019psr, Miller:2019cac, Miller:2021qha, raaijmakers2021constraints}. The detection of the binary neutron star merger GW170817 \cite{abbott2017gw170817}, in particular, has spurred further investigations in this field \cite{abbott2019tests, abbott2019properties,abbott2018gw170817}.

The groundbreaking GW170817 event, marked by the first detection of gravitational waves from a neutron star merger, has been central to numerous investigations. Beyond its historical significance, it has spurred inquiries into the internal aspects of NSs. Studies have explored the influence of the NS crust on the equation of state (EOS) \cite{suleiman2021influence, fortin2016neutron}, the impact of introducing an isovector-scalar meson into the quark-meson coupling description of nuclear matter \cite{motta2019isovector, machleidt1989meson}, and various Skyrme-like parametrizations \cite{zahed1986skyrme, chabanat1997skyrme, bartel1982towards}. Nonparametric inference has indicated a preference for a soft EOS \cite{landry2019nonparametric,essick2020nonparametric}.

The event has also prompted critical evaluations of the EOS for dense matter, incorporating nuclear physics within the chiral effective field theory framework \cite{machleidt2011chiral, kruger2013neutron}. Despite progress, certain regions of the EOS, particularly at high densities, remain not fully understood. Identifying an electromagnetic counterpart to the event enabled the first joint gravitational wave-electromagnetic constraints on the NS EOS. Using the binary's tidal deformability parameter, simulations of electromagnetic observations within numerical relativity and kilonova models were conducted, ruling out extreme EOS models.

Statistical Bayesian methods were employed in the context of GW170817, with microscopic models of cold NSs using chiral effective models. Recent gravitational wave events with X-ray sources were also combined and studied with relativistic mean-field models \cite{ring1996relativistic, lourencco2019consistent, Thakur:2024scc}. Additionally, NASA's Neutron Star Interior Composition Explorer (NICER) made significant electromagnetic measurements, constraining the mass-radius relationship of the pulsar PSR J0030+0451 \cite{miller2019psr, Miller:2019cac, vinciguerra2024updated}. In this case, nonparametric inference suggested a preference for stiffer EOS.

Meanwhile, Earth-based experiments like the Lead Radius Experiment (PREX-2) have provided valuable insights into nuclear matter around saturation density, with implications for the neutron star crust. Extrapolating the data has constrained stellar radii to specific ranges, indicating a softening in the intermediate region and a stiffening at high densities, potentially leading to a phase transition in the stellar core. The expanding pool of observational data has been crucial in establishing further constraints on the EOS for dense matter, creating fertile ground for statistical and machine-learning models \cite{PhysRevLett.126.172503}.

The composition of neutron star (NS) cores has been a topic of debate and mystery in recent years. While astrophysical observations provide some insight, determining the exact nature of these dense interiors from observational data alone remains elusive. Hypotheses in the literature suggest the presence of free quarks, hyperons, delta isobars, or kaon condensation within NSs \cite{KAPLAN1988273, PhysRevLett.34.1353, PhysRevC.36.1019}. The presence of such exotic particles tends to soften the equation of state (EOS), affecting observables like the mass and tidal deformability of the star \cite{PhysRevLett.67.2414, PhysRevC.85.065802}. Besides standard exotic particles, theoretical studies also propose the presence of dark matter (DM) particles within NSs. The ultra-dense profile and particle-antiparticle asymmetry in DM could lead to gravitational trapping of DM without self-annihilation in NS cores \cite{PhysRevD.40.3221, PhysRevD.77.023006}. Various DM candidates, such as neutralinos, axions, and technibaryons, have been theoretically explored \cite{BERTONE2005279, Duffy_2009, PhysRevC.89.025803}. Though the nature and interactions of DM with standard model particles are weak and uncertain, theoretical efforts have examined DM's effects on astral bodies \cite{X.Y.Li_2012, PhysRevD.93.083009}. Gravitational wave events like GW170817 and GW190814 have provided new avenues to study DM's impact on the universe \cite{ELLIS2018607, LIGOScientific:2020zkf}.

The universe is known to contain only about 6 \% visible matter, with the remaining 94 \% comprising DM  and dark energy. Zwicky's mass estimation of the universe revealed this discrepancy, leading to the identification of DM (approximately 26 \%) and dark energy (around 68 \%) \cite{RePEc:nat:nature:v:573:y:2019:i:7772:d:10.1038_d41586-019-02603-7, Kouvaris_2015}. Numerous theoretical, experimental, and observational efforts have been dedicated to unravelling the mysteries of DM and dark energy. Several DM candidates have been hypothesized, including weakly interacting massive particles (WIMPs), feebly interacting massive particles (FIMPs), neutralinos, and axions \cite{Kouvaris:2011gb, galaxies11050101}. WIMPs, produced in the early universe and annihilating in pairs, are considered thermal relics. Various experiments, such as DAMA \cite{bernabei2020dama}, Xenon \cite{aprile2007xenon}, and CDMS \cite{ahmed2009dark}, aim to detect the interactions between WIMPs and nucleons, while indirect detection experiments like Fermi large area telescopes \cite{atwood2009large} and imaging air Cherenkov telescopes \cite{mirzoyan1994first} also contribute to DM research.

It is known that NSs, the densest known objects in the universe, have garnered attention because of their potential to provide insight into DM. With increasingly precise observational data, NSs are becoming valuable for studying DM properties. Theoretical models of DM-admixed NSs (DNSs) integrate nuclear matter (NM) equations of state (EOS) with hypothetical DM EOS using a general-relativistic two-fluid approach \cite{Giangrandi:2022wht, Karkevandi:2021ygv}. A major challenge in understanding DM in NSs is the complex mass-radius relationship influenced by DM content alongside the uncertain high-density NM EOS. This complexity makes it difficult to draw definitive conclusions about either DM or NM. For instance, observing a very massive 'NS' could indicate either a DM presence or a very stiff NM EOS. Therefore, new theoretical methods are needed to clearly identify and quantify DM in NSs.

Currently, it is generally believed that most observed NSs are standard hadronic stars with minimal DM admixture \cite{Bramante:2023djs}. Estimates suggest that NSs acquire negligible amounts of DM through accretion over their lifetimes, typically less than 10$^{-10}$ solar masses, which is unobservable. Significant DM fractions in DNSs, proposed in this study, would require exotic capture or formation mechanisms, which remain speculative. 

Recent research has intensively investigated the effects of DM on NM and NS properties \cite{Collier:2022cpr}. This includes DM's fundamental impact on NS mass and radius, the possibility of DM acquisition by NSs, and phenomena like heating, internal black hole formation, and collapse. Quantitative studies have examined DM's effects on NM properties and the validity of universal relations between NS compactness, moment of inertia, tidal deformability, and quadrupole moment in the presence of DM \cite{Collier:2022cpr, Scordino:2024ehe, Bramante:2023djs, Thakur:2023aqm, Thakur:2024mxs}. Some studies have considered the possibility that LIGO/Virgo events GW170817  and GW190425 might involve DNSs \cite{Thakur:2024btu}. Other research has focused on DM's effects on NS tidal deformability and related observables, which are measurable by current and future gravitational wave detectors. Additionally, recent studies have explored the impact of DM on pulsar X-ray profiles and NS cooling processes \cite{Liu:2024swd}.

Understanding DM's influence on NSs and other cosmic phenomena requires a multidisciplinary approach, incorporating various aspects of physics and leveraging both theoretical and observational advancements. The potential presence of exotic and DM particles within NSs highlights the need for continued exploration to unravel these cosmic mysteries. In summary, understanding DM's influence on NSs requires integrating various fields of physics and considering multiple DM candidates \cite{Barbat:2024yvi}.

Anisotropic NSs represent a nuanced departure from the conventional TOV stars \cite{Das:2022ell}. While their isotropic counterparts assume uniform pressure throughout their interiors, anisotropic stars permit pressure anisotropy, introducing an additional layer of complexity to their behaviour. This departure from isotropy is not merely theoretical; it has profound consequences for the stability, equilibrium, and observability of these cosmic bodies. Very high magnetic field \cite{Yazadjiev:2011ks}, pion condensation \cite{Sawyer:1972cq}, phase transitions \cite{Carter:1998rn}, relativistic nuclear interaction \cite{Canuto:1974ft}, superfluid core \cite{kippenhahn1990stellar}, etc., are the main causes of the anisotropy inside a star. At very high pressures and densities, nuclear matter shows some irregularity in the tangential and radial components of pressure, which creates an anisotropic fluid  \cite{Mondal:2023wwo}. One of the primary motivations for our exploration of anisotropic stars is their potential to serve as natural laboratories for the study of exotic states of matter. Within their ultra-dense cores, where nuclear densities prevail, exotic phases of matter, such as quark matter or hyperonic matter, may exist. The existence and properties of these phases hold profound implications for our comprehension of the fundamental forces governing the universe. Anisotropic stars \cite{Silva:2014fca}, with their altered pressure profiles, may offer unique insights into the presence and properties of these elusive forms of matter. The complexity of anisotropic NS interiors and the potential presence of exotic and DM particles highlight the need for continued theoretical, observational, and experimental efforts to unravel these cosmic mysteries.

The paper is organized as follows. In Section~\ref{formalism}, we briefly introduce the construction of the Nuclear Equation of State (EOS) and fermionic DM EOS to be studied here. Section~\ref{TOV} emphasizes the equation of stellar structure defining the hydrostatic equilibrium condition and anisotropic TOV equation in two fluid approaches along with the tidal love number equation. In Section~\ref{results & discussion}, we provide a detailed examination of our results with a detailed discussion considering the GW170817 data, NICER x-ray data, and all pulsars data as our observational constraints for physical observables of NSs like mass, radius, and tidal deformability in subsection~\ref{obs constr}, then in subsection~\ref{fixed g} taking a fermionic DM at the core of the NS where the baryonic matter (BM) is assumed to be anisotropic, and the DM form is taken to be isotropic\& it's validation with respect to the given observation constraints, followed by the neutron star properties with variation of the coupling strength g impacting core-halo transition in subsection ~\ref{varying g-value} and in subsection~\ref{corelation} we provide the correlation analysis of the key physical parameters involved in our admixed-NS study, using both Pearson and Kendall correlation coefficients for the considered EOS taken. Finally, in Section~\ref{conclusion}, we present our concluding remarks. 

\section{FORMALISM} \label{formalism}

\subsection{Equations of State}
\subsubsection{Nuclear EOS}

Numerous models have been suggested to describe the properties of nuclear matter in highly dense environments, particularly at densities exceeding the nuclear saturation density $n_0 \sim 0.16 \,{\rm fm}^{-3}$. These models can be based either on a Hamiltonian framework (potential models) or a Lagrangian framework (field-theoretical models).

 Below are the key characteristics and assumptions of the EOS analyzed in this study:
 
\begin{itemize}
    \item In this work, we utilize the Brussels-Montreal functional BSk22 \cite{Pearson:2018tkr}, and for comparison, details of BSk19-21 can be found in Ref.~\cite{PhysRevC.88.024308}. This nuclear EOS is parametrized primarily by fitting the measured masses of atomic nuclei with $Z$, $N \geq 8$, based on the 2012 Atomic Mass Evaluation \cite{Audi_2012}. BSk22 is characterized by its relative stiffness, allowing it to support heavy NSs in alignment with observational data.
    \item The relatively stiff MPA1 equation of state (EOS) is derived using the relativistic Dirac–Brueckner–Hartree–Fock (DBHF) formalism. This EOS incorporates the contributions to the energy arising from the exchange interactions between the $\pi$ and $\rho$ mesons \cite{Muther:1987xaa}.
    
    \item We have also considered the well-known AP3 EOS, which utilizes the Argonne 18 potential along with the inclusion of the delta-meson ($\delta$) interaction, but without incorporating the UIX three-nucleon potential, leading to a slightly stiffer EOS compared to AP4 \cite{Akmal:1998cf}.

\end{itemize}
 It should be noted that this EOS ($P_{B}(\rho_{B})$; where $\rho_{B}$ = Baryonic Energy density \& $P_{B}$ = Pressure of BM) serves as a benchmark to illustrate the effects of new physics and comes with its own constraints; future research might favor alternative EOSs \cite{doi:10.1142/S021830131330018X}.
 
\begin{figure}[] 
    \centering
    \begin{minipage}{0.48\textwidth}
        \centering
        \includegraphics[width=\linewidth]{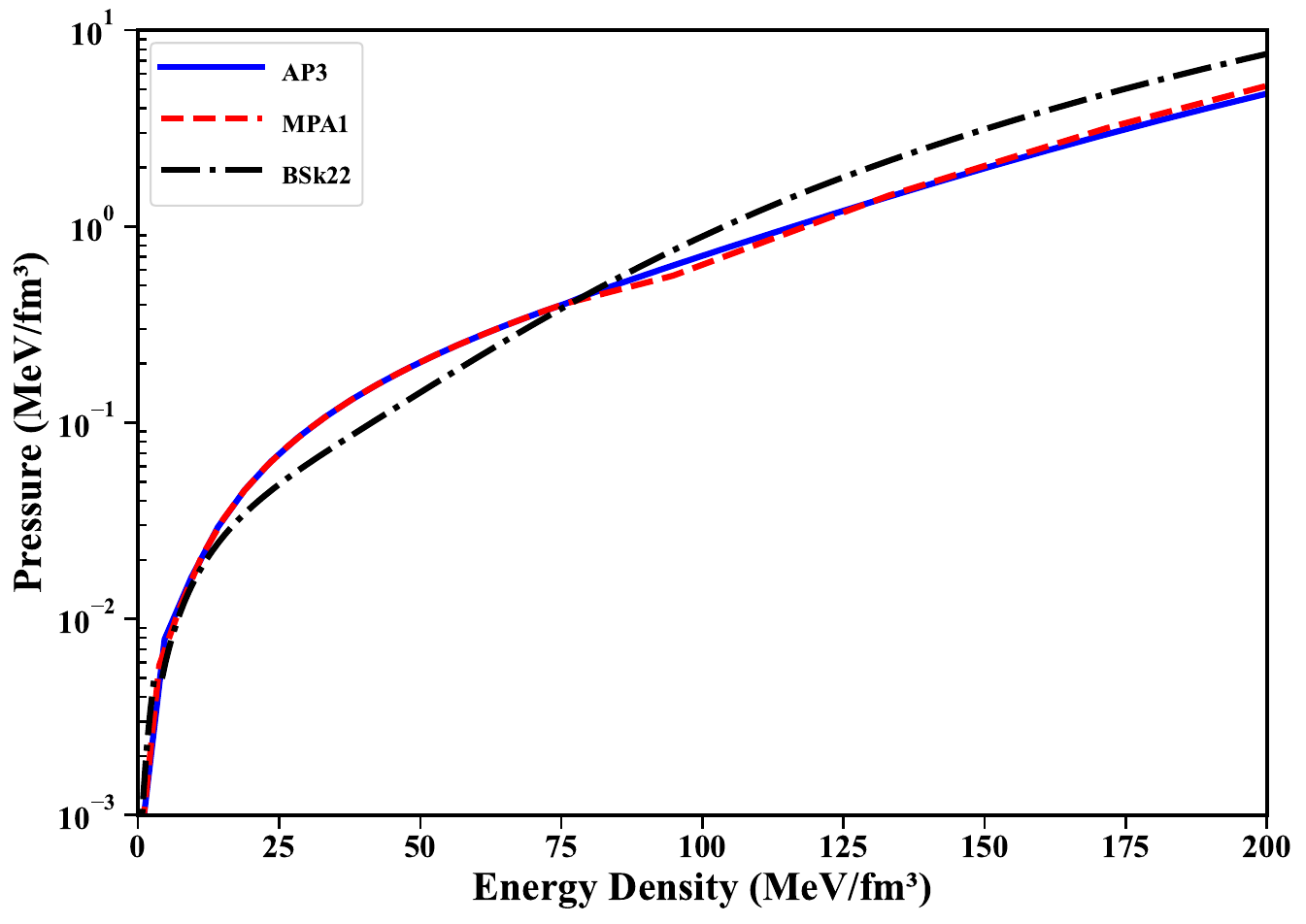}
        \caption{\justifying The plot compares the pressure (MeV/fm$^3$) as a function of energy density (MeV/fm$^3$) for three EOS models: AP3 (solid blue line), MPA1 (dashed red line), and BSk22 (dash-dotted black line), highlighting the variation in the stiffness of the isotropic BM across different models.}
        \label{fig:Press_VS_Energy}
    \end{minipage}
    \hfill
    \begin{minipage}{0.48\textwidth}
        \centering
        \includegraphics[width=\linewidth]{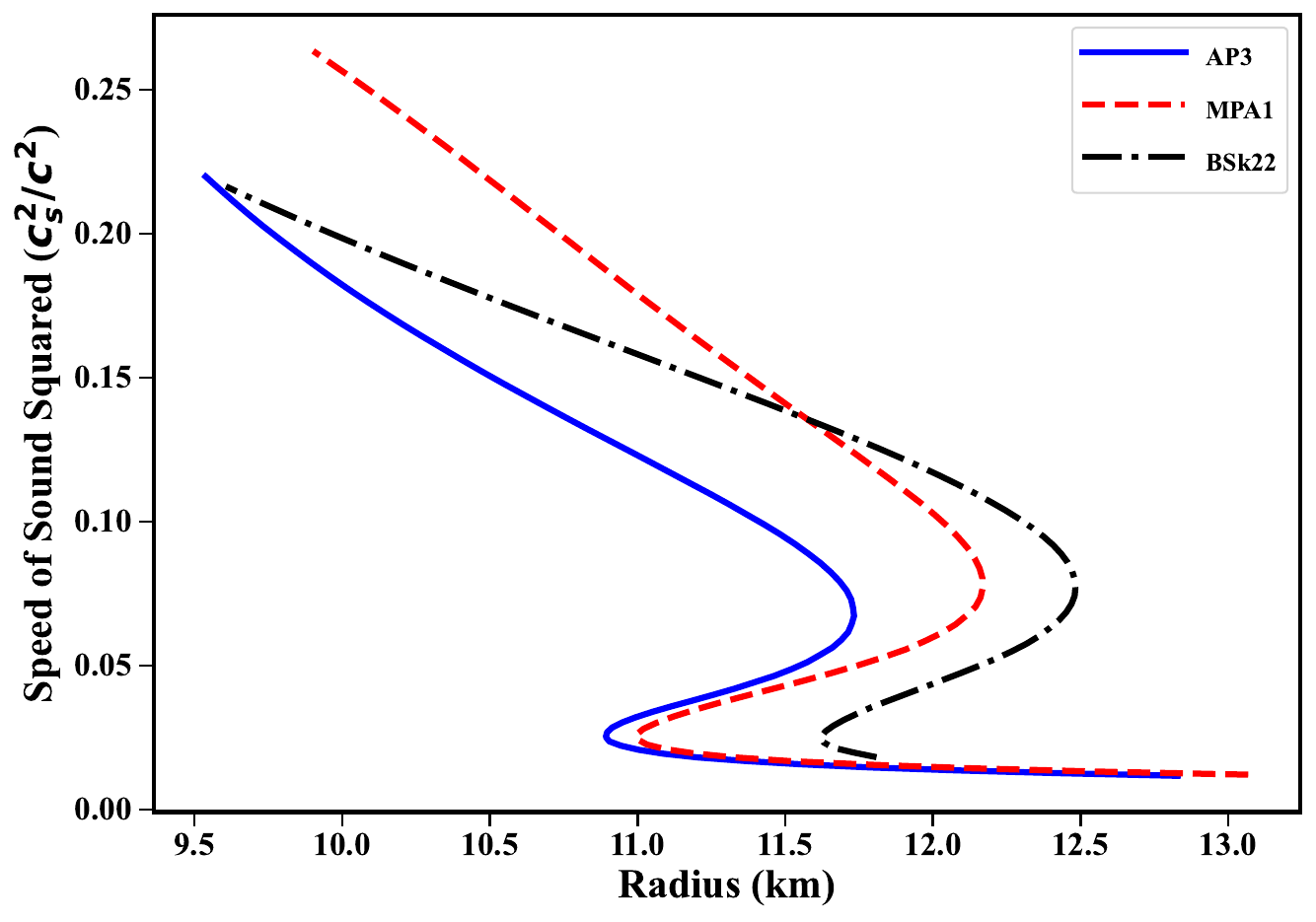}
        \caption{\justifying The profile shows the speed of sound squared normalized by the squared speed of light as a function of radius (Km) for three EOS models (as seen in Fig. \ref{fig:Press_VS_Energy}). This illustrates the differences in the speed of sound within an isotropic NS, indicating how the internal structure varies depending on the chosen EOS.}
        \label{fig:c2byc_VS_R}
    \end{minipage}
\end{figure}

 Fig. \ref{fig:Press_VS_Energy} shows the pressure (MeV/fm$^3$) as a function of the energy density (MeV/fm$^3$) for the three chosen equations of state (EOS): AP3, BSk22, and MPA1, in the context of an isotropic neutron star. These EOS models provide insights into the behavior of matter under extreme conditions, such as those found in NSs. Furthermore, we have calculated the squared speed of sound, normalized by the squared speed of light ($c^2$), to ensure that it remains below the causality limit ($c_s^2 / c^2 = 1$) for all EOS, as shown in Fig. \ref{fig:c2byc_VS_R}.

\subsubsection{Fermions with Yukawa Interactions}

To explore the effects of novel particles, we consider strongly self-interacting asymmetric fermionic dark matter with an equation of state described by \cite{Kouvaris:2015rea, Mukhopadhyay:2016dsg};
\begin{equation}
\label{eq:ferm}
    \begin{split}
       \rho_{\chi} =& \rho_{kin}(x) + \rho_{Y}(x) =  \frac{m_\chi^4}{8 \pi^2} \left( x \sqrt{1+x^2} (1 + 2 x^2) - \ln\left(x + \sqrt{1+x^2} \right) \right) + \frac{g^2 x^6 m_\chi^6}{2 (3 \pi^2)^2 m_\phi^2} \\
        P_{\chi} =& P_{kin}(x) + P_Y(x)  =  \frac{m_\chi^4}{8 \pi^2} \left( x \sqrt{1+x^2} \left( \frac{2 }{3} x^2 -1\right) + \ln\left(x + \sqrt{1+x^2} \right) \right) + \frac{g^2 x^6 m_\chi^6}{2 (3 \pi^2)^2 m_\phi^2}
    \end{split}
\end{equation}
Here $\rho_{kin}(x)$ and $\rho_Y(x) $ are the kinetic energy density and Yukawa potential (arising due to dark matter self-interaction) energy density and $P_{kin}(x)$ and $P_Y(x) $ are the kinetic and Yukawa contribution to the pressure density, respectively. Here $ x = \frac{P_{\chi}}{m_\chi }$ (where $P_{\chi}$ denotes the Fermi momentum and $m_\chi$ is the mass of the dark matter fermion). Here, we use natural units $\hbar = c = 1$. The potential arising due to the self-interaction(repulsive) of two dark matter fermions $1$ and $2$ is considered to be of Yukawa type(repulsive in nature) -  
$$V_{12} = \frac{g^2}{4 \pi} \frac{e^{-m_\phi r_{12}}}{r_{12}}$$ 
where $r_{12}$ is the distance of separation of two dark matter fermions, 
$g$ is the Yukawa coupling constant (of the dark matter and the scalar mediator interaction term $g \overline{\chi} \chi \phi$), and $m_\phi$ is the mass of the scalar mediator.
It is important to note that this EOS is consistent for repulsive self-interactions, which we consider; however, it may be inconsistent for relativistic fermions in the case of attractive self-interactions mediated by scalars~\cite{WALECKA1974491, schmitt2014introductionsuperfluidityfieldtheoretical}.

\section{TOV Equations}  \label{TOV}
\subsection{Anisotropic TOV}

We start with the action governing our system given by:
\begin{equation} \label{eq: S}
    S = S_G + S_M,
\end{equation}
where \( S_G \) represents the gravitational component, defined by the standard Einstein-Hilbert action, and \( S_M \) encapsulates the matter components, modeled as a perfect fluid with properties determined by the energy density \( \rho \), radial pressure \( P_r \), tangential pressure \( P_\perp \), and an equation of state described by \( F(\rho, P_r) = 0 \) \cite{2019EPJP..134..454L}.

By varying this action with respect to the metric, we derive the Einstein field equations (without a cosmological constant):
\begin{equation}
\label{eq: EEq}
G_{\mu\nu} = R_{\mu\nu} - \frac{1}{2}g_{\mu\nu}R = 8 \pi T_{\mu\nu},
\end{equation}
where \( G \), Newton's constant, is set to unity. The stress-energy tensor for the system is:
\begin{equation} \label{eq: Tmn}
    T^\mu_\nu = \mathrm{Diag}\left(-\rho, P_r, P_\perp, P_\perp\right).
\end{equation}

In the comoving frame, the matter content is an anisotropic fluid with energy density \( \rho \), radial pressure \( P_r \), and tangential pressure \( P_\perp \).

The general form of the metric is given by:
\begin{equation} \label{eq:g}
ds^2 = -e^\nu dt^2 + e^\lambda dr^2 + r^2 d\Omega^2,
\end{equation}
where \( \nu(r) \) and \( \lambda(r) \) are functions of the radial coordinate \( r \), and \( d\Omega^2 \equiv d\theta^2 + \sin^2\theta \, d\phi^2 \) denotes the solid angle element \cite{2007MNRAS.375.1265S}. Solving Einstein's equations, we find:
\begin{equation} \label{eq: B}
    e^{\lambda(r)} = \frac{1}{1 - \frac{2m}{r}},
\end{equation}
and
\begin{equation} \label{eq: lam}
    \nu'(r) = \frac{2m + 8 \pi r^3 P_r}{r^2 - 2mr}.
\end{equation}

The hydrostatic equilibrium equation, known as the generalized Tolman–Oppenheimer–Volkoff (TOV) equation, is given by \citep{1939PhRv...55..374O, 1939PhRv...55..364T}:
\begin{equation} \label{eq:TOV}
    \frac{dP_r}{dr} = -\frac{(\rho + P_r)(m + 4\pi r^3 P_r)}{r(r - 2m)} + \frac{2\Pi}{r},
\end{equation}
where \( \Pi \equiv P_\perp - P_r = \) pressure anisotropy. For isotropic matter, \( \Pi = 0 \), which implies \( P_\perp = P_r \). Thus, the anisotropic term vanishes, simplifying the TOV equation to describe hydrostatic equilibrium in isotropic stars. The expression for $P_\perp$ is given by,
\begin{align}
    P_\perp = P_{r} + \frac{\alpha}{3} \dfrac{(\rho + 3 P_{r}) (\rho +  P_{r}) r^2}{1 - 2m / r}
\end{align}
where the factor $\alpha$ measures the degree of anisotropy in the fluid, similar to the Bowers-Liang (BL) model of anisotropy \cite{1974ApJ...188..657B}.

Interpreting the terms in equation \eqref{eq:TOV}, we see that the hydrostatic force \( F_h = -dP_r/dr \), gravitational force \( F_g = -[(\rho + P_r)(m + 4\pi r^3 P_r)]/[r(r - 2m)] \), and anisotropic force \( F_a = 2\Pi/r \) sum to zero:
\begin{equation} \label{eq: fsum}
    F_g + F_h + F_a = 0.
\end{equation}

The mass function \( m(r) \) is related to the density by the differential equation:
\begin{equation} \label{eq: dmdr}
    \frac{dm}{dr} = 4\pi r^2 \rho.
\end{equation}

In our analysis, we assume that the interaction between BM and DM is negligible, allowing us to treat them as separate components: BM (\( B \)) and non-baryonic DM (\( \chi \)). This leads to the separation of the fluid's energy density into \( \rho = \rho_B + \rho_\chi \) and similarly for pressures \( P_r = P_{rB} + P_{r\chi} \).

Consequently, the equations \eqref{eq:TOV} and \eqref{eq: dmdr} are split into two sets for baryonic (\( B \)) and DM (\( \chi \)) components:
\begin{equation} \label{eq:TOVi}
    \frac{dP_{ri}}{dr} = -\frac{(\rho_i + P_{ri})(m + 4\pi r^3 P_r)}{r(r - 2m)} + \frac{2\Pi_i}{r},
\end{equation}
where \( \Pi_i = P_{ri} - P_{\perp i} \), and \( P_{r} = P_{rB} + P_{r \chi} \)
\begin{equation} \label{eq: dmdri}
    \frac{dm_i}{dr} = 4\pi r^2 \rho_i.
\end{equation}

Here, the subscript \( i = B \) or \( \chi \) indicates whether the quantity refers to BM or DM, respectively. For darkmatter matter, we assume isotropy, so \( P_{\perp \chi} = P_{r \chi} \) and hence \( \Pi_\chi = 0 \).

The total gravitational mass \( m(r) \) is the sum of the masses of both components:
\begin{equation}
    m(r) = m_B(r) + m_\chi(r).
\end{equation}

We solve equations \eqref{eq:TOVi} and \eqref{eq: dmdri} using appropriate central and boundary conditions. The central conditions specify the initial densities of \( B \) and \( \chi \). At the boundaries, hydrostatic equilibrium requires that the pressures vanish:
\begin{equation} \label{eq:pR}
    P_i(R_i) = 0,
\end{equation}
where \( R_B \) and \( R_\chi \) are the radii containing all the mass of baryonic and DM components, respectively. Typically, \( R_B \neq R_\chi \).

Here are two possible scenarios for a neutron star (NS) mixed with dark matter (DM):

\begin{enumerate}
    \item \textbf{DM Condensed in the NS Core}: \\
    In this case, the dark matter is concentrated within the core of the neutron star, resulting in the radius of the DM component ($R_{\chi}$) being smaller than the radius of the baryonic matter (BM) component ($R_{B}$), meaning $R_{B} > R_{\chi}$.

    \item \textbf{DM Forming an Extended Halo}: \\
    Alternatively, the dark matter can form a halo around the neutron star, where the radius of the DM component ($R_{\chi}$) exceeds that of the baryonic matter component ($R_{B}$), so $R_{\chi} > R_{B}$.
\end{enumerate}

These configurations are valid when the interaction between baryonic matter and dark matter is purely gravitational, allowing for both components to mix within the core and possibly leading to one component extending into the outer regions of the combined object.

Since only the BM is directly observable, we identify the observed radius \( R \) of the neutron star with \( R_B \). This allows our results to be compared with observational data, where \( R \) is the visible radius of the neutron star, and the additional mass is due to the DM component. 

The total gravitational mass and the DM subfraction within the neutron star are given by:
\begin{equation} \label{eq:Mt}
    M_T = M_B(R_B) + M_\chi(R_\chi),
\end{equation}
and
\begin{equation} \label{eq:fchi}
    f_\chi = \frac{M_\chi(R_\chi)}{M_T}.
\end{equation}

By varying the central densities of baryonic and DM, we can explore different combinations of total gravitational mass \( M_T \) and neutron star radius \( R \) for a given DM subfraction \( f_\chi \).

\subsection{Tidal Deformability} \label{tidef}

Apart from mass and radius, the tidal deformability of NSs plays a pivotal role in defining their structural attributes. During the final phases of inspiral in a binary neutron star system, the gravitational field exerted by one neutron star onto its companion induces quadrupole deformations. These deformations, resulting from tidal forces, quantify the extent of distortion experienced by the neutron star, known as ``tidal deformability". This parameter provides insights into how much the neutron star is affected by the gravitational forces exerted by its companion.

The tidal deformability parameter $\lambda$ can be expressed as \cite{Hinderer_2010,PhysRevD.109.083007,Malik_2018},
\begin{equation}
\lambda = - \frac{Q_{ij}}{ {\cal E}_{ij}},
\end{equation}
where $Q_{ij}$ represents the components of the induced quadrupole moment tensor and ${\cal E}_{ij}$ denotes the components of the tidal field tensor. In terms of the Love  number $k_2$, the  mass normalized dimensionless tidal deformability parameter is given by
\begin{equation}
\Lambda \equiv \frac{\lambda}{M^{5}} = \frac{2}{3}k_2\left(\frac{R}{M}\right)^{5}
\equiv \frac{2}{3}k_2 C^{-5} ,
\label{eq:Lambda}
\end{equation}
where $R$ and $M$ are the radius and mass of the star, and $C \equiv M/R$ is its compactness.

The tidal Love number $k_2$ depends on the underlying EOS of the star, and it can be expressed in terms of the dimensionless compactness parameter $C$ as~\cite{Damour_2009, Flanagan_2008},
\begin{align}
k_2 &= \frac{8C^5}{5}\left(1-2C\right)^2
\left[2+2C\left(y_R-1\right)-y_R\right] \times \big\{ 2C\left[6-3 y_R+3 C(5y_R-8)\right] \nonumber\\
& +4C^3\left[13-11y_R+C(3 y_R-2) + 2 C^2(1+y_R)\right] +3(1-2C)^2\left[2-y_R+2C(y_R-1)\right]\ln\left(1-2C\right) \big\}^{-1}.
\label{eq:k2}
\end{align}
 The function $y_R \equiv y(r)|_{r=R}$ is related to the metric perturbation and satisfies the following differential equation:
\begin{align}
r \frac{d y(r)}{dr} + {y(r)}^2 + y(r) F(r) + r^2 Q(r) = 0
\label{TidalLove2} ,
\end{align}

with
\begin{align*}
F(r) &= \frac{r-4 \pi r^3 \left( \varepsilon(r) - P(r)\right) }{r-2 m(r)}, \\
Q(r) &= \frac{4 \pi r \left(5 \varepsilon(r) +9 P(r) + \sum\limits_{i} \frac{\varepsilon_{i}(r) + P_{i}(r)}{\partial P_{i}(r)/\partial \varepsilon_{i}(r)} - \frac{6}{4 \pi r^2}\right)}{r-2m(r)} \quad -  4\left[\frac{m(r) + 4 \pi r^3 P(r)}{r^2\left(1-2m(r)/r\right)}\right]^2.
\end{align*}

For a two-fluid system composed of DM and BM, the parameters $\epsilon$, $p$ and $M$ are defined as
\begin{eqnarray}
p=\sum_{i} p_{i},\ \  \epsilon=\sum_{i} \epsilon_{i}, \ \ M=\sum_{i} M_{i}, \ \ \text{i=BM, DM}.~~~~~
\end{eqnarray}
To calculate the Love number and the tidal deformability parameter $\Lambda$ for a spherically symmetric star, one can solve Eq. (\ref{TidalLove2}) and the Tolman-Oppenheimer-Volkoff (TOV) equations (Eqs. \eqref{eq:TOVi}-\eqref{eq: dmdri}) simultaneously. This process involves utilizing boundary conditions such as $P(0) = P_{c}$ and $m(0) = 0$, along with $y(0) = 2$ obtained from the perturbative expansion of the deformed metric up to the second order.

\section{Results and discussion} \label{results & discussion}

The introduction of additional degrees of freedom to a celestial body can alter its physical characteristics. Without exploring the specific production mechanisms, we will examine the properties of DNSs under varying degrees of anisotropy using three different EOSs (AP3, BSk22, MPA1). These DNSs incorporate a DM particle with a mass of 1 GeV, interacting through a light mediator particle with a mass of 1 keV, governed by a coupling constant of $g=10^{-5}$. The detection of gravitational waves from GW170817 has constrained the DM fraction to $10\%$  for soft EOS and $1\%$ for stiff nuclear EOS \cite {ellis2018search, barbat2024comprehensive}.  In line with this, and as highlighted in \cite{PhysRevD.97.123007, Collier:2022cpr}, a $5\%$ DM subfraction can significantly influence the interpretation of NS properties. Consequently, we examine a $5\%$ DM subfraction alongside two additional subfractions $1\%$, $0.1\%$, and compared them with the no DM scenario. The anisotropic parameter is varied from -2 to 2 in increments of $0.25$. In our analysis, we evaluate $M_{\text{max}}$, $R_{\text{max}}$, $R_{1.4}$, and $\Lambda_{1.4}$, and compare the resulting parametric values with observational constraints as shown in \ref{obs constr}.

\subsection{Observational Constraints}  \label{obs constr}

We compare the prediction of the model with recent observational data from various sources to validate our findings. Table \ref{tab:my-table} summarizes these constraints, which include measurements from the GW170817 event \cite{Nathanail:2021tay}, NICER X-ray observations \cite{Miller:2019cac, Riley:2019yda, Miller:2021qha, Riley:2021pdl}, and mass-radius data from specific pulsars \cite{Zhao:2016rfv, Takisa:2014sva, Romani_2022}. Our analysis adheres to these observational constraints to ensure that the proposed neutron star configurations, particularly those involving DM, remain consistent with empirical data.

\begin{table}[]
\caption{\justifying Observational constraints}
\label{tab:my-table}
\renewcommand{\arraystretch}{2}
\setlength{\tabcolsep}{40pt}
\noindent\rule{\linewidth}{0.5mm}
\begin{tabular}{lccc}
\textbf{Measurements} & \textbf{Radius} \textbf{(Km)}& \textbf{Mass} \textbf{(M$_{\bm\odot}$)}& \textbf{Ref.} \\ \hline
\renewcommand{\arraystretch}{2} 
                                                               & $13.02^{+1.24}_{-1.06}$               & $1.44^{+0.15}_{-0.14}$                    & \cite{Miller:2019cac}    \\
\multirow{-2}{*}{\textbf{PSR J0030+0451}}                      & $12.71^{+1.14}_{-1.19}$               & $1.34^{+0.15}_{-0.16}$                    & \cite{Riley:2019yda}    \\
\multirow{2}{*}{\textbf{PSR J0740+6620}}                       & $13.7^{+2.6}_{-1.5}$                  & $2.08\pm0.07$                             & \cite{Miller:2021qha}    \\
                                                               & $12.39^{+1.30}_{-0.98}$               & $2.072^{+0.067}_{-0.066}$                 & \cite{Riley:2021pdl}    \\
\textbf{PSR J0348+0432}                                        & $12.957\sim12.246$                    & $2.01\pm0.04$                             & \cite{Zhao:2016rfv}    \\
\textbf{PSR J1614-2230}                                        &  $10.30\sim9.67$                                     & $1.97\pm0.04$                             & \cite{Takisa:2014sva}    \\
\textbf{PSR J0952-0607}                                        &  $12.96\sim13.39$                                     & $2.35\pm0.17$                             & \cite{Romani_2022}    \\ \hline
\multirow{2}{*}{\textbf{GW170817}}                             & \textbf{Radius} \textbf{(Km)}         & $\bm\Lambda_{\bm 1.36}$                   &            \\
                                                               & $10.62-12.83$                         & $720$                                     &\multirow{-2}{*}{\cite{Nathanail:2021tay}} \\
\end{tabular}
\noindent\rule{\linewidth}{0.5mm}
\end{table}

\subsection{Neutron Stars properties with various DM subfraction for a fixed value of $g$} \label{fixed g}

In this subsection, we examine how varying subfractions of DM within the NS, specifically at its core, influence its physical properties. In the two-fluid model, we consider the visible radius $R_B$ as the star's outer radius, while the DM radius $R_{\chi}$ lies within the BM fluid. We analyze the effects of DM fractions of 5\%, 1\%, 0.1\%, and 0\% across three different EOS models (AP3, BSk22, and MPA1), with the anisotropic parameter $\alpha$ ranging from $-2$ to $+2$ (where $\alpha = 0$ represents the isotropic BM case).

\begin{figure}[]
\centering
\begin{minipage}[b]{0.49\linewidth}
  \centering
  \includegraphics[width=\linewidth]{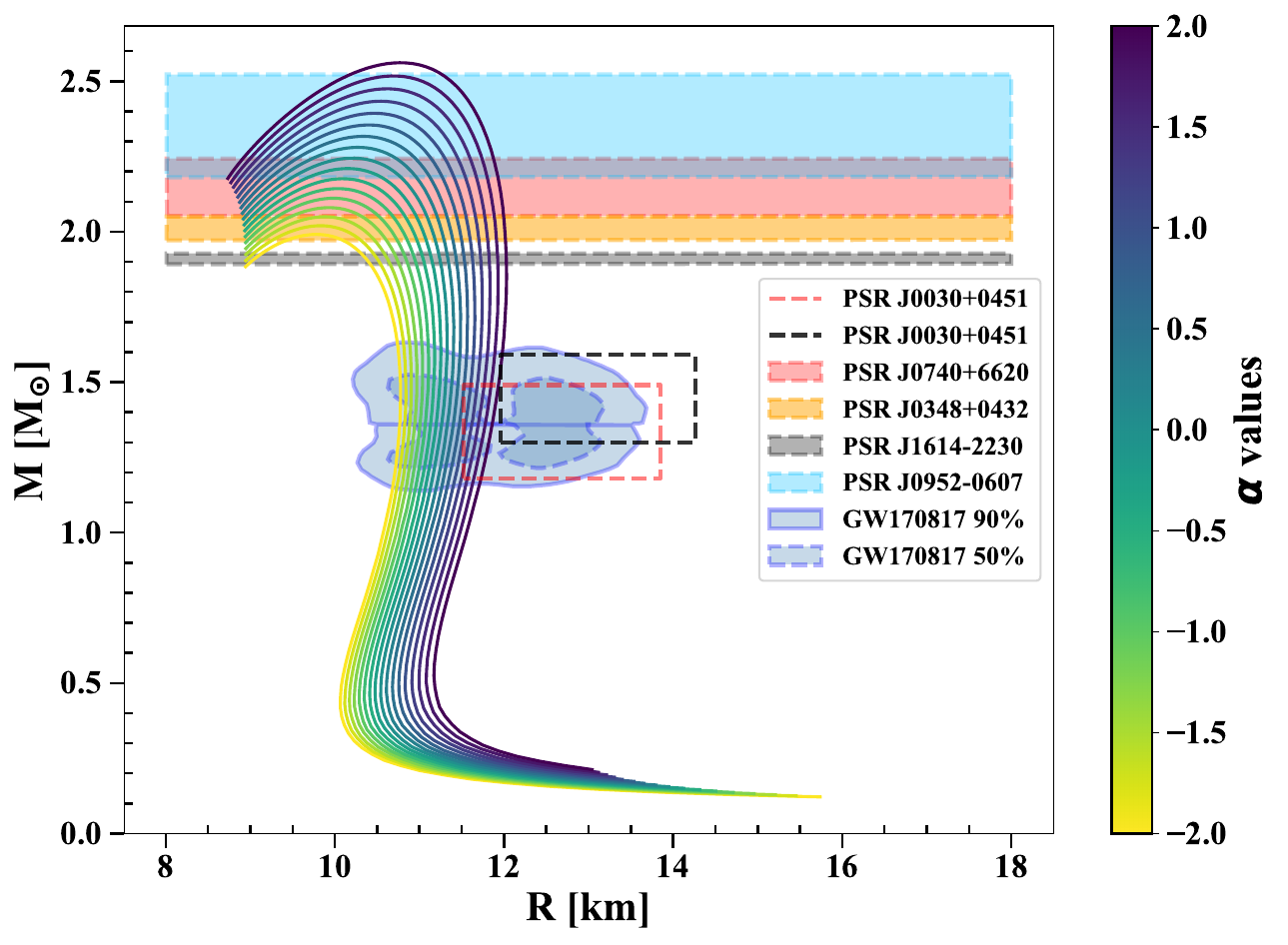} 
  \caption*{(a) $5\%$ DM}
 \label{fig:AP3_RvM5DM}
\end{minipage}%
\hfill
\begin{minipage}[b]{0.49\linewidth}
  \centering
  \includegraphics[width=\linewidth]{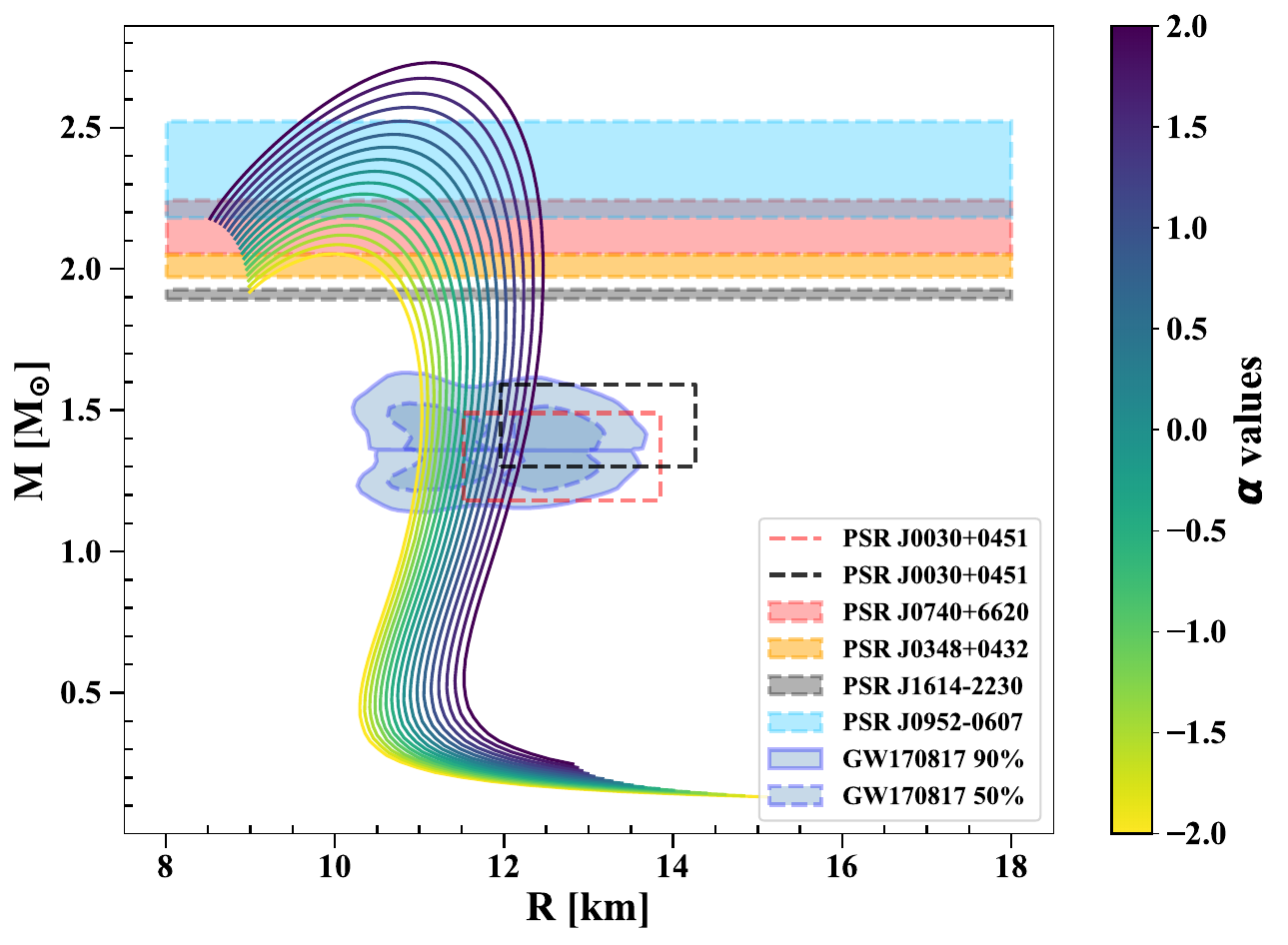}
  \caption*{(b) $1\%$ DM}
  \label{fig:AP3_RvM1DM}
\end{minipage}

\vspace{0.2cm} 

\begin{minipage}[b]{0.49\linewidth}
  \centering
  \includegraphics[width=\linewidth]{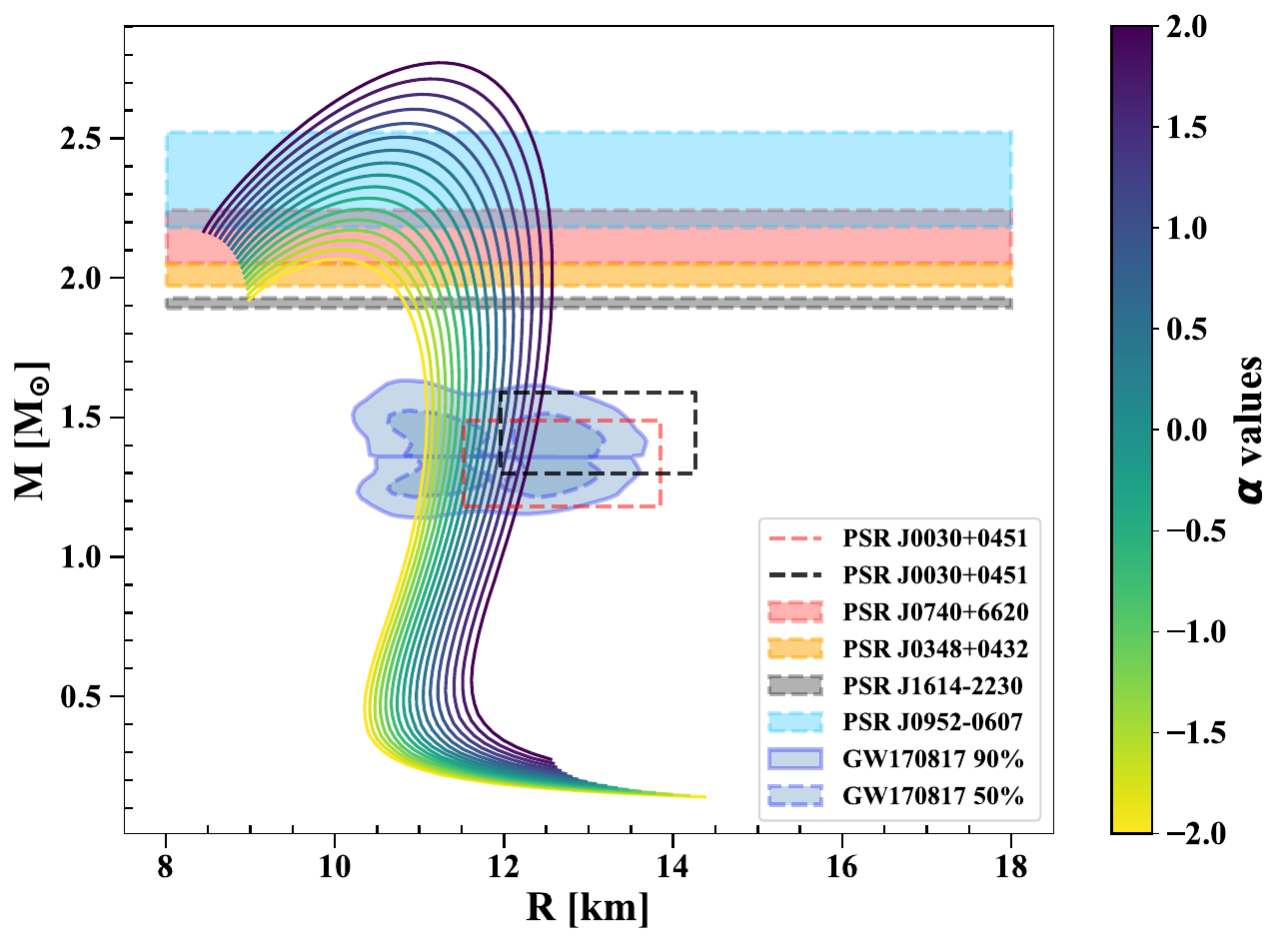}
  \caption*{(c) $0.1\%$ DM}
  \label{fig:AP3_RvM0p1DM}
\end{minipage}%
\hfill
\begin{minipage}[b]{0.49\linewidth}
  \centering
  \includegraphics[width=\linewidth]{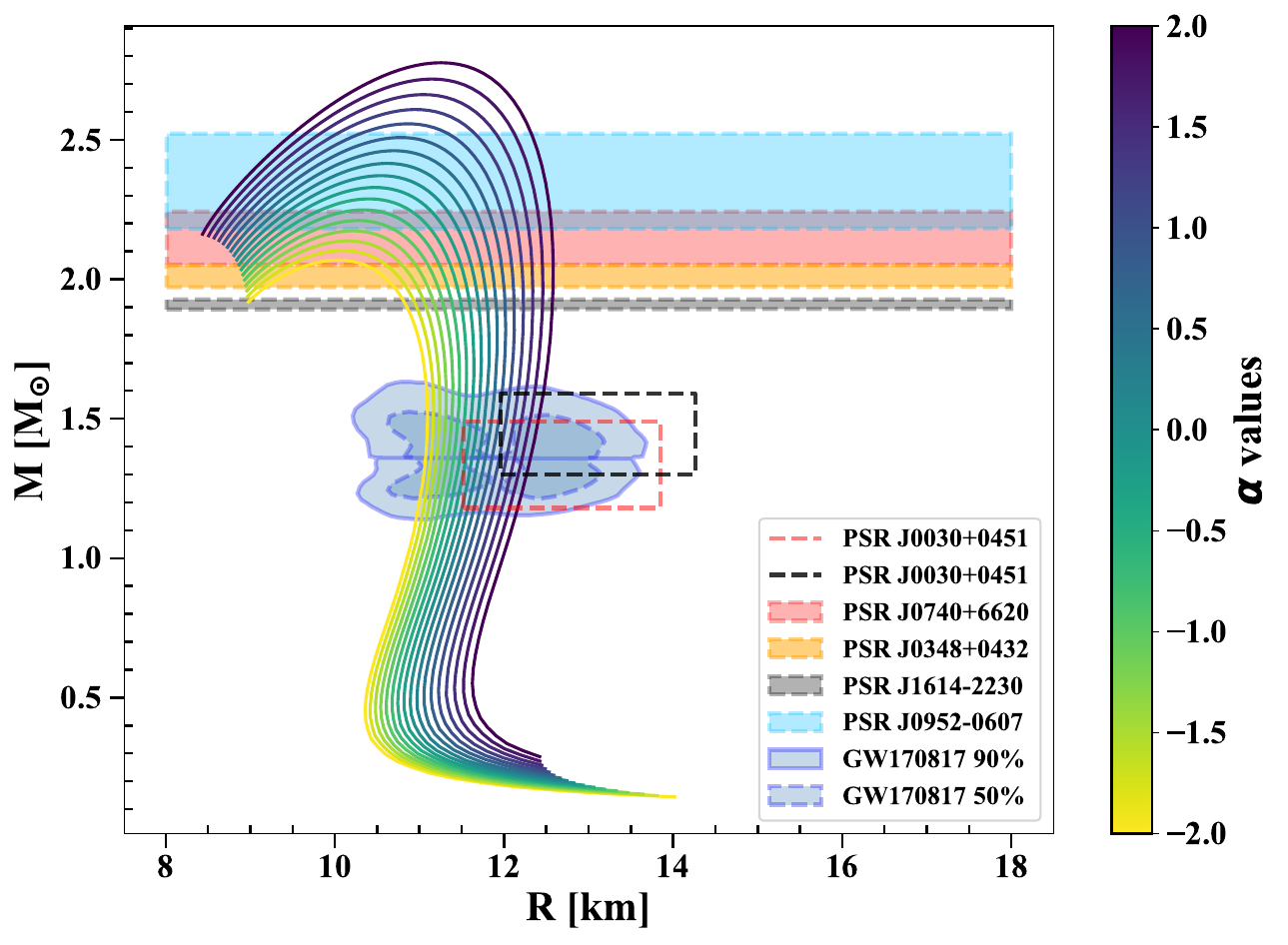}
  \caption*{(d) No DM}
  \label{fig:AP3_RvM0DM}
\end{minipage}
\caption{\justifying Mass-radius stability profile for DM-admixed anisotropic NS with $g=10^{-5}$, $m_\chi=1$ GeV, and $m_\phi=1$ KeV using the AP3 EOS. The color contour highlights the range of $\alpha$-values from $-2$ to $2$. The subplots illustrate different DM fractions: (a) $5\%$ DM, (b) $1\%$ DM, (c) $0.1\%$ DM, and (d) No DM. Observational constraints are shown with deep sky blue (PSR J0952-0607), red (PSR J0740+6620), orange (PSR J0348+0432), black (PSR J1614-2230), steel-blue solid patch (GW170817 90\%), steel-blue dashed patch (GW170817 50\%), red and black dashed line for PSR J0030+0451.}
\label{fig;AP3_MvR}
\end{figure}


As shown in Fig. \ref{fig;AP3_MvR} and Table \ref{tab: AP3 MvR}, the mass-radius stability profiles of DM-admixed anisotropic NSs, characterized by parameters $g=10^{-5}$, $m_\chi=1$ GeV, and $m_\phi=1$ KeV, are governed by the AP3 EOS. It is observed that increasing the value of $\alpha$ leads to an increase in the maximum masses and their corresponding radii, and vice-versa. The corresponding subplots highlights different fractions of DM, 5\% (Fig. 3 (a)), 1\% (Fig. 3 (b)), 0.1\% (Fig. 3 (c)) and no DM 
(Fig. 3 (d)). In the no DM case, only negative values of $\alpha= (-2, -1)$ satisfy the observational constraints, such as the $M_{max}$ values from all pulsar data and the $R_{1.4}$ region provided by the GW170817 and NICER x-ray data. As the DM subfraction increases from 0.1\% to 1\% and 5\%, significant changes occur, with the constraints being satisfied for $\alpha= (-2, -1)$, $(-2, -1, 0)$, and $(-2, -1, 0)$, respectively. Increasing the $\alpha$ value results in stiffer NSs while increasing the DM subfraction softens them.

\begin{figure}[h!]
\centering
\begin{minipage}[b]{0.49\linewidth}
  \centering
  \includegraphics[width=\linewidth]{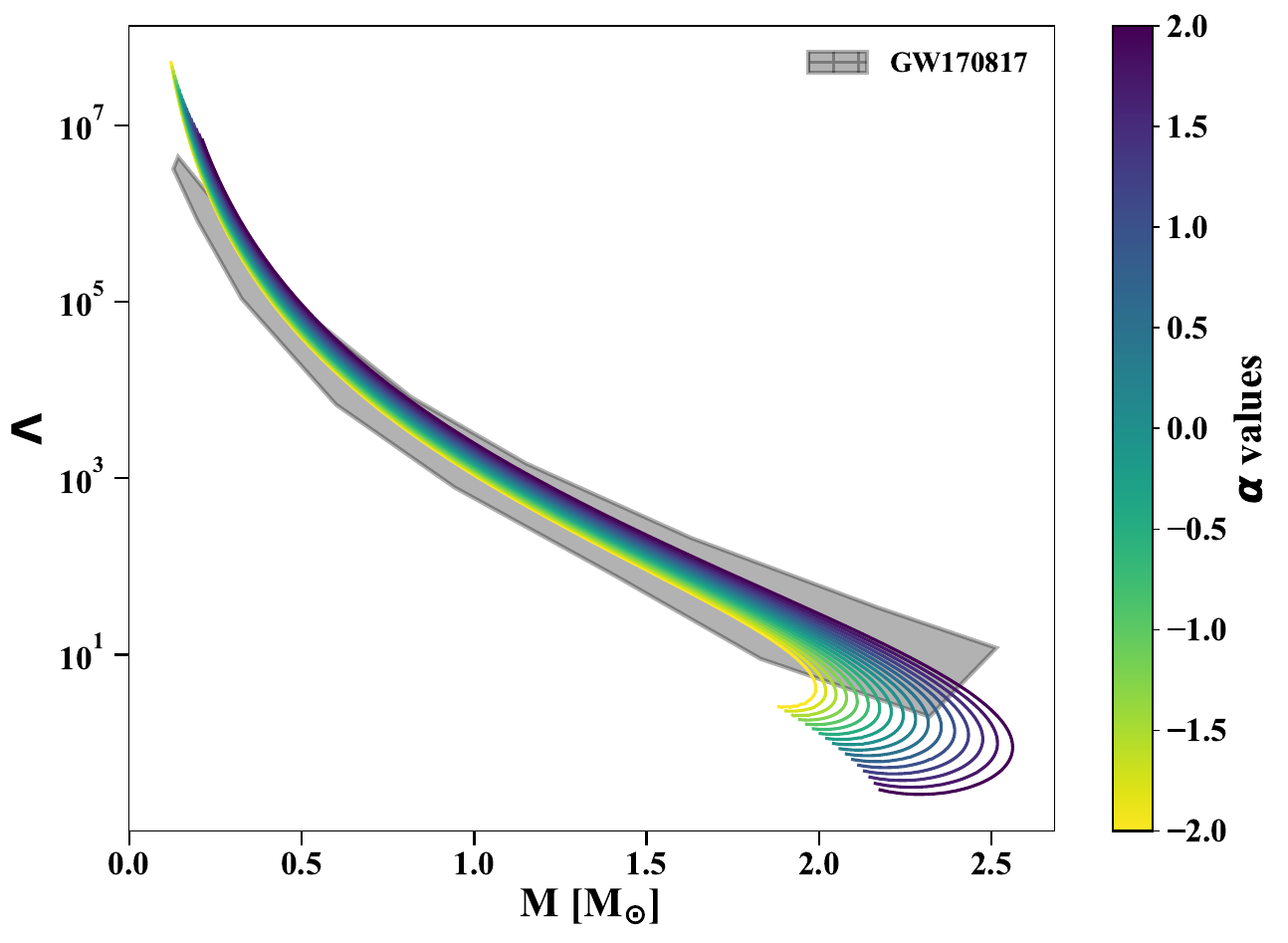}
  \caption*{(a) $5\%$ DM}
  \label{fig:AP3_TDvM5DM}
\end{minipage}%
\hfill
\begin{minipage}[b]{0.49\linewidth}
  \centering
  \includegraphics[width=\linewidth]{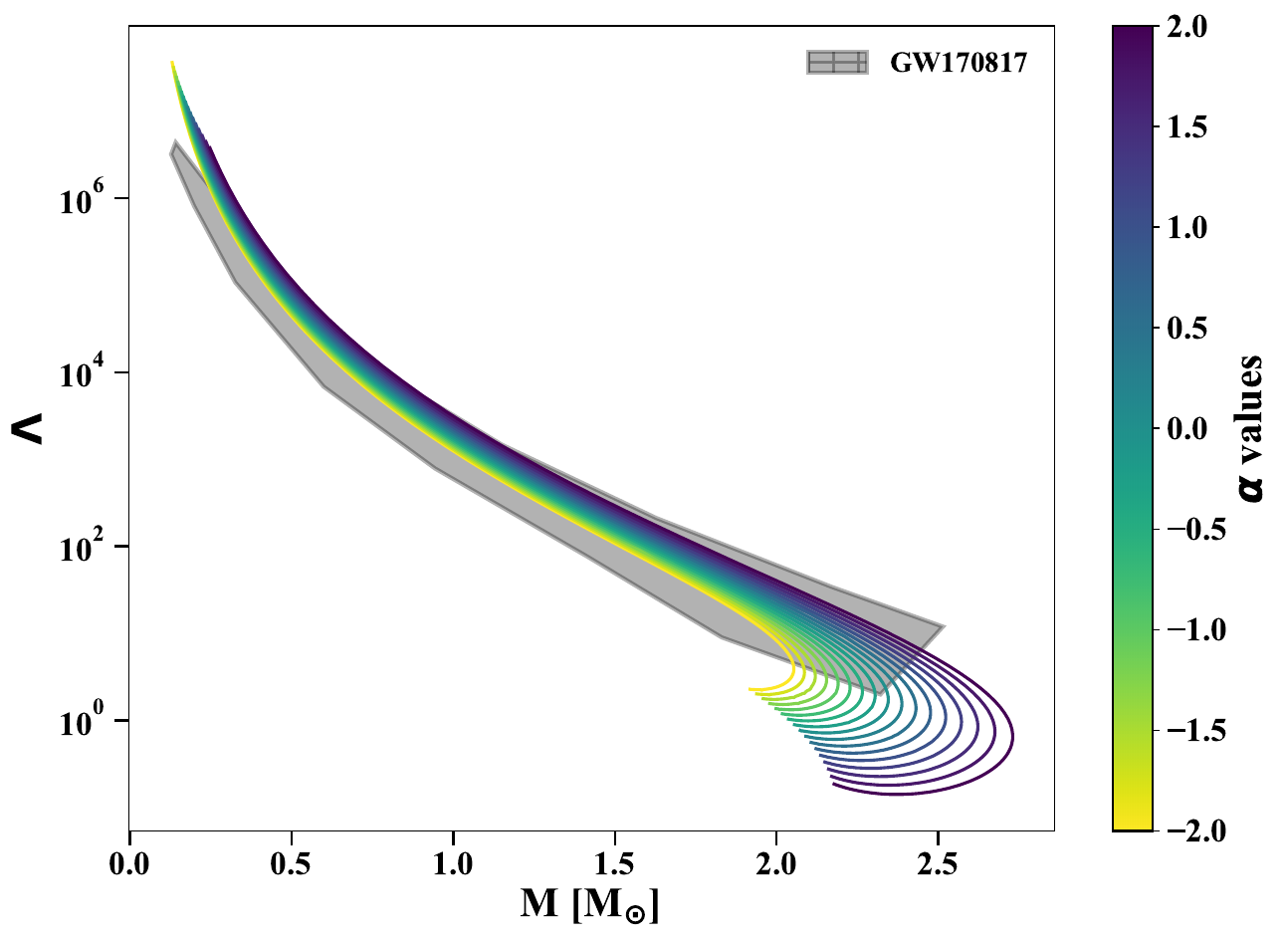}
  \caption*{(b) $1\%$ DM}
  \label{fig:AP3_TDvM1DM}
\end{minipage}

\vspace{0.2cm} 

\begin{minipage}[b]{0.49\linewidth}
  \centering
  \includegraphics[width=\linewidth]{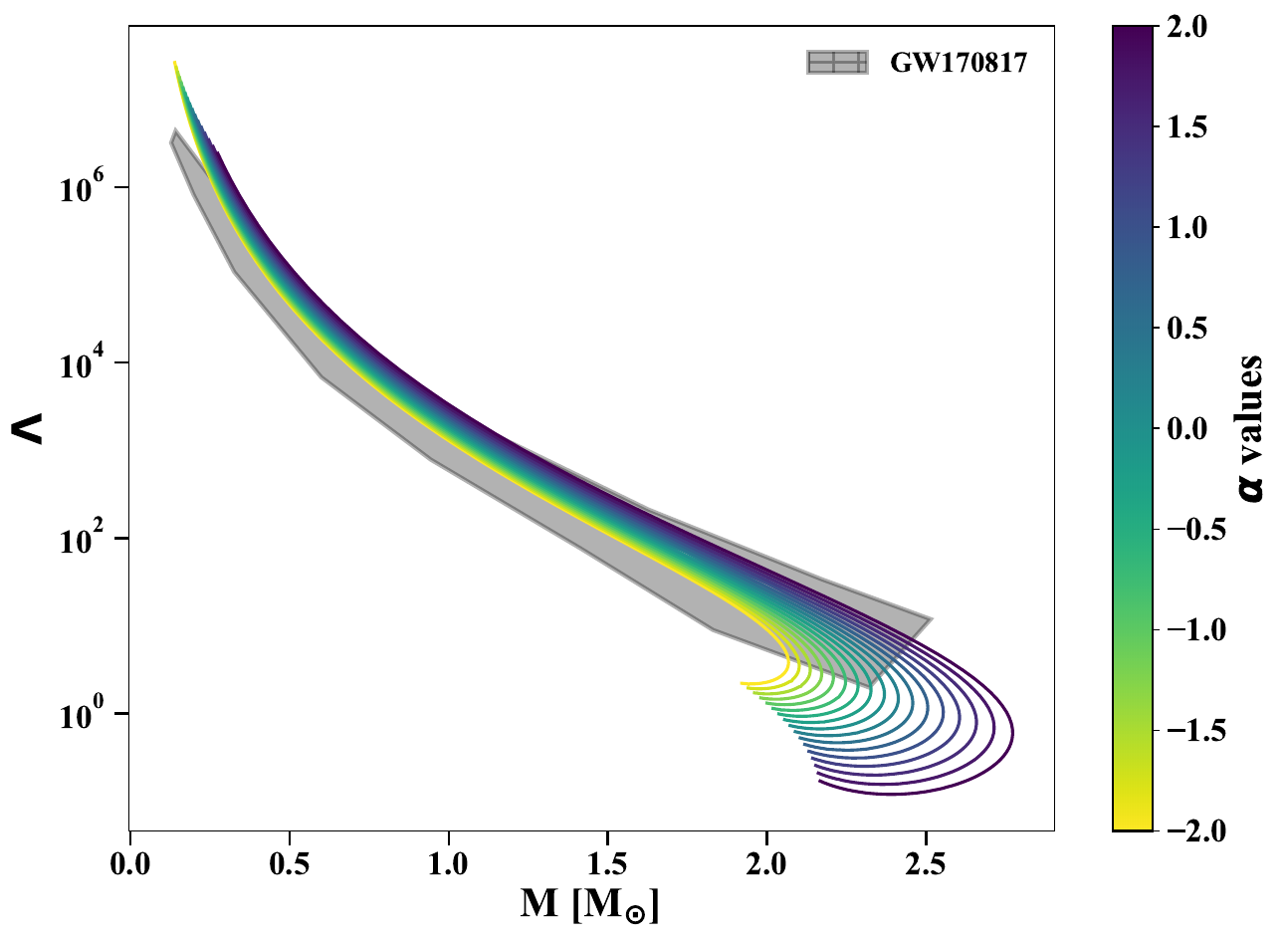}
  \caption*{(c) $0.1\%$ DM}
  \label{fig:AP3_TDvM0p1DM}
\end{minipage}%
\hfill
\begin{minipage}[b]{0.49\linewidth}
  \centering
  \includegraphics[width=\linewidth]{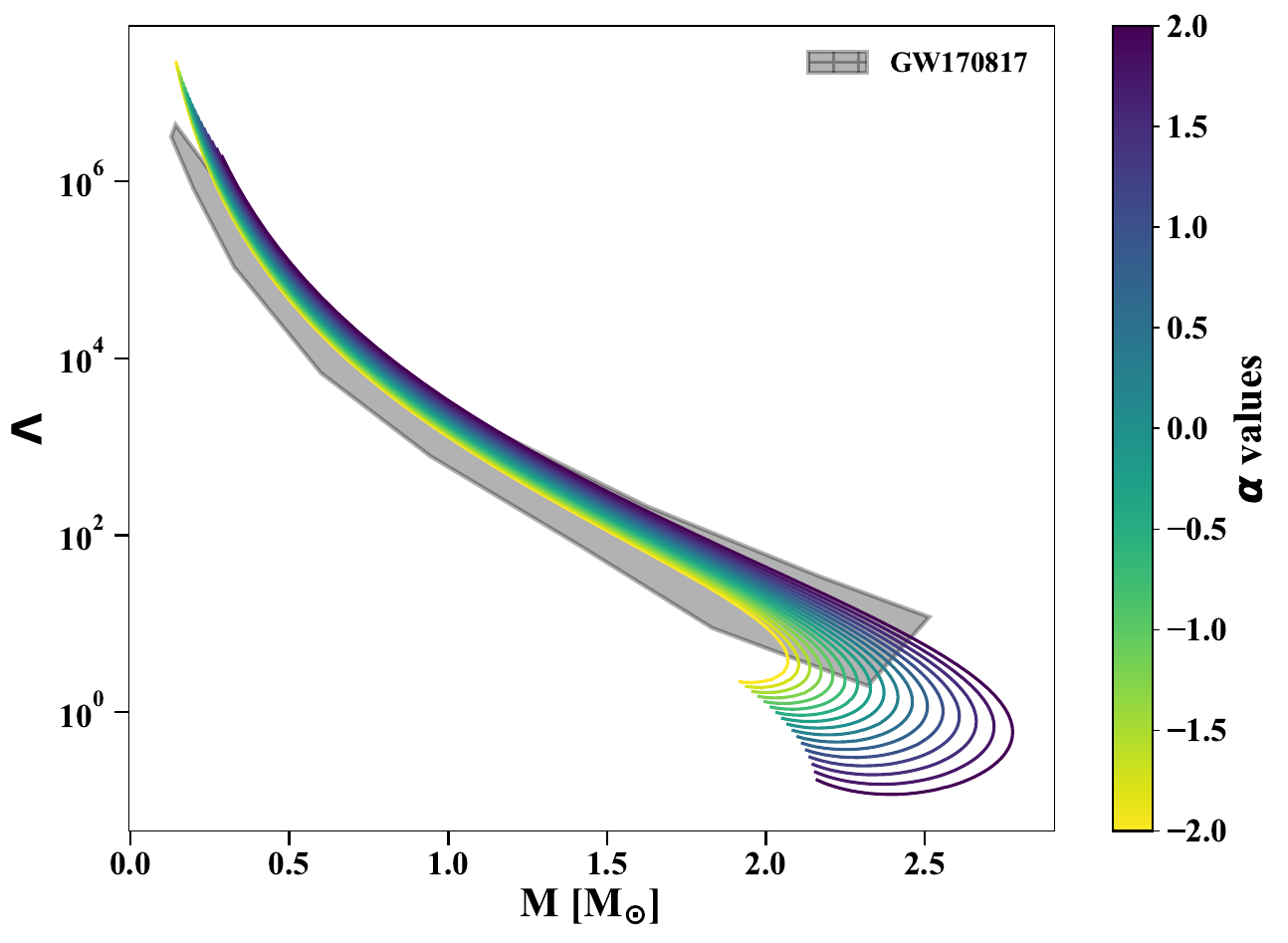}
  \caption*{(d) No DM}
  \label{fig:AP3_TDvM0DM}
\end{minipage}

\caption{\justifying Tidal deformability-mass stability profile for DM-admixed anisotropic NS with $g=10^{-5}$, $m_\chi=1$ GeV, and $m_\phi=1$ KeV using the AP3 EOS. The color contour highlights the range of $\alpha$-values from $-2$ to $2$. The subplots illustrate different DM fractions: (a) $5\%$ DM, (b) $1\%$ DM, (c) $0.1\%$ DM, and (d) No DM. The GW170817 constraint is shown in a black shape.}
\label{AP3_TDvM}
\end{figure}

Tidal deformability is an important physical parameter to define NS properties. We get the value of $\Lambda_{1.4}$ for different $\alpha$ values get reduced corresponding to increased values of DM subfraction from no DM, 0.1\%, 1\%, and 5\% suggesting that DM induces a reduction in tidal deformation between two coalescence bodies as the radius is squeezed as shown in Fig. \ref{AP3_TDvM}. All these results are within the given $\Lambda_{1.36} = 720$ value, defined by aLIGO detector for GW170817.
Similarly, we perform this analysis for BSk22 and MPA1 EOS and get significant changes for \textbf{BSk22} EOS, which has some typical behaviour while increasing (or decreasing) the DM fractions as mentioned below. From Fig. \ref{BSk22_MvR} and table \ref{tab: BSk22 MvR}, the mass-radius profile for no DM case is found to be only fit for $\alpha = -1$ satisfying all the given observational GW170817 and NICER x-ray $R_{1.4}$ region and all Pulsars maximum mass $M_{max}$ data which wasn't satisfied for \textit{isotropic} case($\alpha = 0$) being a stiff EOS. While increasing the DM subfraction from 0.1\%,  1\%, and  5\% a substantial change happens to the $\alpha$ values (like $(-1, 0), (-1,0)$ and $(-1, 0, +1)$ respectively) which satisfy all the constraints. Similarly, the dimensionless tidal deformability value $\Lambda_{1.4}$ which satisfied the $\Lambda_{1.36} \simeq 720$  of  GW170817 data irrespective of the $\alpha$ values with different DM fractions for AP3 EOS, doesn't satisfy the same for BSk22 EOS with no DM case. But, introducing a DM core with 5 \% DM subfraction to the single fluid anisotropic baryonic model depicts that all the $\alpha$ values ($-2$ to $+2$ ) satisfy the aLIGO band as shown in Fig. \ref{Bsk22_TDvM}.
\begin{table}[]
\caption{\justifying Maximum mass $M_{\max}$ (in $M_\odot$), maximum radius $R_{\max}$ (in km), the radius at 1.4 $M_{\odot}$, i.e., $R_{1.4}$ (in km), and dimensionless tidal deformability at 1.4 $M_{\odot}$, i.e., $\Lambda_{1.4}$ for \textbf{AP3} EOS as a function of anisotropic parameter $\alpha$ for different DM fractions with $g=10^{-5}$, $m_\chi=1$GeV, and $m_\phi=1$KeV.}
\label{tab: AP3 MvR}
\renewcommand{\arraystretch}{1.5}
\setlength{\tabcolsep}{11pt}
\begin{threeparttable}
\begin{tabular}{|c|cccccccc|}
\hline\hline
\multirow{2}{*}{$\bm\alpha$} & \textbf{M$\bm_{max}$} & \textbf{R$\bm_{max}$} & \textbf{R$\bm_{1.4}$} & \textbf{$\bm\Lambda_{1.4}$} & \textbf{M$\bm_{max}$} & \textbf{R$\bm_{max}$} & \textbf{R$\bm_{1.4}$} & \textbf{$\bm\Lambda_{1.4}$} \\ \cline{2-9} 
                                       & \multicolumn{4}{c}{\textbf{5\% DM}}                                             & \multicolumn{4}{c|}{\textbf{1\% DM}}                                             \\ \hline\hline
\textbf{-2}                                  & 1.990922$^{\bm a}$   & 9.792874   & 10.77445$^{\bm b}$   & \multicolumn{1}{c||}{145.3217$^{\bm b}$}      & 2.053282$^{\bm a}$   & 9.973753   & 11.01953$^{\bm b}$   & 171.535$^{\bm b}$           \\
\textbf{-1}                                     & 2.111048$^{\bm a}$   & 9.978438   & 11.02011$^{\bm b}$   & \multicolumn{1}{c||}{175.6372$^{\bm b}$}      & 2.19029$^{\bm a}$   & 10.2095    & 11.28282$^{\bm b}$   & 210.1295$^{\bm b}$          \\
\textbf{0}                                      & 2.244332$^{\bm a}$   & 10.23414   & 11.28083$^{\bm b}$   & \multicolumn{1}{c||}{217.2828$^{\bm b}$}      & 2.345324$^{\bm a}$   & 10.47342   & 11.57057$^{\bm b}$   & 261.9516$^{\bm b}$          \\
\textbf{1}                                      & 2.393476   & 10.45414   & 11.56579$^{\bm b}$   & \multicolumn{1}{c||}{270.6123$^{\bm b}$}                & 2.522969   & 10.76735   & 11.88777$^{\bm b}$   & 339.553$^{\bm b}$           \\
\textbf{2}                                      & 2.561424   & 10.76539   & 11.88496$^{\bm b}$   & \multicolumn{1}{c||}{353.4935$^{\bm b}$}                & 2.729768   & 11.19024   & 12.24607$^{\bm b}$   & 455.6631$^{\bm b}$          \\ \hline\hline
             $\bm\alpha$    & \multicolumn{4}{c}{\textbf{0.1\% DM}}                                           & \multicolumn{4}{c|}{\textbf{0\% DM}}                                             \\ \hline\hline
\textbf{-2}                                     & 2.067496$^{\bm a}$   & 10.01782   & 11.07719$^{\bm b}$   & \multicolumn{1}{c||}{178.1355$^{\bm b}$}    & 2.069149$^{\bm a}$  & 10.04622   & 11.08376$^{\bm b}$   & 178.5406$^{\bm b}$        \\
\textbf{-1}                                     & 2.20847$^{\bm a}$    & 10.25232   & 11.34663$^{\bm b}$   & \multicolumn{1}{c||}{220.2592$^{\bm b}$}    & 2.210571$^{\bm a}$   & 10.2821    & 11.3539$^{\bm b}$    & 221.2384$^{\bm b}$        \\
\textbf{0 }                                     & 2.368799   & 10.51242   & 11.63955$^{\bm b}$   & \multicolumn{1}{c||}{276.158$^{\bm b}$ }          & 2.371583   & 10.54436   & 11.64559$^{\bm b}$   & 278.5179$^{\bm b}$        \\
\textbf{1}                                      & 2.55387    & 10.8849    & 11.96126$^{\bm b}$   & \multicolumn{1}{c||}{353.5615$^{\bm b}$}          & 2.557372   & 10.83569   & 11.96985$^{\bm b}$   & 359.3642$^{\bm b}$        \\
\textbf{2}                                      & 2.771613   & 11.22083   & 12.33604$^{\bm b}$   & \multicolumn{1}{c||}{478.4485$^{\bm b}$}          & 2.776461   & 11.2587    & 12.34343$^{\bm b}$   & 487.6149$^{\bm b}$        \\ \hline\hline
\end{tabular}
\begin{tablenotes}
    \item[$\bm a$] Satisfying PSR (1.97-2.35).
    \item[$\bm b$] Satisfying GW170817 ($R_{1.4} \approx 10.62-12.38$ and $\Lambda_{1.36}\leq720$).
\end{tablenotes}
\end{threeparttable}
\end{table}

Considering MPA1 EOS for no DM case we get a typical behavior of mass-radius profile, being stiff EOS for \textit{isotropic} case, i.e. $\alpha = 0$ the maximum mass value doesn't satisfy the given pulsars $M_{max}$ data, where decrement of $\alpha$ values to $-1$ and $-2$ which gives $M_{max} = 2.29, 2.14 M_{\odot}$ within the pulsars mass data and the radius at $1.4 M_{\odot}$ i.e $R_{1.4}$ are $11.68$ km (and  $11.43$ km) satisfying the GW170817 and NICER x-ray data PSRJ0030+0451 90 \% CI region and varying the DM subfraction from 0.1 \%, 1 \%, and 5 \%, the physical curves satisfied for $\alpha = $ (-2, -1), (-2,-1), and (-2,-1, 0) satisfying all the mass-radius constraints as shown in Fig. \ref{MPA1_MvR}. For this case, $\Lambda_{1.4}$ without DM and with DM always satisfy the aLIGO dimensionless tidal deformability $\Lambda_{1.36}$ band, only a reduction of $\Lambda_{1.4}$ value for greater DM subfraction case seen represented in Fig. \ref{MPA1_TDvM}   and table \ref{tab: MPA1 MvR}.

An excellent observation from this analysis is that we get the presence of DM inside the core of the neutron star \textbf{soften} in the EOS model, which gives a maximum mass and maximum radius with values not satisfying any observational constraints, suggesting a more compactness and lesser mass-radius value. As the tidal deformability depends on the $\mathcal{O}(5)$ power of the radius value significantly, its value at $1.4 M_{\odot}$, i.e., $\Lambda_{1.4}$ is reduced even if we increase the DM subfraction inside the core of the star.

Fig. \ref{LvsR}  depicts the relation between dimensionless tidal deformability and the radius of a 1.4 $M_\odot$ i.e $\Lambda_{1.4}$ and $R_{1.4}$ for DM-admixed anisotropic NSs is investigated for all the 3 taken  EOS (AP3, BSk22, MPA1) with  $g=10^{-5}$, $m_\chi=1$GeV, and $m_\phi=1$KeV. The color contour represents the range of $\alpha$-values from $-2$ to $2$. The subplots illustrate different DM subfractions: (a) $5\%$ DM, (b) $1\%$ DM, (c) $0.1\%$ DM, and (d) No DM. The GW170817 constraint (deep sky blue) reflects observational data from the binary neutron star merger. An important observation we get, for \textbf{BSk22} EOS the color contours for zero DM case with positive values of $\alpha$ are going beyond the observational bounds, but increasing DM subfraction up to 5\% gives a \textbf{\textit{unique and robust}} relation satisfying between $\Lambda_{1.4}$ and $R_{1.4}$ in the GW170817 band. The other 2 EOS don't respond much to increment of DM subfraction; irrespective of those values, they consistently satisfy the LIGO tidal deformability band.

From table \ref{tab:Results_satisfying}, we represent the range of anisotropic parameter ($\alpha$) satisfying the observational constraints are identified for different DM subfractions under the conditions of $g=10^{-5}$, $m_\chi=1$GeV, and $m_\phi=1$KeV. The $\alpha$-values ranged from $[-2, 2]$, with increments of $0.25$. The symbol (\ding{51}) indicates that all $\alpha$-values within this range satisfy the specified constraint. On the other hand, the symbol (\ding{55}) denotes that none of the $\alpha$-values meet the constraints. Notably, NICER observations of PSR J0030+0451 rule out AP3, and MPA1 EOS, even when anisotropic configurations are taken into consideration. By increasing the DM subfraction up to 5 \%, dimensionless tidal deformability $\Lambda_{1.4}$ of BSk22 EOS satisfies the range given by GW170817 $\Lambda_{1.36}$ value that wasn't satisfied for no DM case, so suggesting a DM admixed NS. If the value of anisotropic parameters $\alpha$ is negative, it means that the corresponding model satisfies the given pulsar maximum mass constraint with 5 \% DM inside the core of the star.

\begin{widetext}

\begin{figure}[h!]
\centering
\begin{minipage}[b]{0.49\linewidth}
  \centering
  \includegraphics[width=\linewidth]{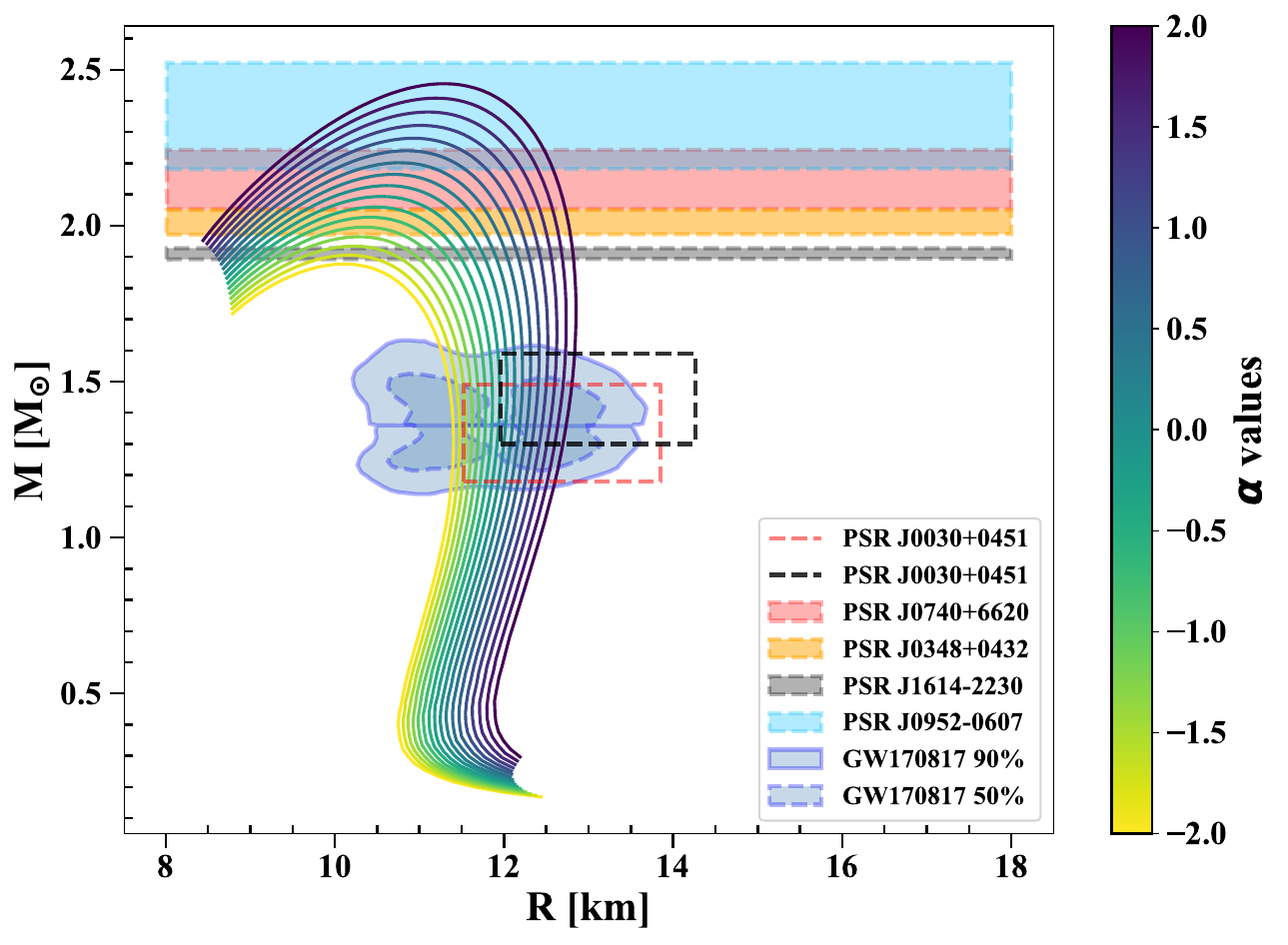}
  \caption*{(a) $5\%$ DM}
  \label{fig:RvM5DM}
\end{minipage}%
\hfill
\begin{minipage}[b]{0.49\linewidth}
  \centering
  \includegraphics[width=\linewidth]{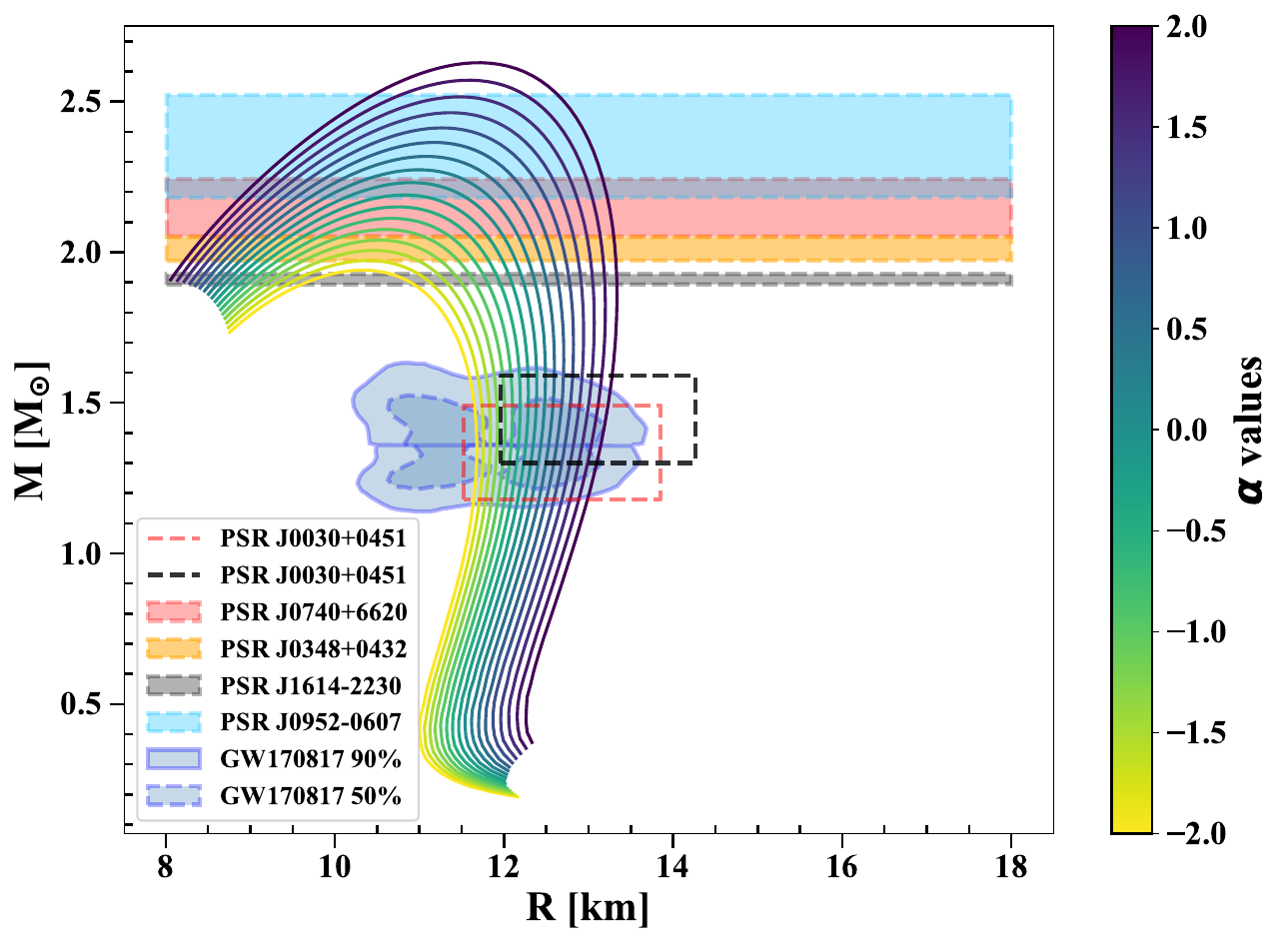}
  \caption*{(b) $1\%$ DM}
  \label{fig:RvM1DM}
\end{minipage}

\vspace{0.2cm} 

\begin{minipage}[b]{0.49\linewidth}
  \centering
  \includegraphics[width=\linewidth]{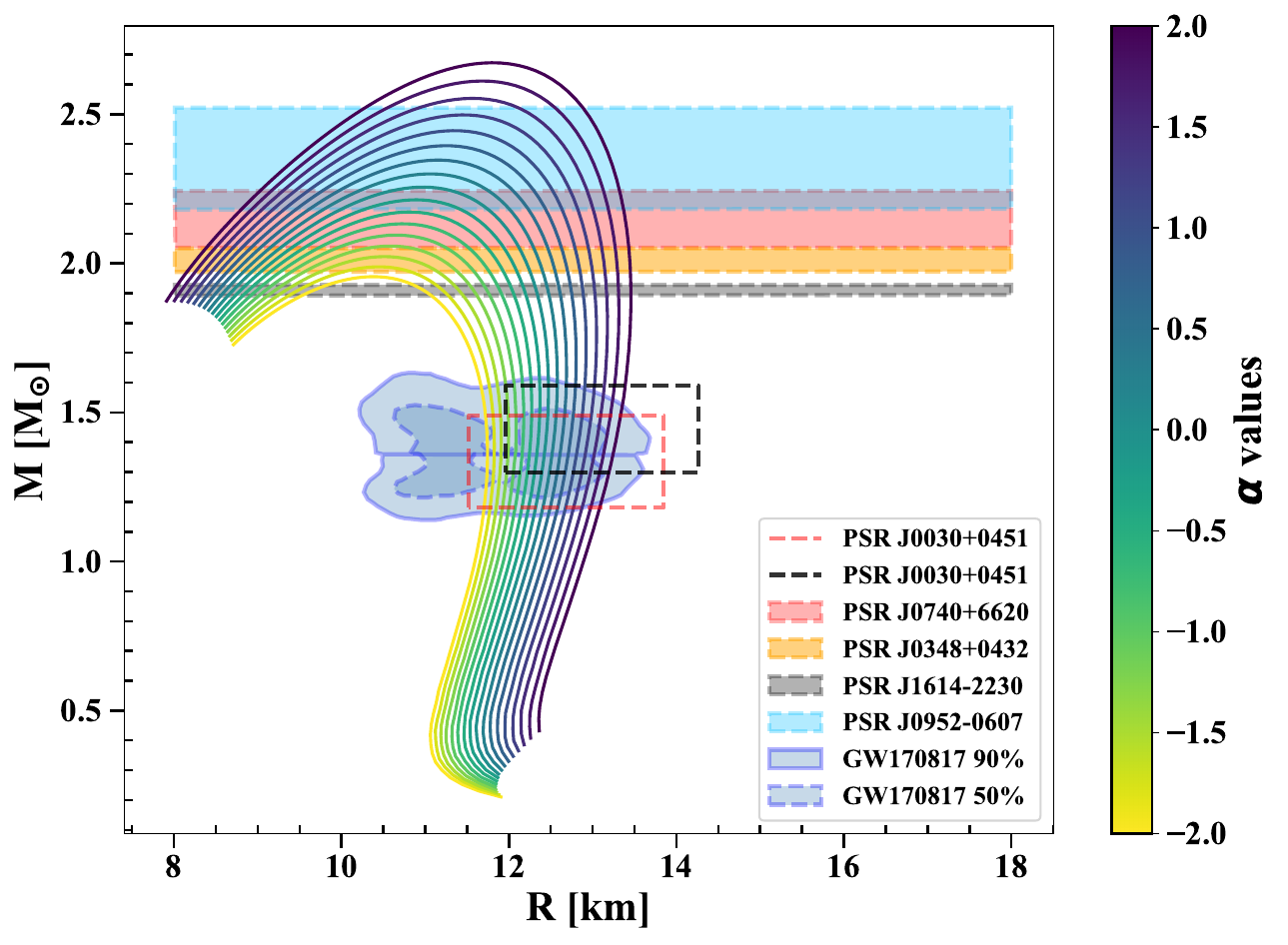}
  \caption*{(c) $0.1\%$ DM}
  \label{fig:RvM0p1DM}
\end{minipage}%
\hfill
\begin{minipage}[b]{0.49\linewidth}
  \centering
  \includegraphics[width=\linewidth]{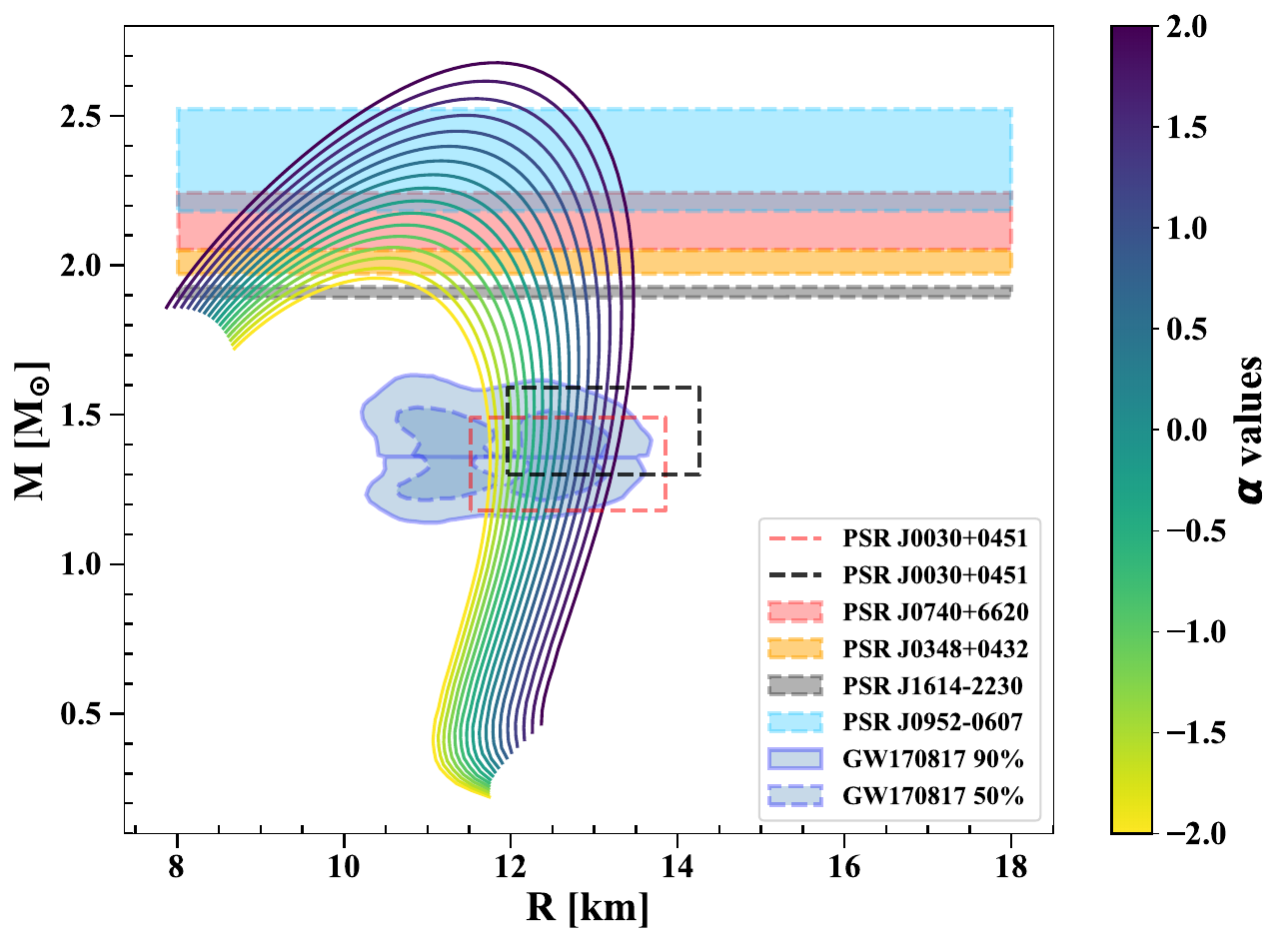}
  \caption*{(d) No DM}
  \label{fig:RvM0DM}
\end{minipage}

\caption{\justifying Same as Fig. \ref{fig;AP3_MvR}, but for BSk22 EOS.}
\label{BSk22_MvR}
\end{figure}


\begin{figure}[h!]
\centering
\begin{minipage}[b]{0.49\linewidth}
  \centering
  \includegraphics[width=\linewidth]{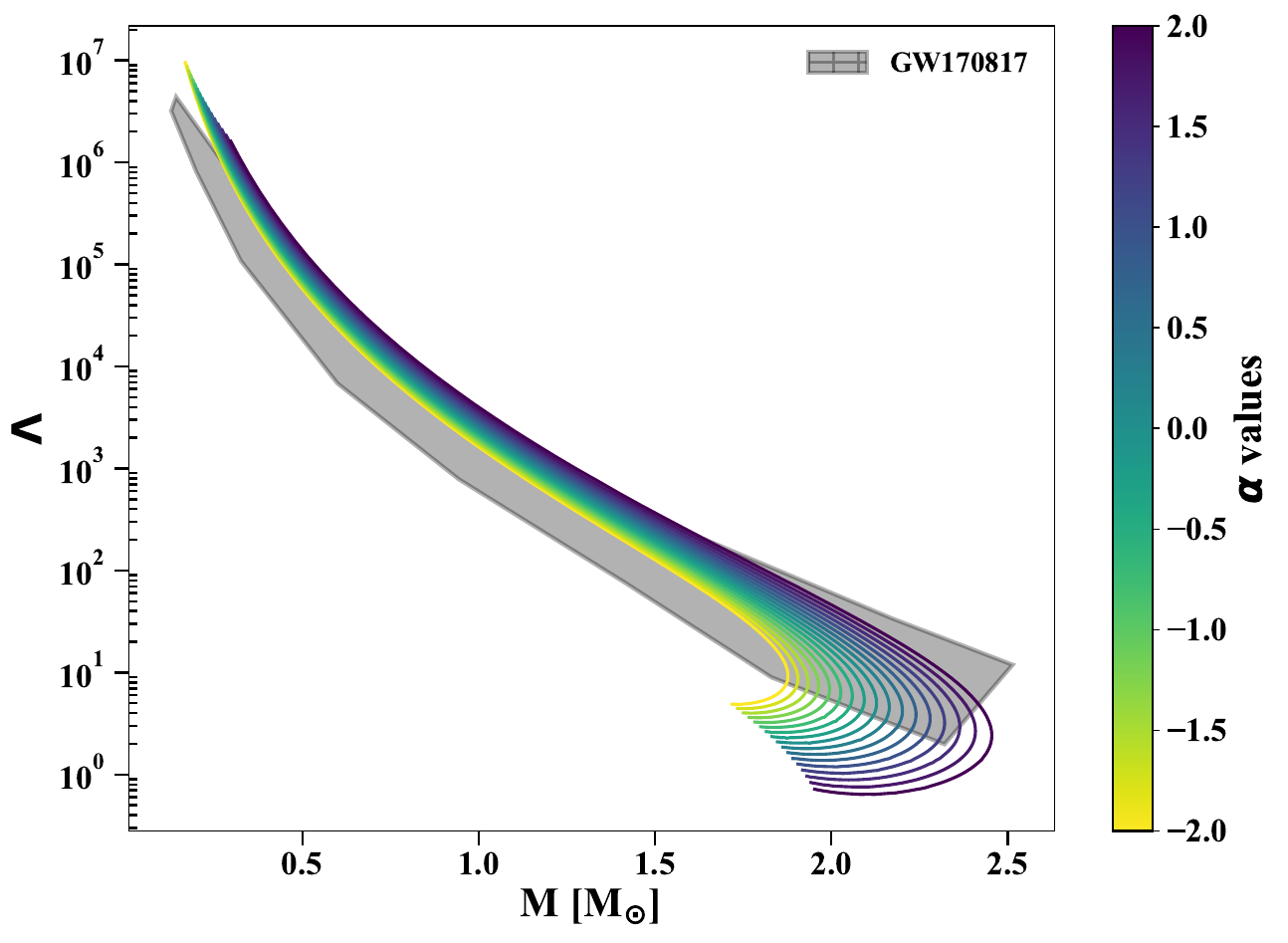}
  \caption*{(a) $5\%$ DM}
  \label{fig:TDvM5DM}
\end{minipage}%
\hfill
\begin{minipage}[b]{0.49\linewidth}
  \centering
  \includegraphics[width=\linewidth]{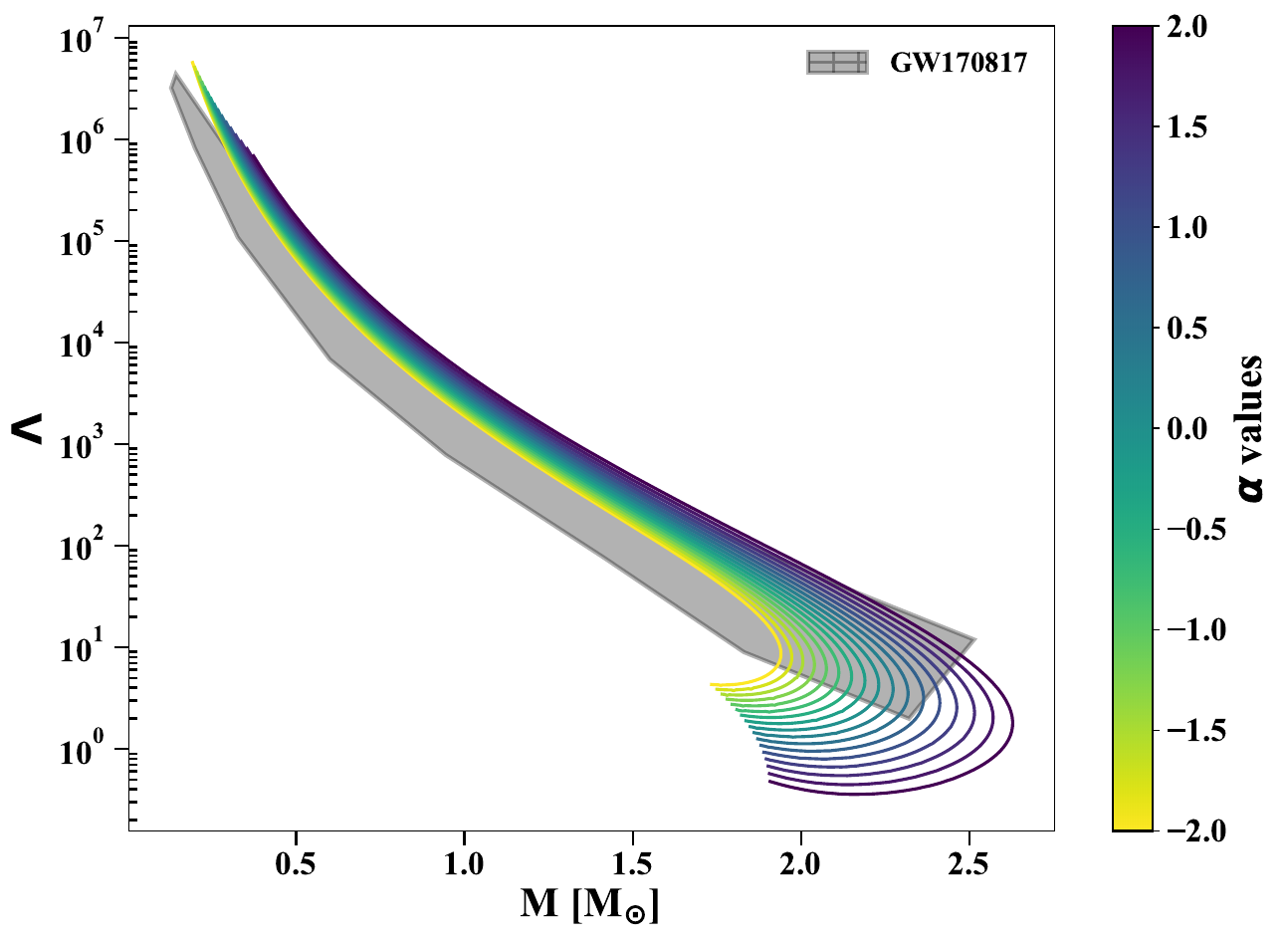}
  \caption*{(b) $1\%$ DM}
  \label{fig:TDvM1DM}
\end{minipage}

\vspace{0.2cm} 

\begin{minipage}[b]{0.49\linewidth}
  \centering
  \includegraphics[width=\linewidth]{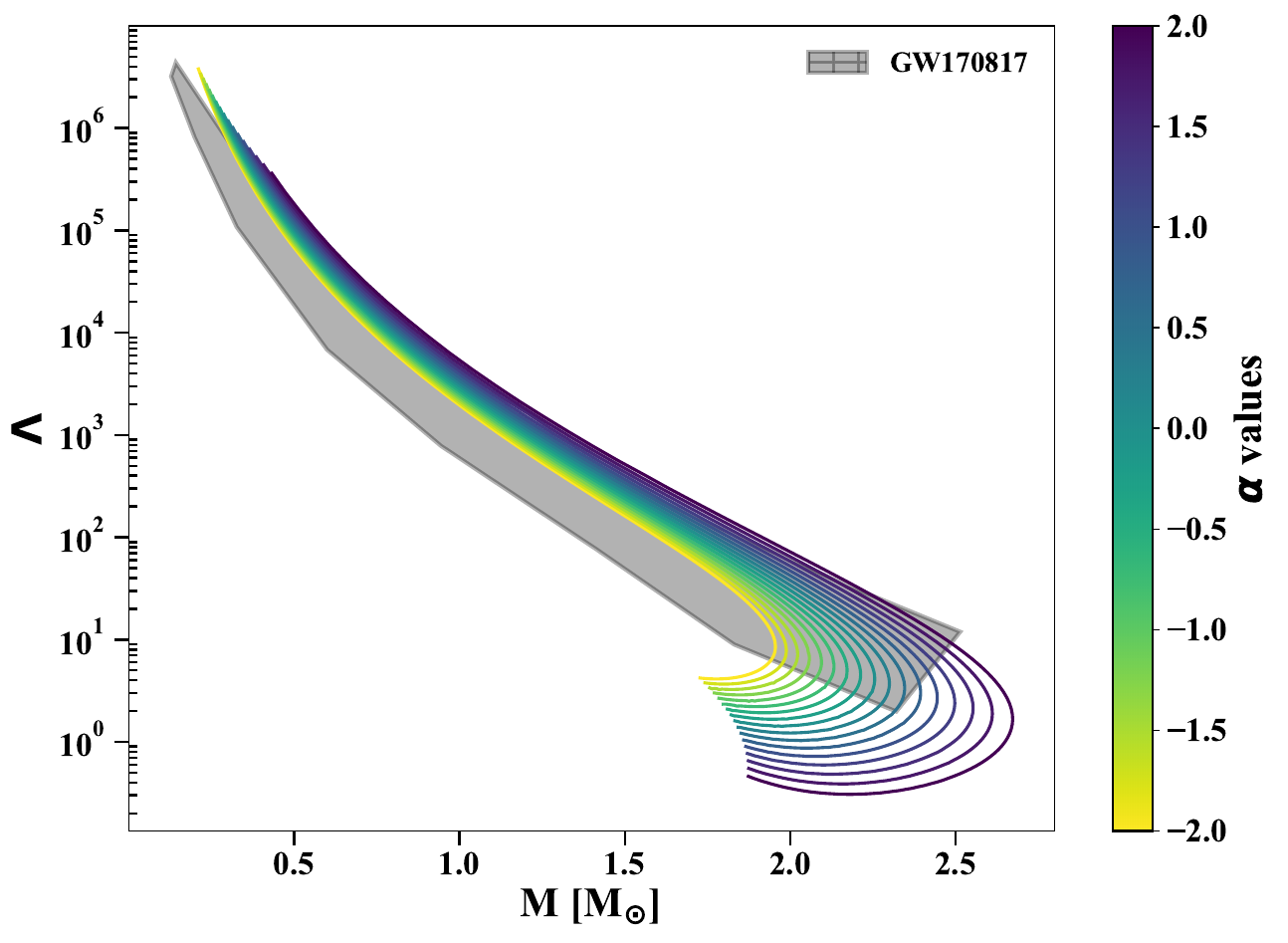}
  \caption*{(c) $0.1\%$ DM}
  \label{fig:TDvM0p1DM}
\end{minipage}%
\hfill
\begin{minipage}[b]{0.49\linewidth}
  \centering
  \includegraphics[width=\linewidth]{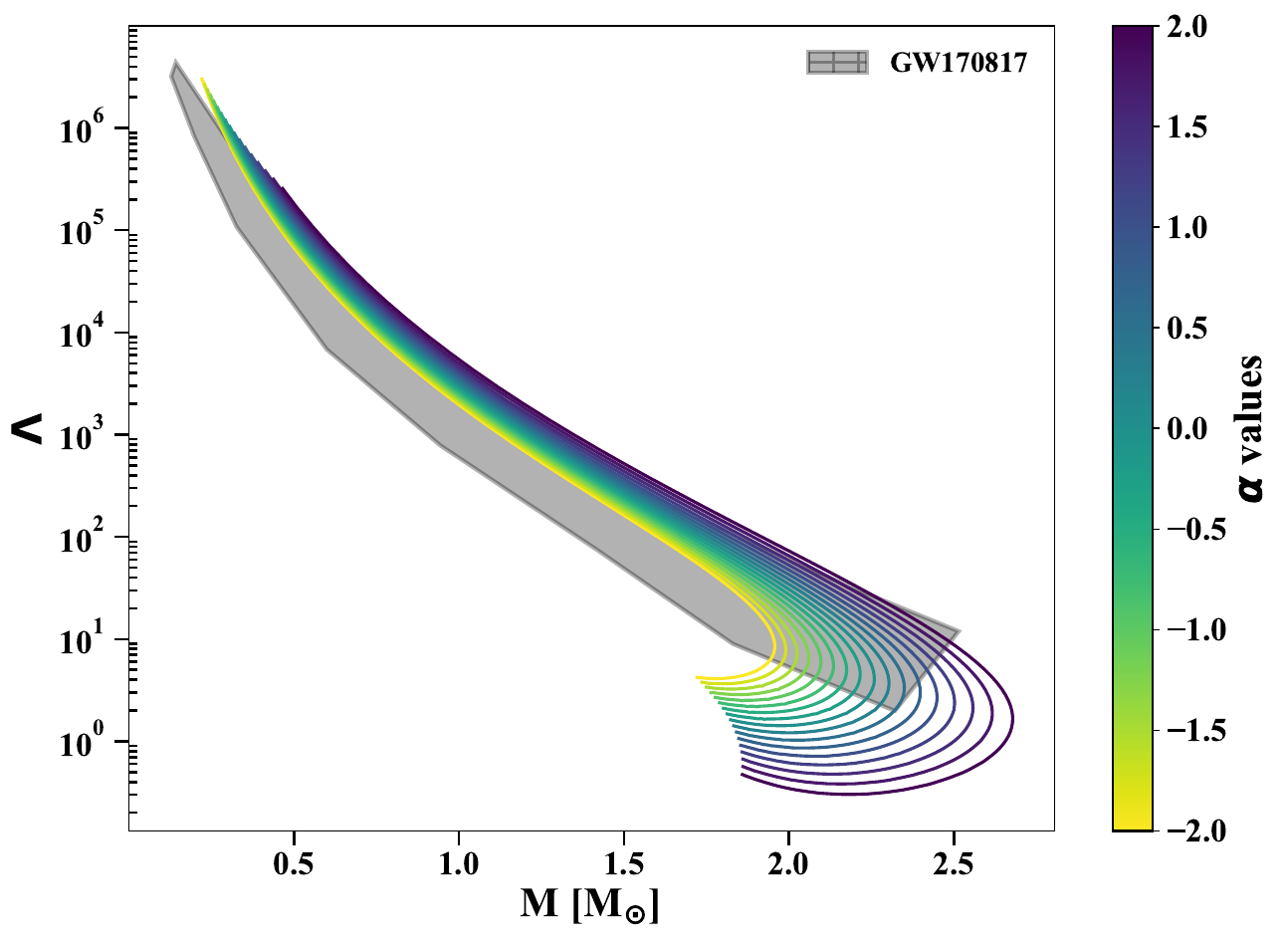}
  \caption*{(d) No DM}
  \label{fig:TDvM0DM}
\end{minipage}

\caption{\justifying Same as Fig. \ref{AP3_TDvM}, but for BSk22 EOS.}
\label{Bsk22_TDvM}
\end{figure}


\begin{table}[]
\caption{\justifying Same as Table. \ref{tab: AP3 MvR}, but for BSk22 EOS}
\label{tab: BSk22 MvR}
\renewcommand{\arraystretch}{1.5}
\setlength{\tabcolsep}{11pt}
\begin{threeparttable}
\begin{tabular}{|c|cccccccc|}
\hline\hline
\multirow{2}{*}{$\bm\alpha$} & \textbf{M$\bm_{max}$} & \textbf{R$\bm_{max}$} & \textbf{R$\bm_{1.4}$} & \textbf{$\bm\Lambda_{1.4}$} & \textbf{M$\bm_{max}$} & \textbf{R$\bm_{max}$} & \textbf{R$\bm_{1.4}$} & \textbf{$\bm\Lambda_{1.4}$} \\ \cline{2-9} 
                                       & \multicolumn{4}{c}{\textbf{5\% DM}}                                             & \multicolumn{4}{c|}{\textbf{1\% DM}}                                          \\ \hline\hline
\textbf{-2}                                     & 1.87695    & 10.10278   & 11.39167$^{\bm b}$   & \multicolumn{1}{c||}{208.1634$^{\bm b}$}                        & 1.940311   & 10.34318   & 11.6754$^{\bm b}$    & 249.7727$^{\bm b}$                        \\
\textbf{-1}                                     & 1.995051$^{\bm a}$   & 10.37499   & 11.7025$^{\bm b}$    & \multicolumn{1}{c||}{261.6699$^{\bm b}$}                        & 2.07548$^{\bm a}$    & 10.61815   & 12.01361$^{\bm b}$   & 318.463$^{\bm b}$                         \\
\textbf{0}                                      & 2.128373$^{\bm a}$   & 10.60844   & 12.0283$^{\bm b}$    & \multicolumn{1}{c||}{331.7732$^{\bm b}$}                        & 2.230956$^{\bm a}$   & 10.92327   & 12.36004$^{\bm b}$   & 408.7559$^{\bm b}$                        \\
\textbf{1}                                      & 2.280089$^{\bm a}$   & 10.93761   & 12.37649   & \multicolumn{1}{c||}{428.3349$^{\bm b}$}                        & 2.412579   & 11.26015   & 12.74221   & 538.232$^{\bm b}$                         \\
\textbf{2}                                      & 2.455013   & 11.30635   & 12.7595    & \multicolumn{1}{c||}{567.3526$^{\bm b}$}                        & 2.628923   & 11.73643   & 13.17269   & 748.9832                        \\ \hline\hline
                                     $\bm\alpha$  & \multicolumn{4}{c}{\textbf{0.1\% DM}}                                           & \multicolumn{4}{c|}{\textbf{0\% DM}}                                 \\ \hline\hline
\textbf{-2}                                     & 1.95529    & 10.35763   & 11.74302$^{\bm b}$   & \multicolumn{1}{c||}{259.494$^{\bm b}$ }                        & 1.957056   & 10.40461   & 11.74965$^{\bm b}$   & 261.4628$^{\bm b}$                        \\
\textbf{-1 }                                    & 2.094536$^{\bm a}$   & 10.69284   & 12.08369$^{\bm b}$   & \multicolumn{1}{c||}{334.849$^{\bm b}$ }                        & 2.096857$^{\bm a}$   & 10.67387   & 12.09296$^{\bm b}$   & 336.6464$^{\bm b}$                        \\
\textbf{0}                                      & 2.25565$^{\bm a}$    & 10.99473   & 12.43954   & \multicolumn{1}{c||}{427.4345$^{\bm b}$}                        & 2.258508$^{\bm a}$   & 10.96848   & 12.44901   & 431.3186$^{\bm b}$                        \\
\textbf{1}                                      & 2.444952   & 11.32662   & 12.8311    & \multicolumn{1}{c||}{574.5156$^{\bm b}$}                        & 2.448716   & 11.38221   & 12.84116   & 578.3342$^{\bm b}$                        \\
\textbf{2}                                      & 2.672799   & 11.79638   & 13.27431   & \multicolumn{1}{c||}{791.5682}                        & 2.677767   & 11.8534    & 13.28553   & 805.8132                        \\ \hline\hline
\end{tabular}
\begin{tablenotes}
    \item[$\bm a$] Satisfying PSR (1.97-2.35).
    \item[$\bm b$] Satisfying GW170817 ($R_{1.4} \approx 10.62-12.38$ and $\Lambda_{1.36}\leq720$).
\end{tablenotes}
\end{threeparttable}
\end{table}

\begin{figure}[]
\centering
\begin{minipage}[b]{0.49\linewidth}
  \centering
  \includegraphics[width=\linewidth]{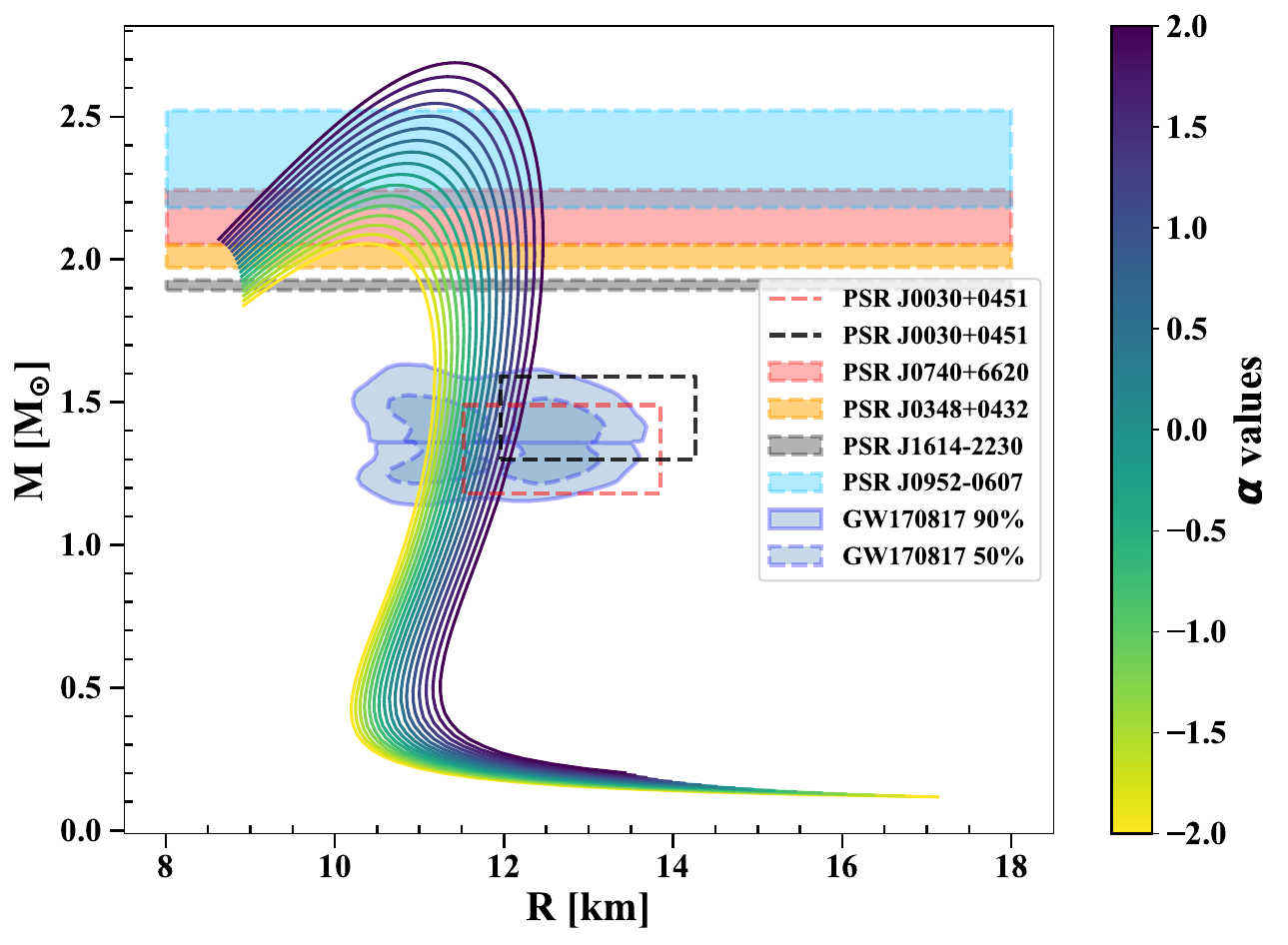}
  \caption*{(a) $5\%$ DM}
  \label{fig:MPA1_RvM5DM}
\end{minipage}%
\hfill
\begin{minipage}[b]{0.49\linewidth}
  \centering
  \includegraphics[width=\linewidth]{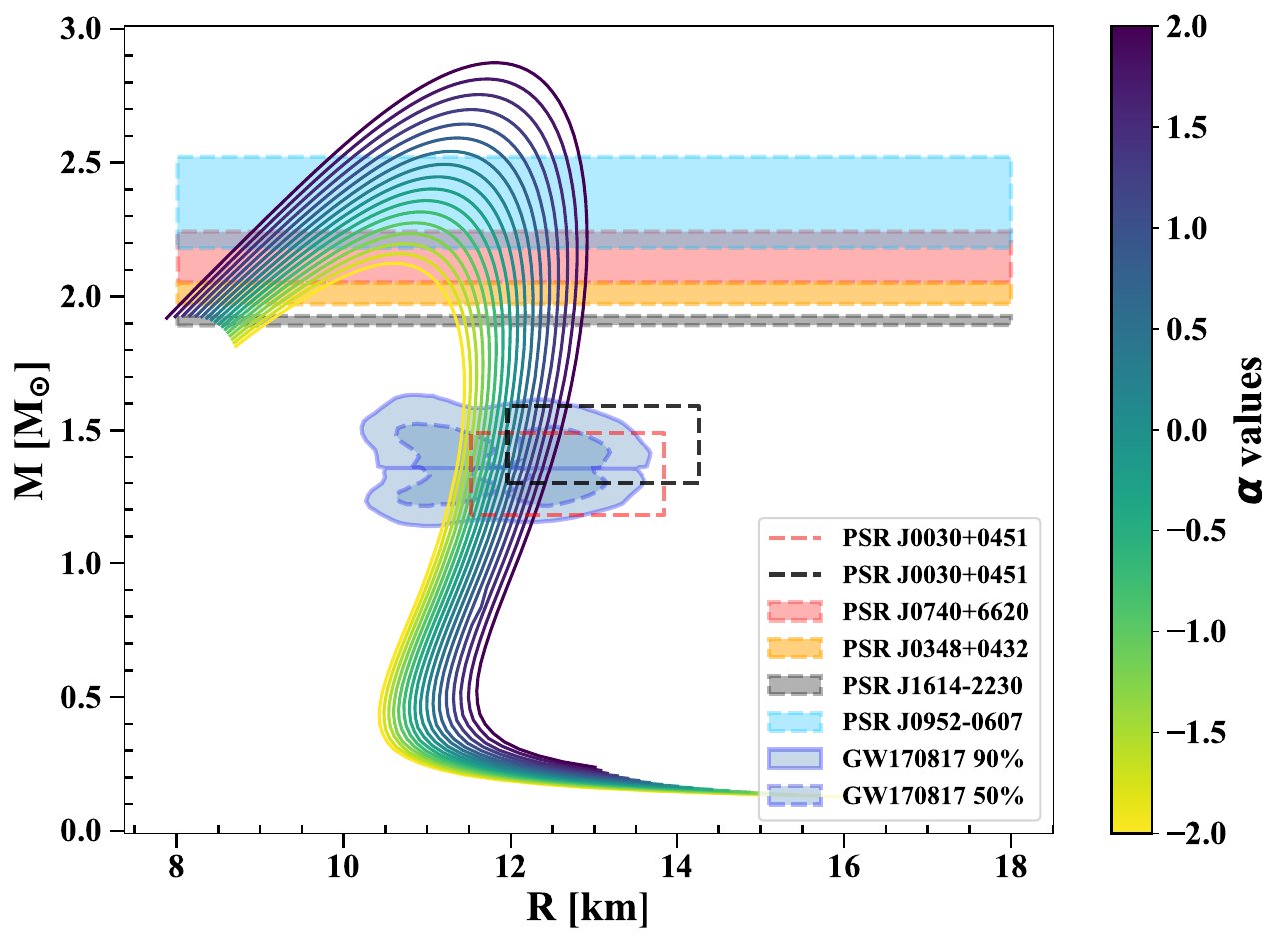}
  \caption*{(b) $1\%$ DM}
  \label{fig:MPA1_RvM1DM}
\end{minipage}

\vspace{0.2cm} 

\begin{minipage}[b]{0.49\linewidth}
  \centering
  \includegraphics[width=\linewidth]{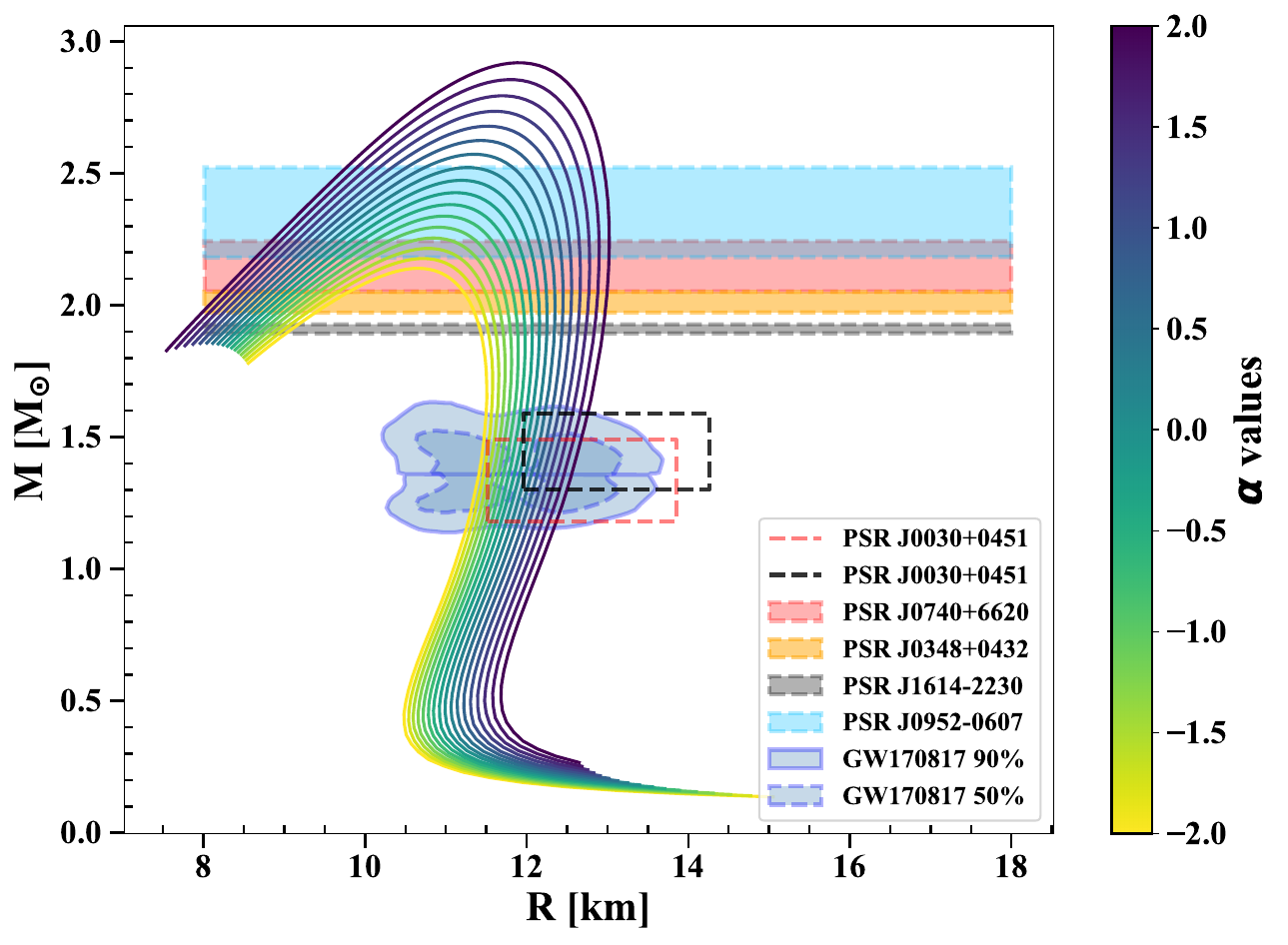}
  \caption*{(c) $0.1\%$ DM}
  \label{fig:MPA1_RvM0p1DM}
\end{minipage}%
\hfill
\begin{minipage}[b]{0.49\linewidth}
  \centering
  \includegraphics[width=\linewidth]{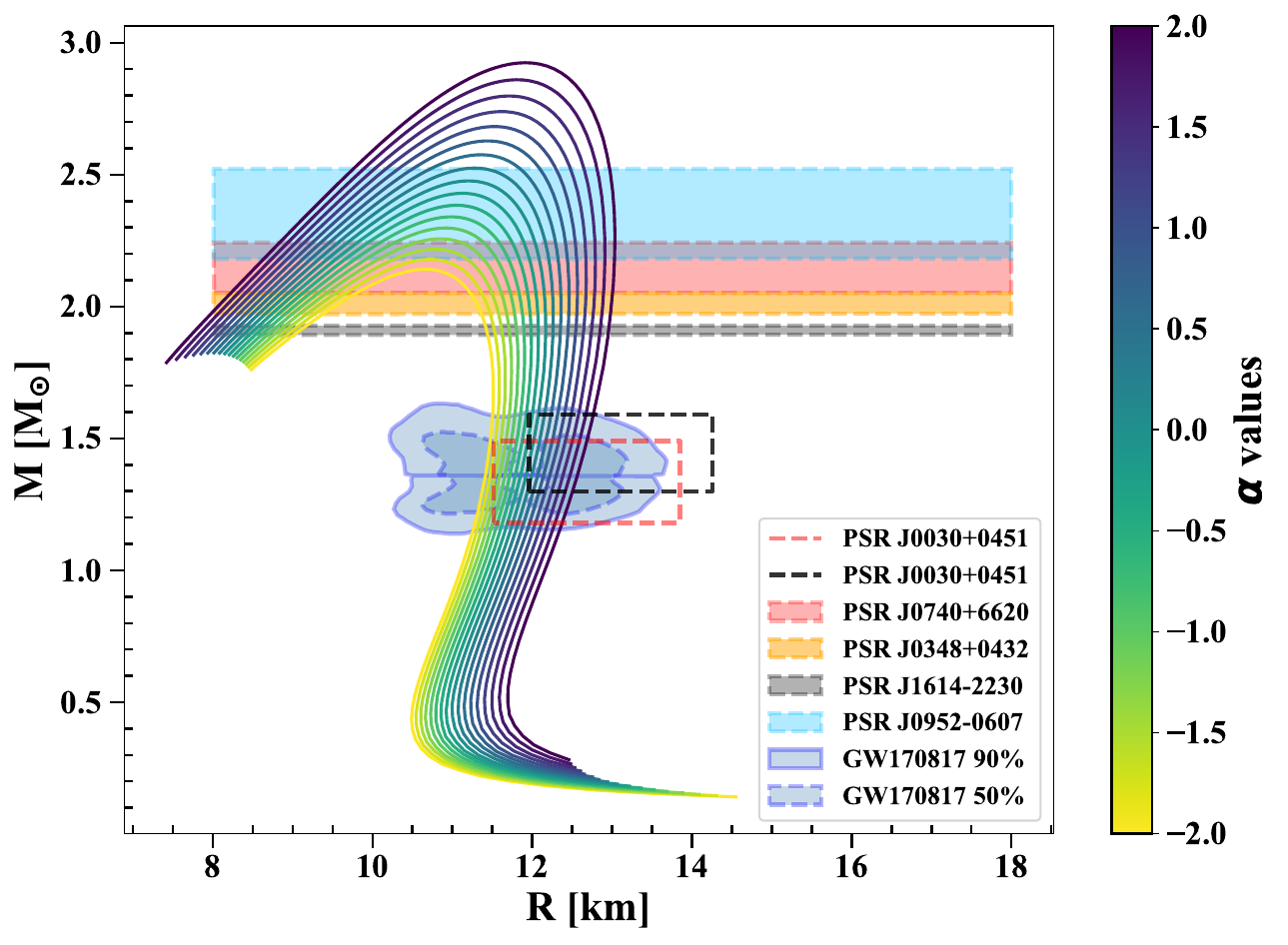}
  \caption*{(d) No DM}
  \label{fig:MPA1_RvM0DM}
\end{minipage}

\caption{\justifying Same as Fig. \ref{fig;AP3_MvR}, but for MPA1 EOS.}
\label{MPA1_MvR}
\end{figure}


\begin{figure}[]
\centering
\begin{minipage}[b]{0.49\linewidth}
  \centering
  \includegraphics[width=\linewidth]{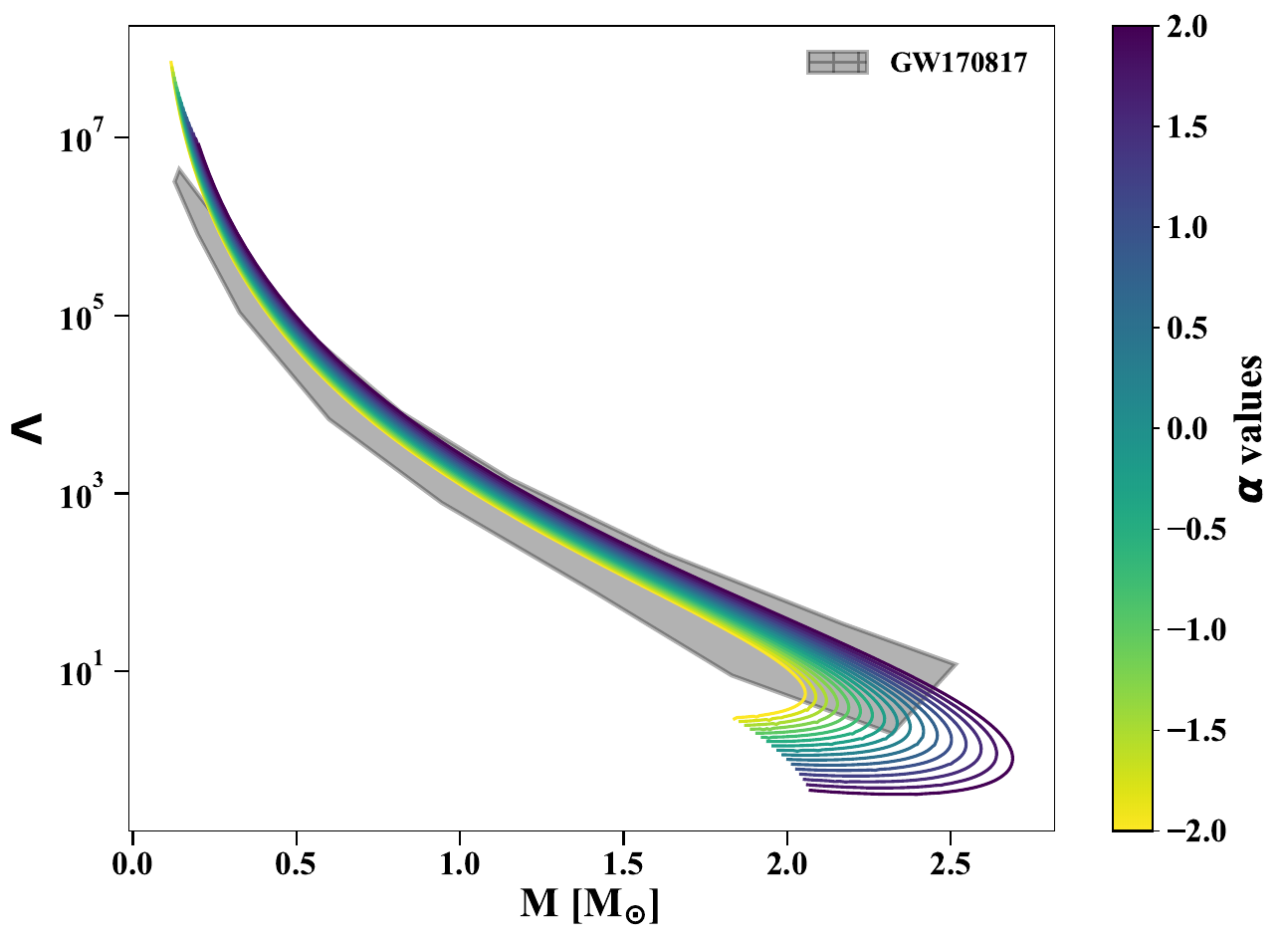}
  \caption*{(a) $5\%$ DM}
  \label{fig:MPA1_TDvM5DM}
\end{minipage}%
\hfill
\begin{minipage}[b]{0.49\linewidth}
  \centering
  \includegraphics[width=\linewidth]{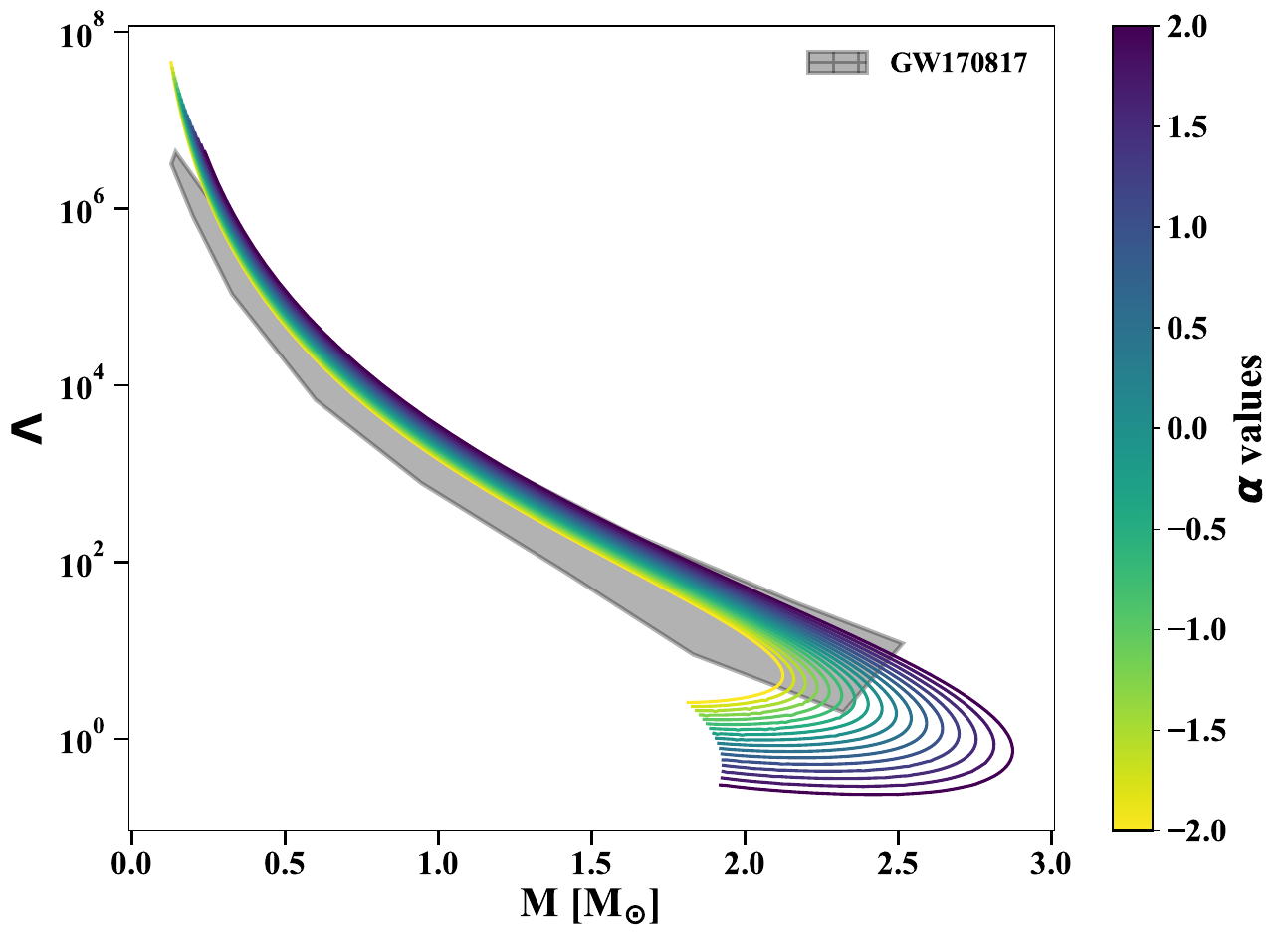}
  \caption*{(b) $1\%$ DM}
  \label{fig:MPA1_TDvM1DM}
\end{minipage}

\vspace{0.2cm} 

\begin{minipage}[b]{0.49\linewidth}
  \centering
  \includegraphics[width=\linewidth]{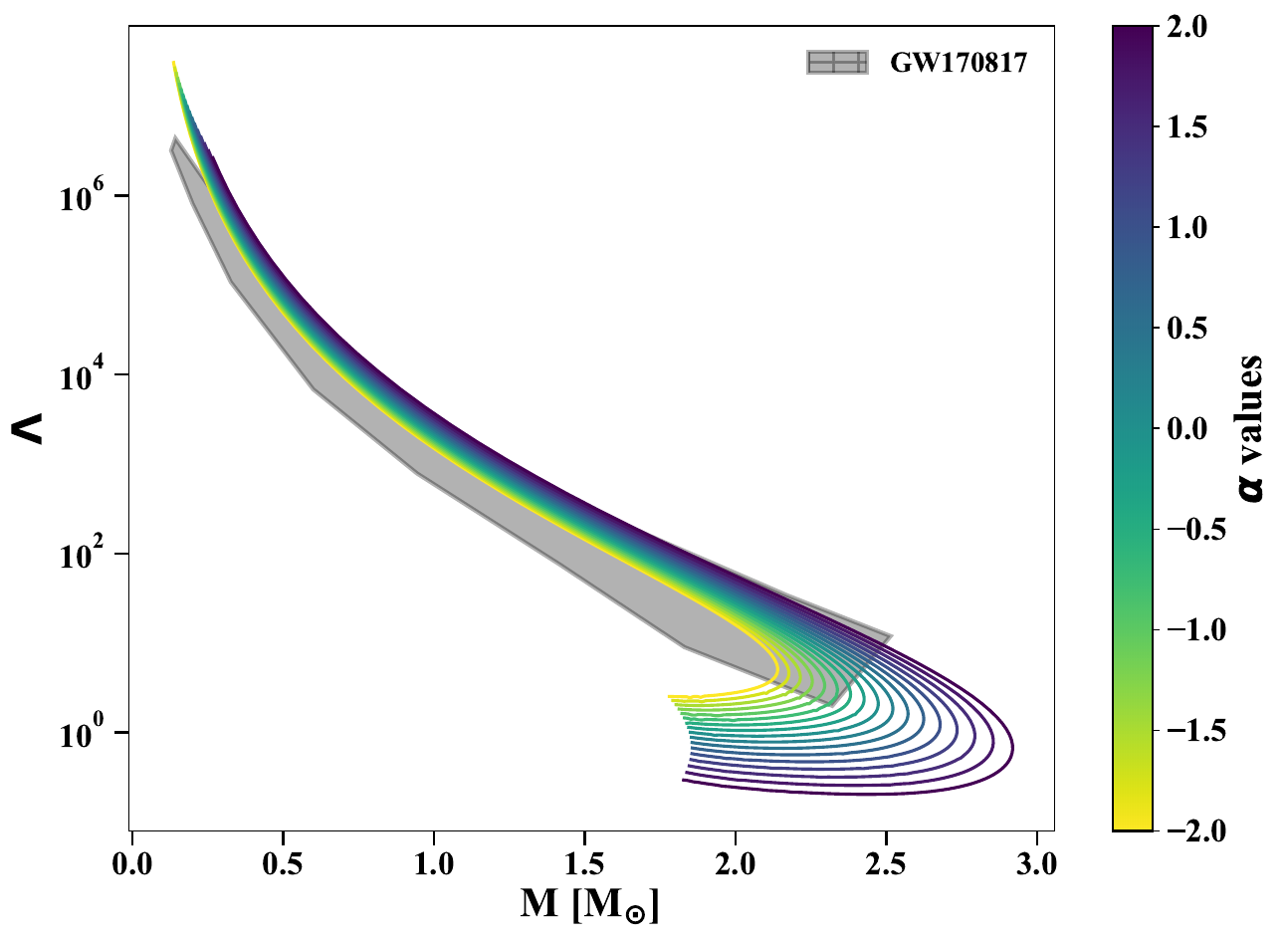}
  \caption*{(c) $0.1\%$ DM}
  \label{fig:MPA1_TDvM0p1DM}
\end{minipage}%
\hfill
\begin{minipage}[b]{0.49\linewidth}
  \centering
  \includegraphics[width=\linewidth]{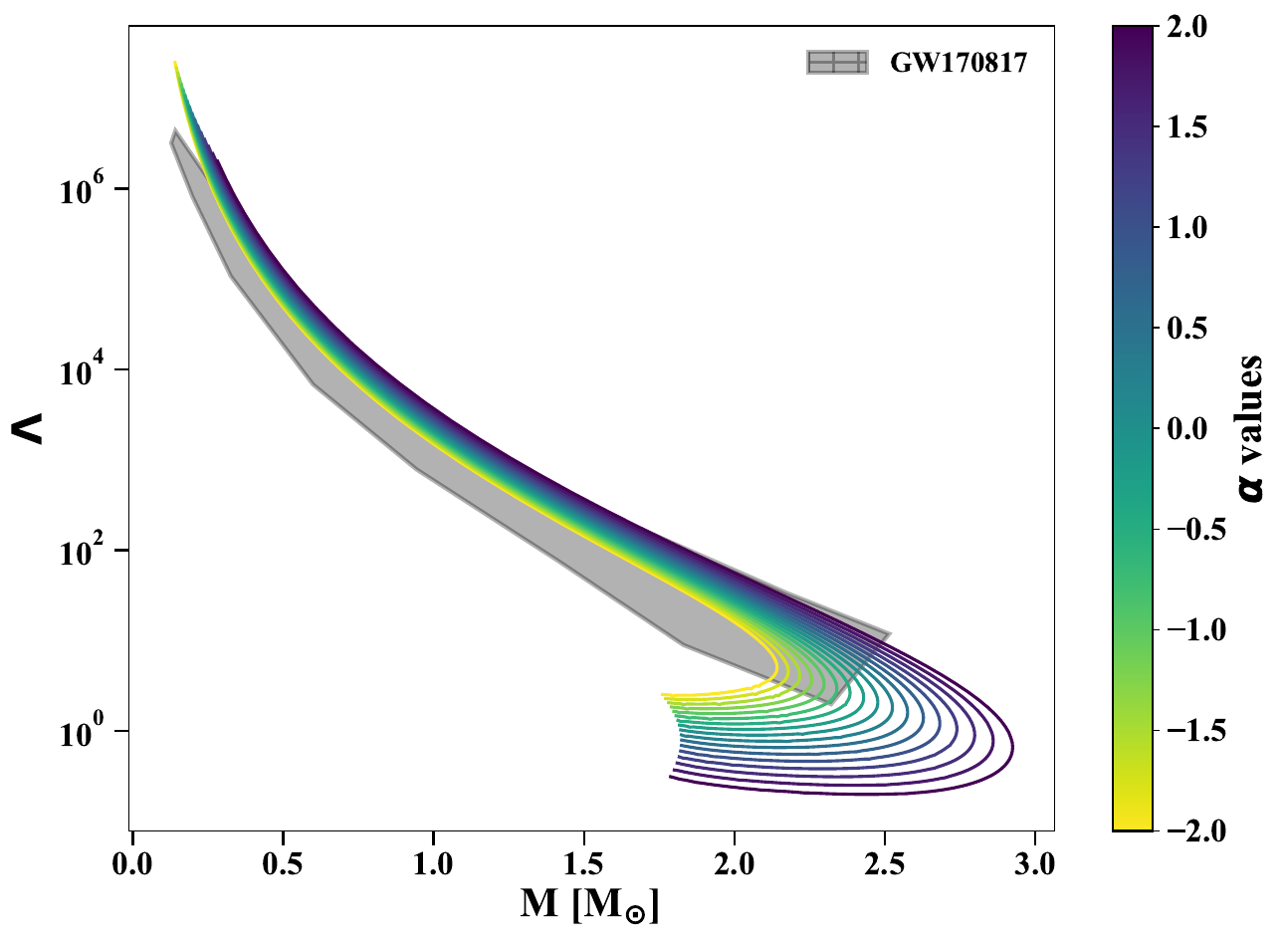}
  \caption*{(d) No DM}
  \label{fig:MPA1_TDvM0DM}
\end{minipage}

\caption{\justifying Same as Fig. \ref{AP3_TDvM}, but for MPA1 EOS.}
\label{MPA1_TDvM}
\end{figure}


\begin{table}[]
\caption{\justifying Same as Table. \ref{tab: AP3 MvR}, but for MPA1 EOS}
\label{tab: MPA1 MvR}
\renewcommand{\arraystretch}{1.5}
\setlength{\tabcolsep}{11pt}
\begin{threeparttable}
\begin{tabular}{|c|cccccccc|}
\hline\hline
\multirow{2}{*}{$\bm\alpha$} & \textbf{M$\bm_{max}$} & \textbf{R$\bm_{max}$} & \textbf{R$\bm_{1.4}$} & \textbf{$\bm\Lambda_{1.4}$} & \textbf{M$\bm_{max}$} & \textbf{R$\bm_{max}$} & \textbf{R$\bm_{1.4}$} & \textbf{$\bm\Lambda_{1.4}$} \\ \cline{2-9}
                                       & \multicolumn{4}{c}{\textbf{5\% DM}}                                                                                                                             & \multicolumn{4}{c|}{\textbf{1\% DM}}                                                                                                                             \\ \hline\hline
\textbf{-2}                            & 2.055423$^{\bm a}$   & 10.36556   & 11.13353$^{\bm b}$   & \multicolumn{1}{c||}{182.7792$^{\bm b}$}   & 2.123774$^{\bm a}$   & 10.62183   & 11.37583$^{\bm b}$   & 212.2901$^{\bm b}$                        \\
\textbf{-1}                            & 2.188266$^{\bm a}$   & 10.62163   & 11.3597$^{\bm b}$    & \multicolumn{1}{c||}{217.9311$^{\bm b}$}   & 2.275085$^{\bm a}$   & 10.83508   & 11.61788$^{\bm b}$   & 259.0657$^{\bm b}$                        \\
\textbf{0}                             & 2.336213$^{\bm a}$   & 10.84892   & 11.59929$^{\bm b}$   & \multicolumn{1}{c||}{265.2593$^{\bm b}$}   & 2.447075   & 11.11402   & 11.88089$^{\bm b}$   & 316.9473$^{\bm b}$                        \\
\textbf{1}                             & 2.502086   & 11.15157   & 11.86266$^{\bm b}$   & \multicolumn{1}{c||}{326.7851$^{\bm b}$}   & 2.644325   & 11.42273   & 12.17308$^{\bm b}$   & 403.427$^{\bm b}$                         \\
\textbf{2}                             & 2.689011   & 11.41954   & 12.16136$^{\bm b}$   & \multicolumn{1}{c||}{415.2856$^{\bm b}$}   & 2.873231   & 11.76593   & 12.5107    & 525.4387$^{\bm b}$                        \\ \hline\hline
                                    $\bm\alpha$   & \multicolumn{4}{c}{\textbf{0.1\% DM}}                                           & \multicolumn{4}{c|}{\textbf{0\% DM}}                                             \\ \hline\hline
\textbf{-2}                            & 2.139665$^{\bm a}$   & 10.67774   & 11.43212$^{\bm b}$   & \multicolumn{1}{c||}{222.2942$^{\bm b}$}   & 2.141574$^{\bm a}$   & 10.6606    & 11.43508$^{\bm b}$   & 223.5843$^{\bm b}$                        \\
\textbf{-1}                            & 2.295371$^{\bm a}$   & 10.88569   & 11.67931$^{\bm b}$   & \multicolumn{1}{c||}{269.4337$^{\bm b}$}   & 2.29781$^{\bm a}$    & 10.91488   & 11.68292$^{\bm b}$   & 269.0018$^{\bm b}$                        \\
\textbf{0}                             & 2.473188   & 11.22123   & 11.9469$^{\bm b}$    & \multicolumn{1}{c||}{331.4052$^{\bm b}$}   & 2.476272   & 11.18908   & 11.95277$^{\bm b}$   & 335.2912$^{\bm b}$                        \\
\textbf{1}                             & 2.678525   & 11.53429   & 12.24663$^{\bm b}$   & \multicolumn{1}{c||}{424.4247$^{\bm b}$}   & 2.682258   & 11.56355   & 12.25592$^{\bm b}$   & 424.5412$^{\bm b}$                        \\
\textbf{2}                             & 2.919039   & 11.881     & 12.59362   & \multicolumn{1}{c||}{566.6536$^{\bm b}$}   & 2.924349   & 11.91085   & 12.6027    & 572.4514$^{\bm b}$                        \\ \hline\hline
\end{tabular}
\begin{tablenotes}
    \item[$\bm a$] Satisfying PSR (1.97-2.35).
    \item[$\bm b$] Satisfying GW170817 ($R_{1.4} \approx 10.62-12.38$ and $\Lambda_{1.36}\leq720$).
\end{tablenotes}
\end{threeparttable}
\end{table}

\begin{figure}[]
\centering
\begin{minipage}[b]{0.49\linewidth}
  \centering
  \includegraphics[width=\linewidth]{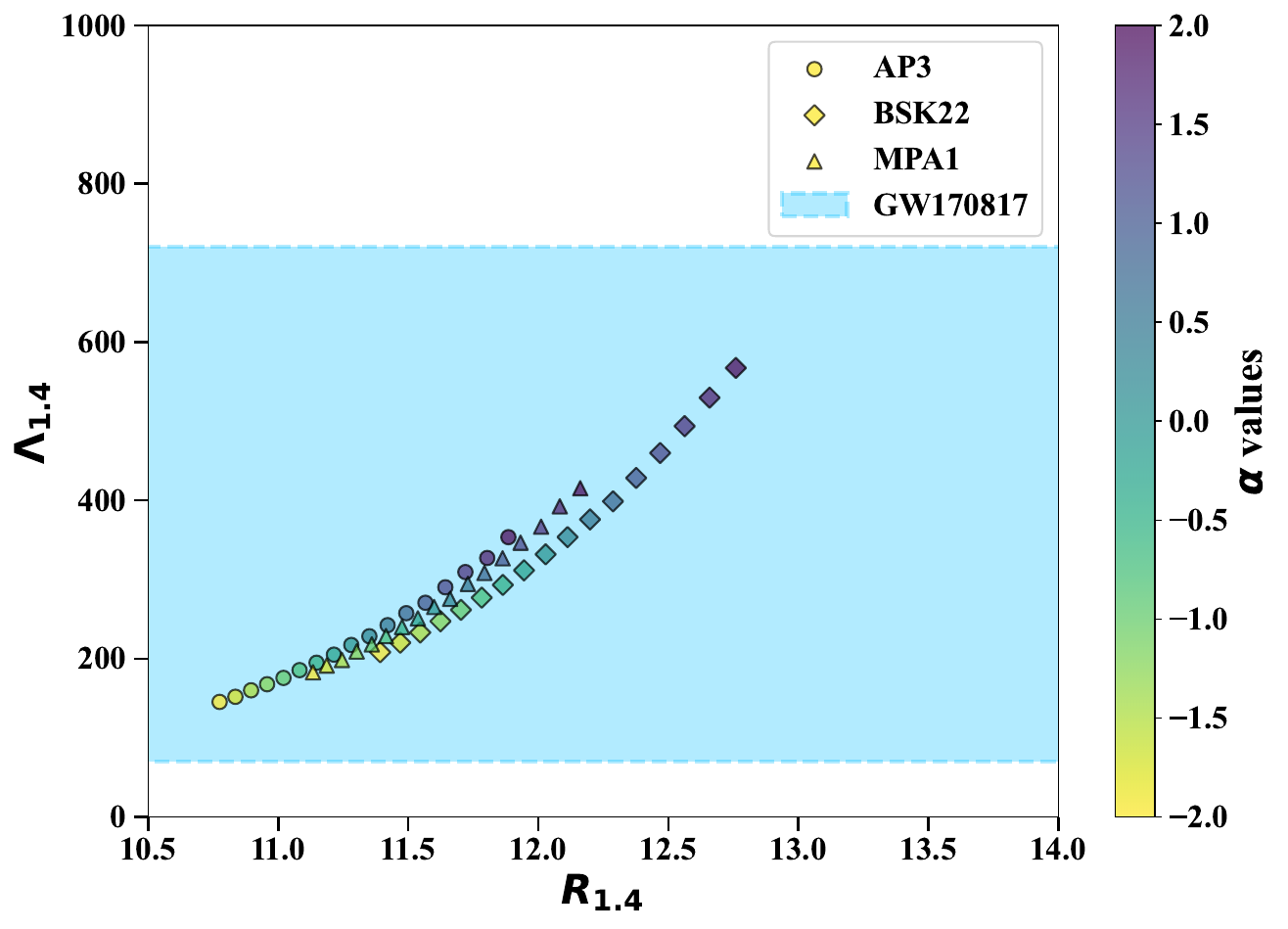}
  \caption*{(a) $5\%$ DM}
  \label{fig:DM5_LvR}
\end{minipage}%
\hfill
\begin{minipage}[b]{0.49\linewidth}
  \centering
  \includegraphics[width=\linewidth]{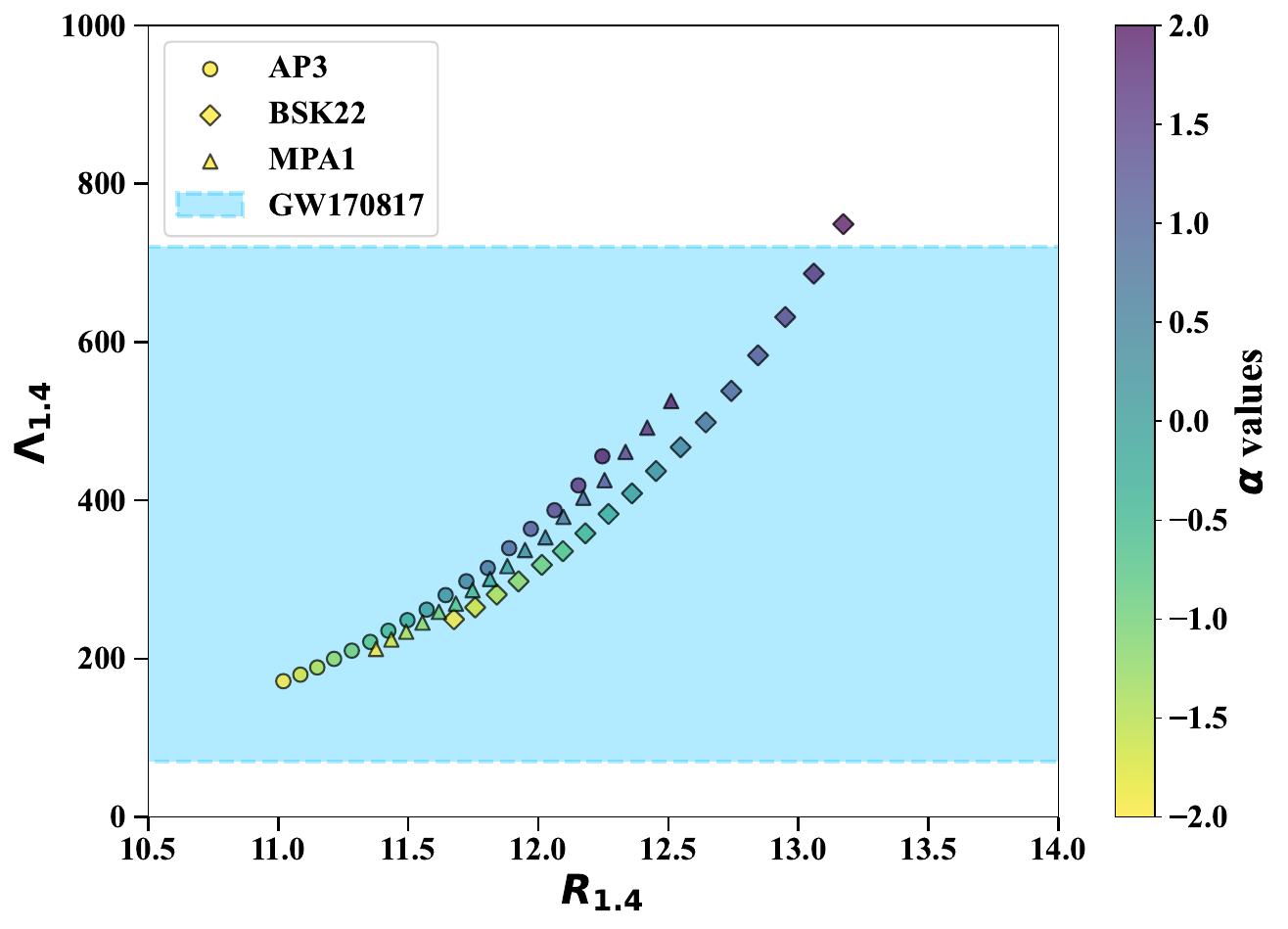}
  \caption*{(b) $1\%$ DM}
  \label{fig:DM1_LvR}
\end{minipage}

\vspace{0.2cm} 

\begin{minipage}[b]{0.49\linewidth}
  \centering
  \includegraphics[width=\linewidth]{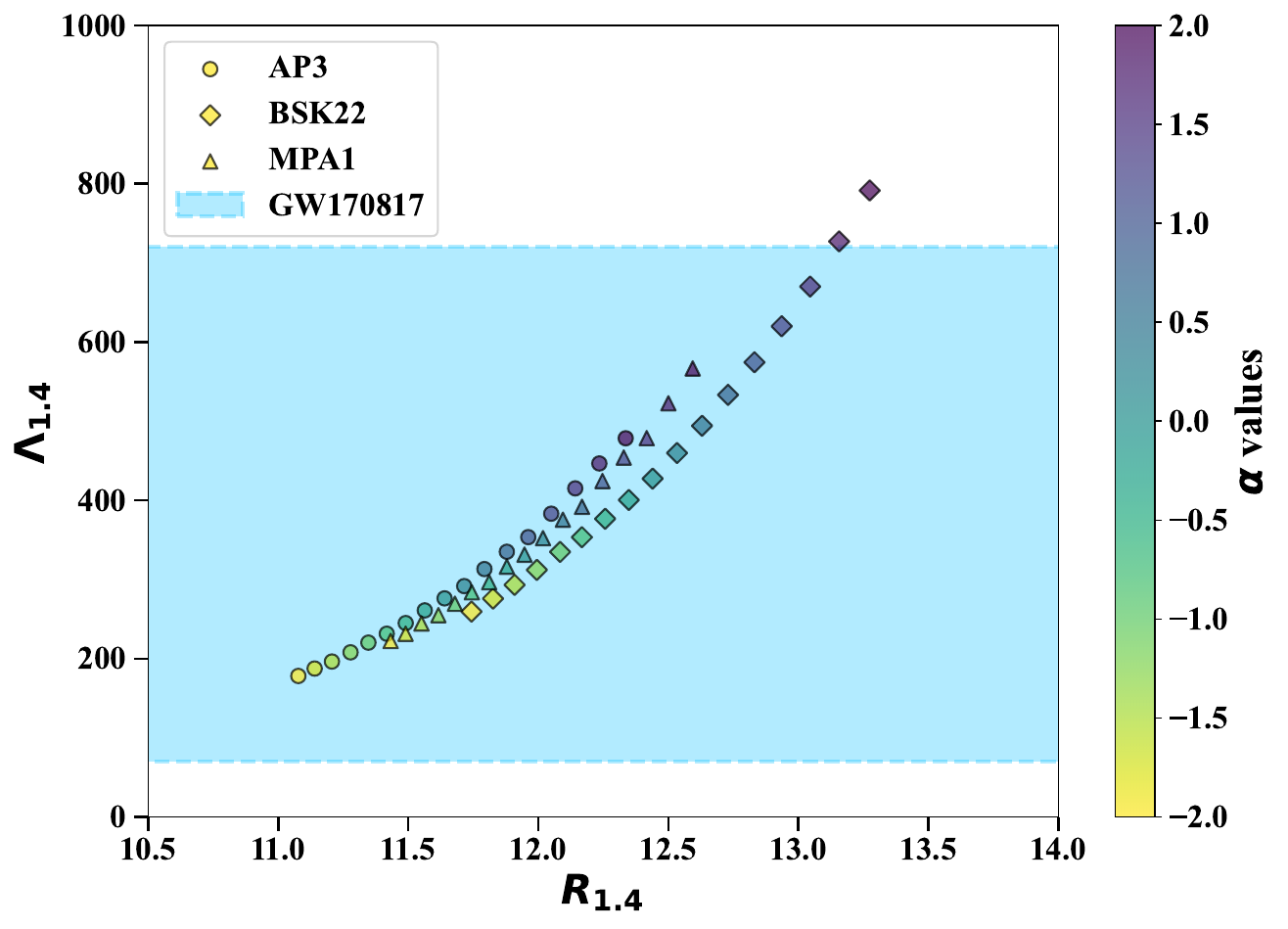}
  \caption*{(c) $0.1\%$ DM}
  \label{fig:DM0p1_LvR}
\end{minipage}%
\hfill
\begin{minipage}[b]{0.49\linewidth}
  \centering
  \includegraphics[width=\linewidth]{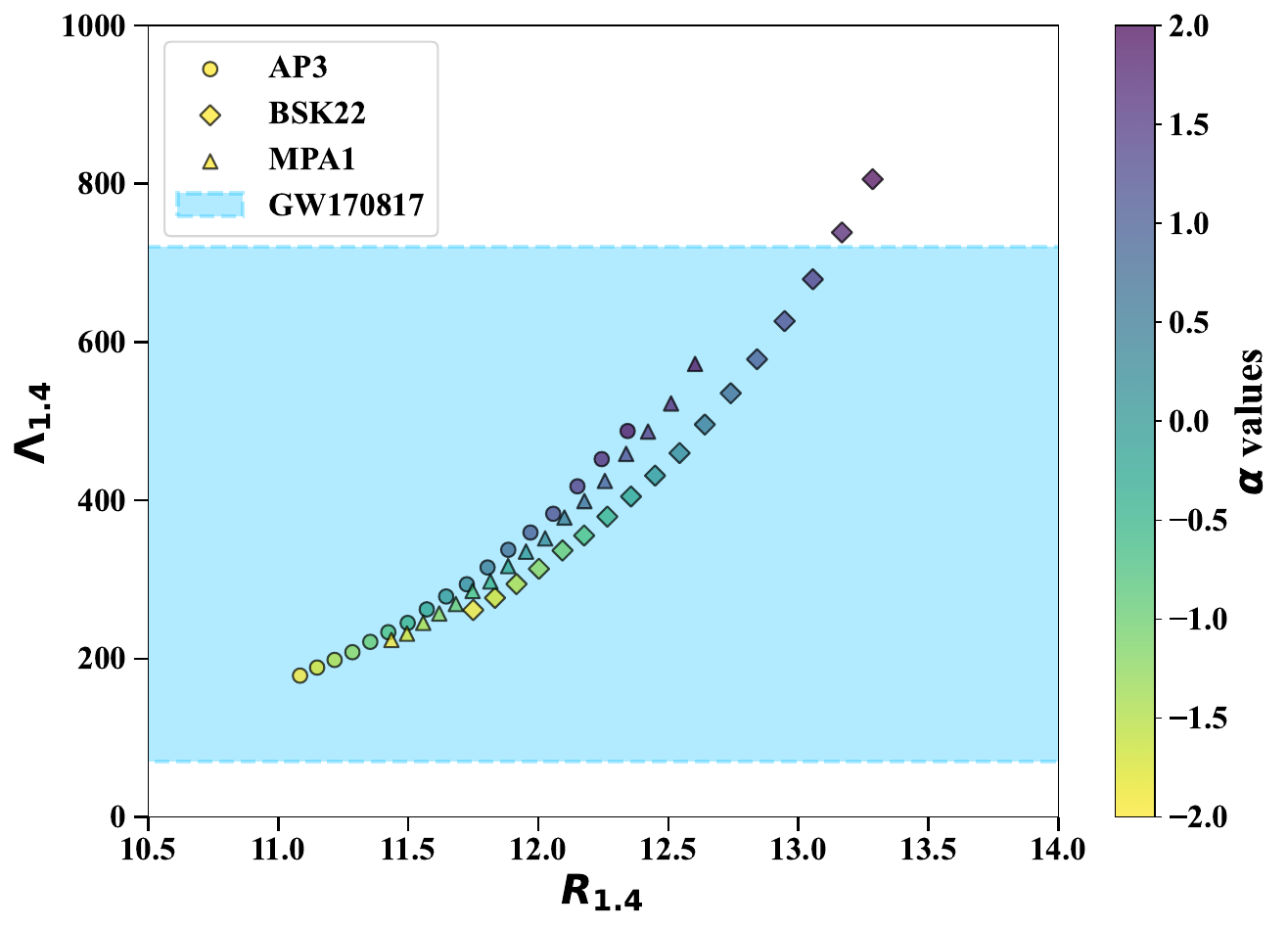}
  \caption*{(d) No DM}
  \label{fig:DM0_LvR}
\end{minipage}

\caption{\justifying The relation between Tidal deformability and the radius of a 1.4 $M_\odot$ DM-admixed anisotropic NS is investigated for the labeled EOS with  $g=10^{-5}$, $m_\chi=1$GeV, and $m_\phi=1$KeV. The color contour represents the range of $\alpha$-values from $-2$ to $2$. The subplots illustrate different DM fractions: (a) $5\%$ DM, (b) $1\%$ DM, (c) $0.1\%$ DM, and (d) No DM. The GW170817 constraint (deep sky blue) reflects observational data from the binary neutron star merger.}
\label{LvsR}
\end{figure}


\begin{table}[]
\caption{\justifying The anisotropic parameter ($\alpha$) satisfying the observational constraints are identified for different DM subfractions under the conditions of $g=10^{-5}$, $m_\chi=1$GeV, and $m_\phi=1$KeV. The $\alpha$-values ranged from $[-2, 2]$, with increments of $0.25$. The symbol (\ding{51}) indicates that all $\alpha$-values within this range satisfy the specified constraint. On the other hand, the symbol (\ding{55}) denotes that none of the $\alpha$-values meet the constraints.}
\renewcommand{\arraystretch}{1.5}
\setlength{\tabcolsep}{7.1pt}
\label{tab:Results_satisfying}
\noindent\rule{\linewidth}{0.6mm}
\begin{tabular}{ccccccc}
\textbf{EOS} & \begin{tabular}[c]{@{}c@{}}\textbf{PSR J0348+0432}\\ $M$\textgreater{}$2.01 M_\odot$\end{tabular} & \begin{tabular}[c]{@{}c@{}}\textbf{PSR J1614-2230}\\ $M$\textgreater{}$1.97 M_\odot$\end{tabular} & \begin{tabular}[c]{@{}c@{}}\textbf{PSR J0952-0607}\\ $M$\textgreater{}$2.35 M_\odot$\end{tabular} & \begin{tabular}[c]{@{}c@{}}\textbf{PSR 0740+6620}\\ $M$\textgreater{}$2.08 M_\odot$\end{tabular} & \begin{tabular}[c]{@{}c@{}}\textbf{PSR J0030+0451}\\ 12.71-13.02 Km\end{tabular} & \begin{tabular}[c]{@{}c@{}}\textbf{GW170817}\\ 10.62-12.83 Km\end{tabular} \\ \hline
\multicolumn{1}{|c}{}   & \multicolumn{6}{c|}{\textbf{5\% DM}}   \\
\multicolumn{1}{|c}{\textbf{AP3}} & $\alpha>-2$ & \ding{51} & $\alpha>0.5$ & $\alpha>-1.25$ & \ding{55} & \multicolumn{1}{c|}{\ding{51}} \\
\multicolumn{1}{|c}{\textbf{BSk22}} & $\alpha>-1$ & $\alpha>-1.25$ & $\alpha>1.25$ & $\alpha>-0.5$ & $\alpha>1.75$ & \multicolumn{1}{c|}{\ding{51}} \\
\multicolumn{1}{|c}{\textbf{MPA1}}  & \ding{51} & \ding{51} & $\alpha>0$ & $\alpha>-2$ & \ding{55} & \multicolumn{1}{c|}{\ding{51}}  \\ \hline \hline

\multicolumn{1}{|c}{\textbf{}}   & \multicolumn{6}{c|}{\textbf{1\% DM}} \\
\multicolumn{1}{|c}{\textbf{AP3}}   & \ding{51} & \ding{51} & $\alpha>0$ & $\alpha>-2$ & \ding{55} & \multicolumn{1}{c|}{\ding{51}} \\
\multicolumn{1}{|c}{\textbf{BSk22}} & $\alpha>-1.5$ & $\alpha>-2$ & $\alpha>0.5$ & $\alpha>-1$ & $\alpha>0.75$ & \multicolumn{1}{c|}{$\alpha<1.25$}  \\
\multicolumn{1}{|c}{\textbf{MPA1}}  & \ding{51} & \ding{51} & $\alpha>-0.75$ & \ding{51} & \ding{55} & \multicolumn{1}{c|}{\ding{51}}  \\ \hline \hline

\multicolumn{1}{|c}{}            & \multicolumn{6}{c|}{\textbf{0.1\% DM}}  \\
\multicolumn{1}{|c}{\textbf{AP3}}   & \ding{51} & \ding{51} & $\alpha>-0.25$ & $\alpha>-2$ & \ding{55} & \multicolumn{1}{c|}{\ding{51}}  \\
\multicolumn{1}{|c}{\textbf{BSk22}} & $\alpha>-1.75$ & $\alpha>-2$ & $\alpha>0.5$ & $\alpha>-1.25$ & $1.5>\alpha>0.5$ & \multicolumn{1}{c|}{$\alpha<1$}  \\
\multicolumn{1}{|c}{\textbf{MPA1}}  & \ding{51} & \ding{51} & $\alpha>-0.75$ & \ding{51} & \ding{55} & \multicolumn{1}{c|}{\ding{51}}  \\ \hline \hline

\multicolumn{1}{|c}{\textbf{}}   & \multicolumn{6}{c|}{\textbf{NO DM}} \\
\multicolumn{1}{|c}{\textbf{AP3}}   & \ding{51} & \ding{51} & $\alpha>-0.25$ & $\alpha>-2$ & \ding{55} & \multicolumn{1}{c|}{\ding{51}}  \\
\multicolumn{1}{|c}{\textbf{BSk22}} & $\alpha>-1.75$ & $\alpha>-2$ & $\alpha>0.5$ & $\alpha>-1.25$ & $1.5>\alpha>0.5$ & \multicolumn{1}{c|}{$\alpha<1$}  \\
\multicolumn{1}{|c}{\textbf{MPA1}}  & \ding{51} & \ding{51} & $\alpha>-0.75$ & \ding{51} & \ding{55} & \multicolumn{1}{c|}{\ding{51}}  \\ \cline{1-7}\noalign{\vskip 0.6ex}\cline{1-5}
\multicolumn{5}{|c|}{\textbf{GW170817} (70 \textless $\Lambda_{1.36}$\textless{}720)} & \multicolumn{2}{c}{}  \\ 
\multicolumn{1}{|c}{}  & \textbf{5\% DM}  & \textbf{1\% DM}  & \textbf{0.1\% DM}  & \multicolumn{1}{c|}{\textbf{NO DM}}  & \multicolumn{2}{c}{}  \\ \cline{1-5}
\multicolumn{1}{|c}{\textbf{AP3}}   &  \ding{51} & \ding{51} & \ding{51} &  \multicolumn{1}{c|}{\ding{51}}   & \multicolumn{2}{c}{}       \\
\multicolumn{1}{|c}{\textbf{BSk22}} & \ding{51} & $\alpha<2$ & $\alpha<1.75$ &  \multicolumn{1}{c|}{$\alpha<1.75$}   & \multicolumn{2}{c}{}  \\
\multicolumn{1}{|c}{\textbf{MPA1}}  &  \ding{51} & \ding{51} & \ding{51} &  \multicolumn{1}{c|}{\ding{51}}   & \multicolumn{2}{c}{} \\ \cline{1-5}
\end{tabular}
{\vskip 0.8ex}
\noindent\rule{\linewidth}{0.6mm}
\end{table}

\end{widetext}

\subsection{Coupling strength $g$ variation effects} \label{varying g-value}

Now, we will see how the variation of the coupling strength $g$ affects the NS properties. For the first case where we fixed the value of g, we took the visible radius to be $R_{B}$ as DM was presented inside the core of the neutron star. Taking the log scale of g for our calculation purpose, we take g values varying from $-5, -4, -3.7,$ and $-3.5$, respectively.

From Fig. \ref{fig:AP3 g vary MvsR} and \ref{fig:AP3 g vary MvsLambda}, it is clear that for AP3 EOS when we increase the coupling strength to $10^{-4}$, for this value core-halo transition occurs, giving a maximum radius of $20.23$ km, $R_{1.4} = 32.04$ km and dimensionless tidal deformability i.e $\Lambda_{1.4} \simeq 13000$ for the isotropic case (i.e $\alpha = 0$). When we try to increase the value of the anisotropy parameter from $-2, 0$ to $+2$, one important observation we get the value of $R_{max}$ \textbf{decrease} from $22.79, 20.23$ to $18.06$ km respectively. Again, further increasing the value of g i.e $10^{-3.7}$ (and $10^{-3.5}$) the characteristics properties of NS such as mass - radius profile \& dimensionless tidal deformability - mass curves gives the maximum radius $R_{max} = 55.17$ km (and $99.41$ km) \& $\Lambda_{1.4} \simeq 3 \times 10^{6}$ (and $\simeq 2.9 \times 10^{8}$ ) respectively for the isotropic case. For this, 2 values of g also $R_{max}$ \textbf{decrease} while changing $\alpha$ values from $-2, 0$ to $+2$ respectively as mentioned in table \ref{tab:g-vary}.

Similarly, the mass-radius plots for BSk22 EOS and MPA1 EOS with g variation are shown in Fig. \ref{fig:BSk22 g vary MvsR} and \ref{fig:MPA1 g vary MvsR}. Putting the anisotropic parameter $\alpha = 0$, the maximum mass $M_{max}$ and maximum radius $R_{max}$ of BSk22 model are found to be $(2.26 M_{\odot}, 22.89 $ km), $(2.33 M_{\odot}, 58.15 $ km), and $(2.36 M_{\odot}, 102.33 $ km) for g values $= 10^{-4}, 10^{-3.7}$, and $10^{-3.5}$ respectively.MPA1 EOS gives the same  $(2.46 M_{\odot}, 20.69 $ km), $(2.55 M_{\odot}, 55.14 $ km), and $(2.58 M_{\odot}, 99.36 $ km) for g values $= 10^{-4}, 10^{-3.7}$, and $10^{-3.5}$. While increasing the values of $\alpha$ from $-2, 0$ to $+ 2$, one important finding we get is the value of $R_{max}$ decrease for increasing value of $\alpha$, while taking this g values to be varied. This unique characteristic of g variation suggests \textbf{a dark matter halo formation} with corresponding core - halo transition g value as the statement is reinforced by the information given in table \ref{tab:g-vary}.

The values of dimensionless tidal deformability $\Lambda_{1.4}$ give a drastic increase up to the order of  $\mathcal{O}(10^{5}$),  $\mathcal{O}(10^{6}$), and  $\mathcal{O}(10^{7}$) for g values $10^{-4}$, $10^{-3.7}$, and $10^{-3.5}$ irrespective of the anisotropic parameter $\alpha$ for all the 3 given EOS models as shown in Fig. \ref{g vary MvsLambda}.

\begin{widetext}

\begin{figure}[]
\centering
\subfigure[AP3]{\includegraphics[width=0.9\linewidth]{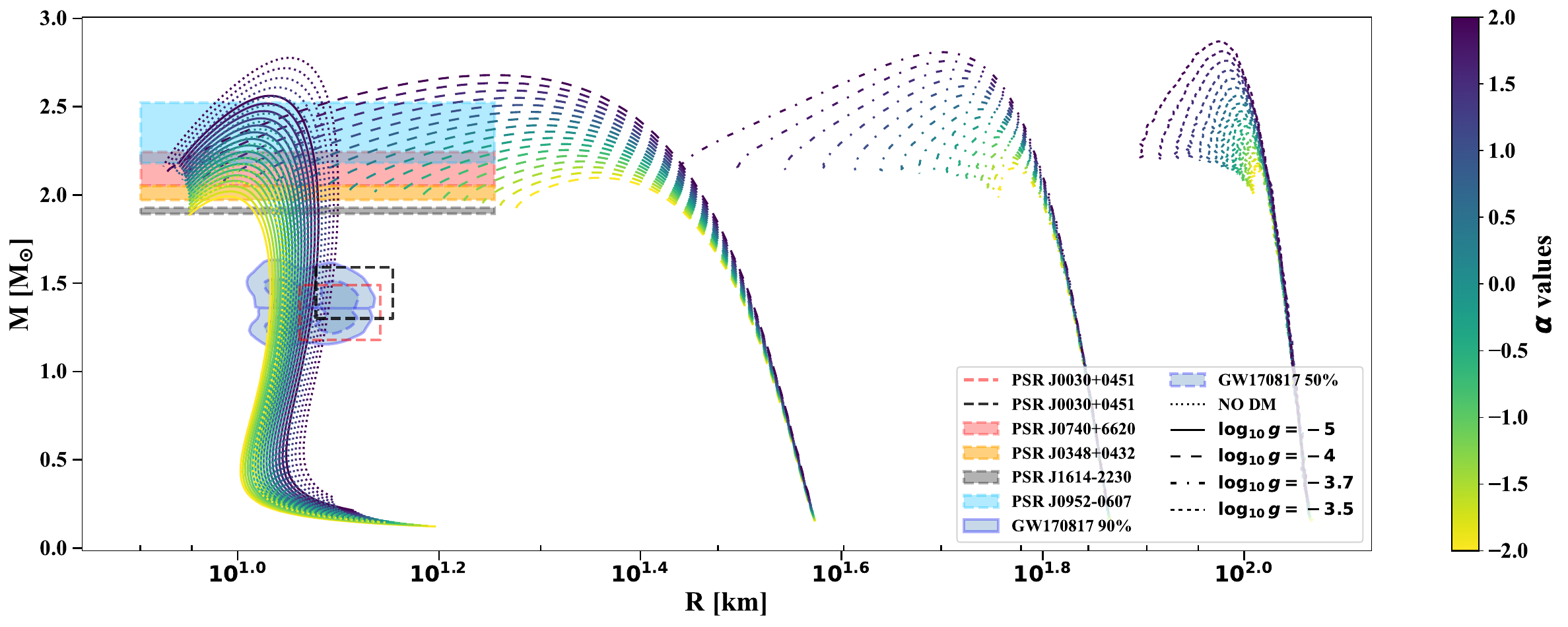}\label{fig:AP3 g vary MvsR}} 
\subfigure[BSk22]{\includegraphics[width=0.9\linewidth]{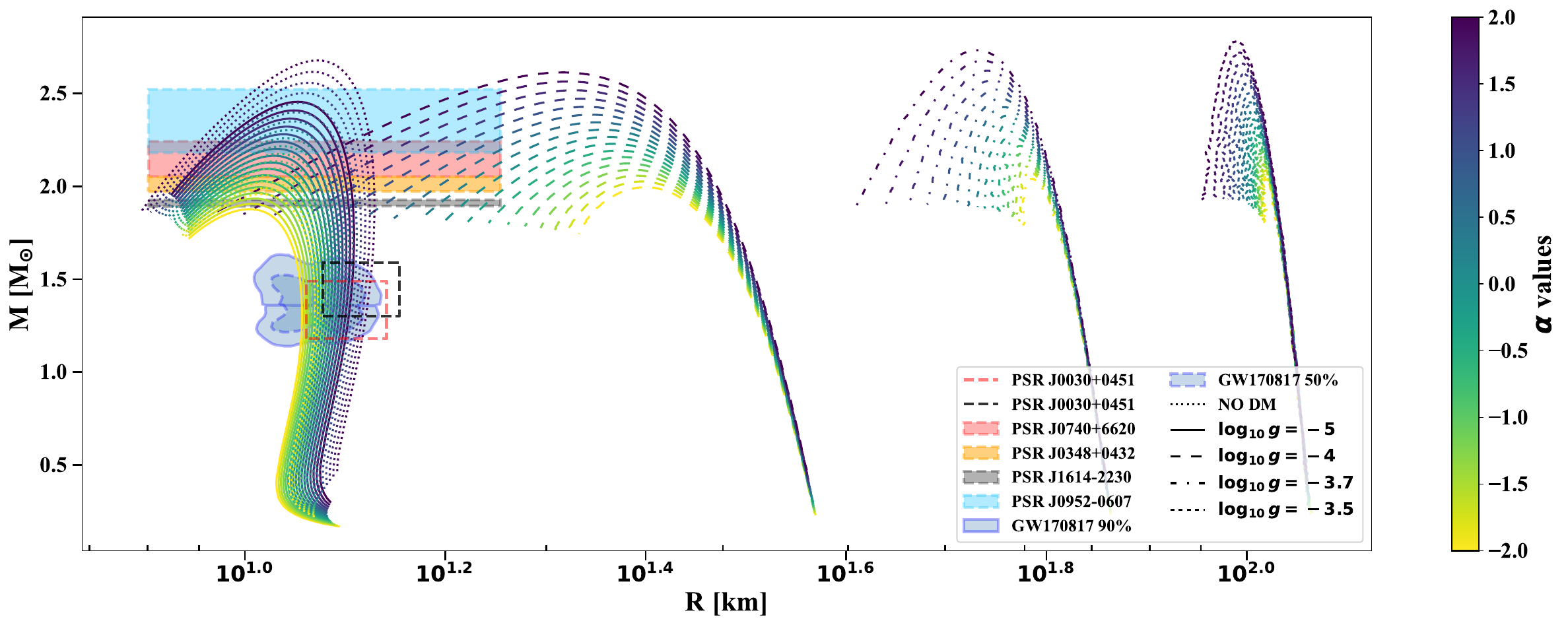}\label{fig:BSk22 g vary MvsR}}
\subfigure[MPA1]{\includegraphics[width=0.9\linewidth]{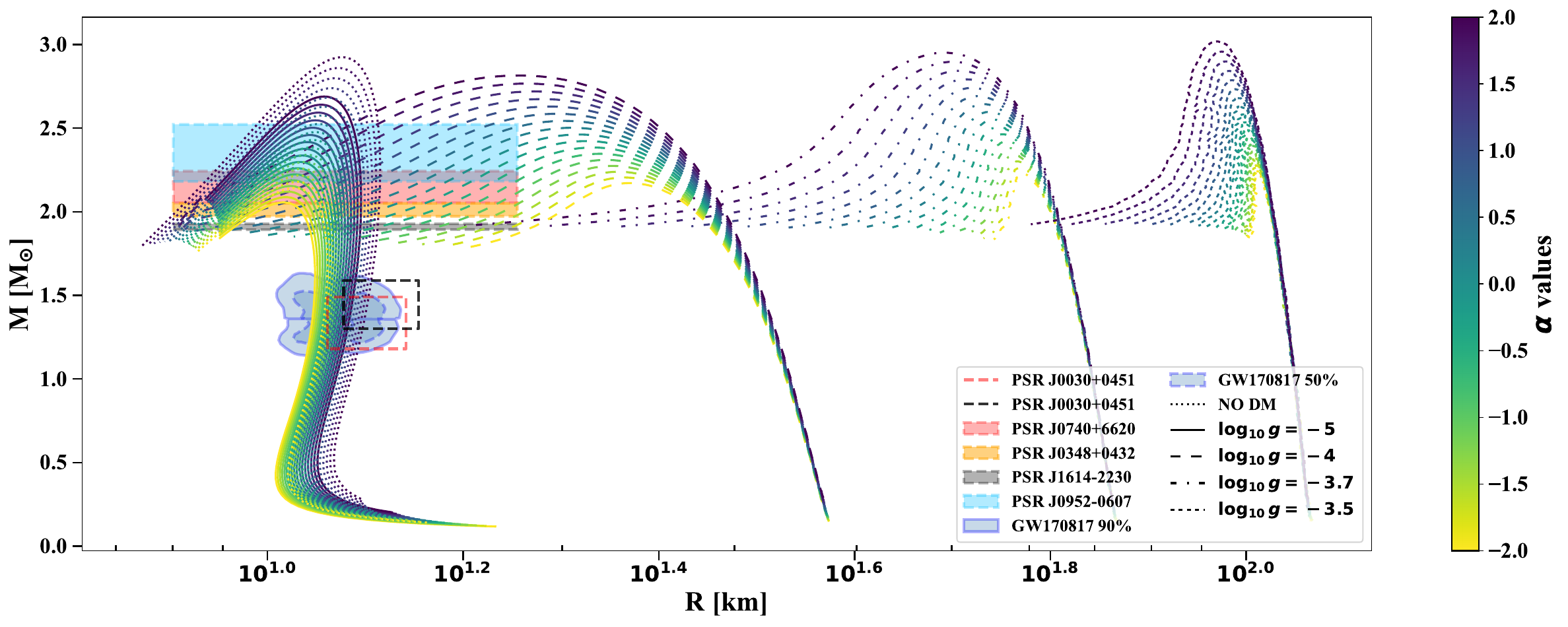}\label{fig:MPA1 g vary MvsR}}
\caption{\justifying Mass-radius stability profile for 5\% DM-admixed anisotropic NS with varying $g$-values, as labeled, and constant $m_\chi=1$GeV, and $m_\phi=1$KeV. The color contour represents the range of $\alpha$-values from $-2$ to $2$. The subplots correspond to (a) AP3 EOS, (b) BSk22 EOS, and (c) MPA1 EOS. Observational constraints are shown with deep sky blue (PSR J0952-0607), red (PSR J0740+6620), orange (PSR J0348+0432), black (PSR J1614-2230), steel-blue solid patch (GW170817 90\%), steel-blue dashed patch (GW170817 50\%), red and black dashed line for PSR J0030+0451.}
\label{g vary MvsR}
\end{figure}
\begin{figure}[]
\centering
\subfigure[AP3]{\includegraphics[width=0.49\linewidth]{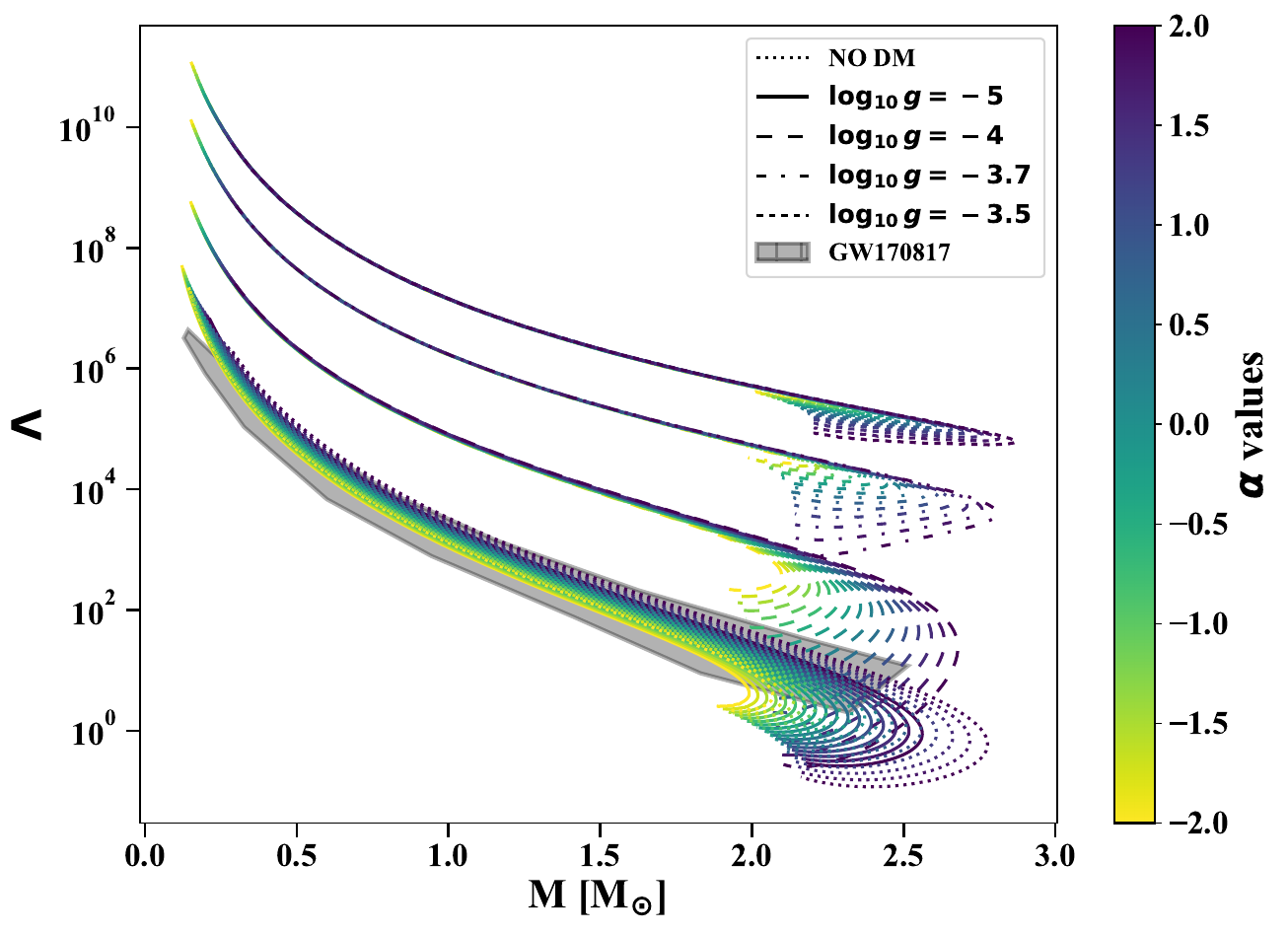}\label{fig:AP3 g vary MvsLambda}} 
\subfigure[BSk22]{\includegraphics[width=0.49\linewidth]{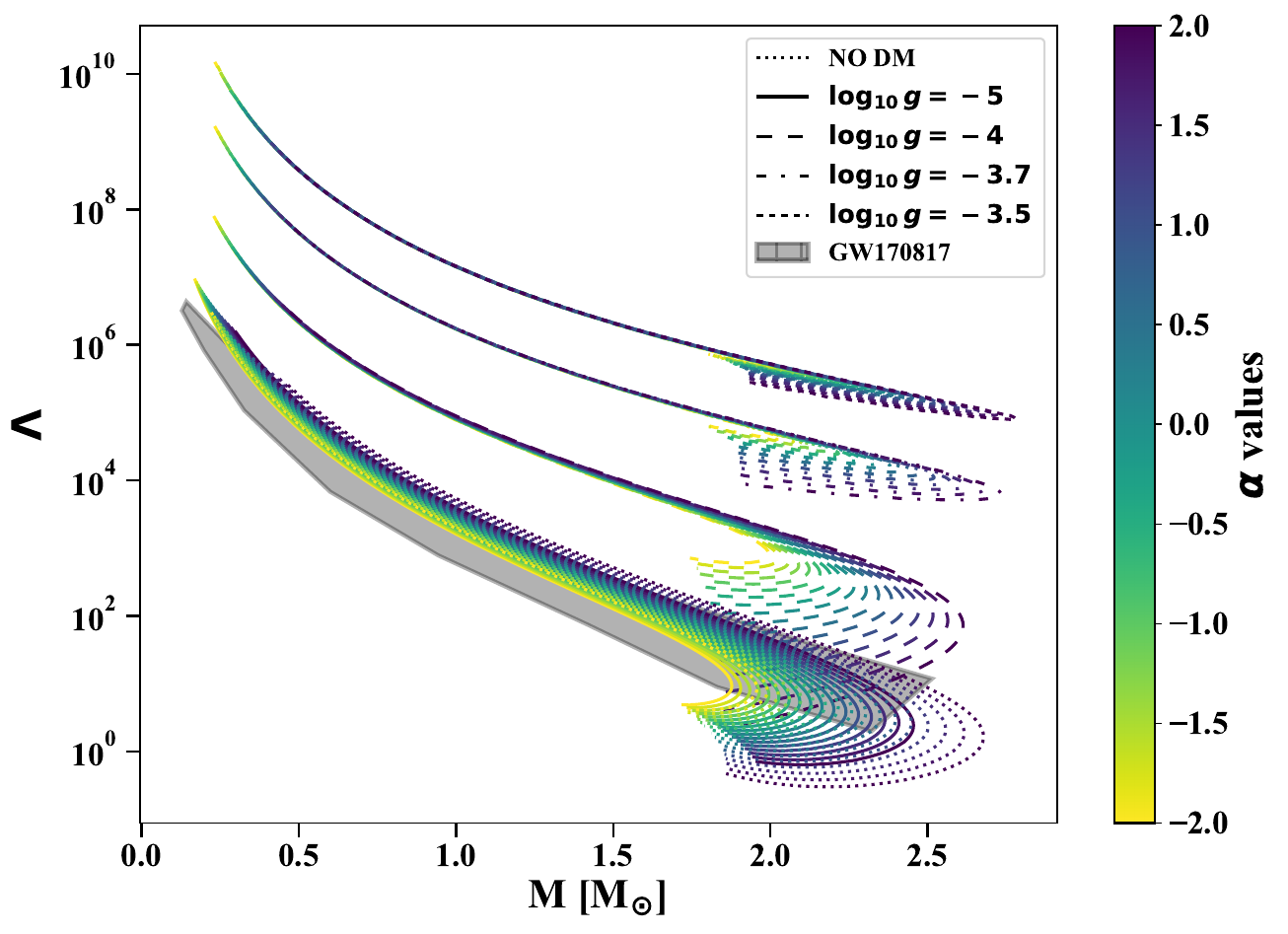}\label{fig:BSk22 g vary MvsLambda}}
\subfigure[MPA1]{\includegraphics[width=0.49\linewidth]{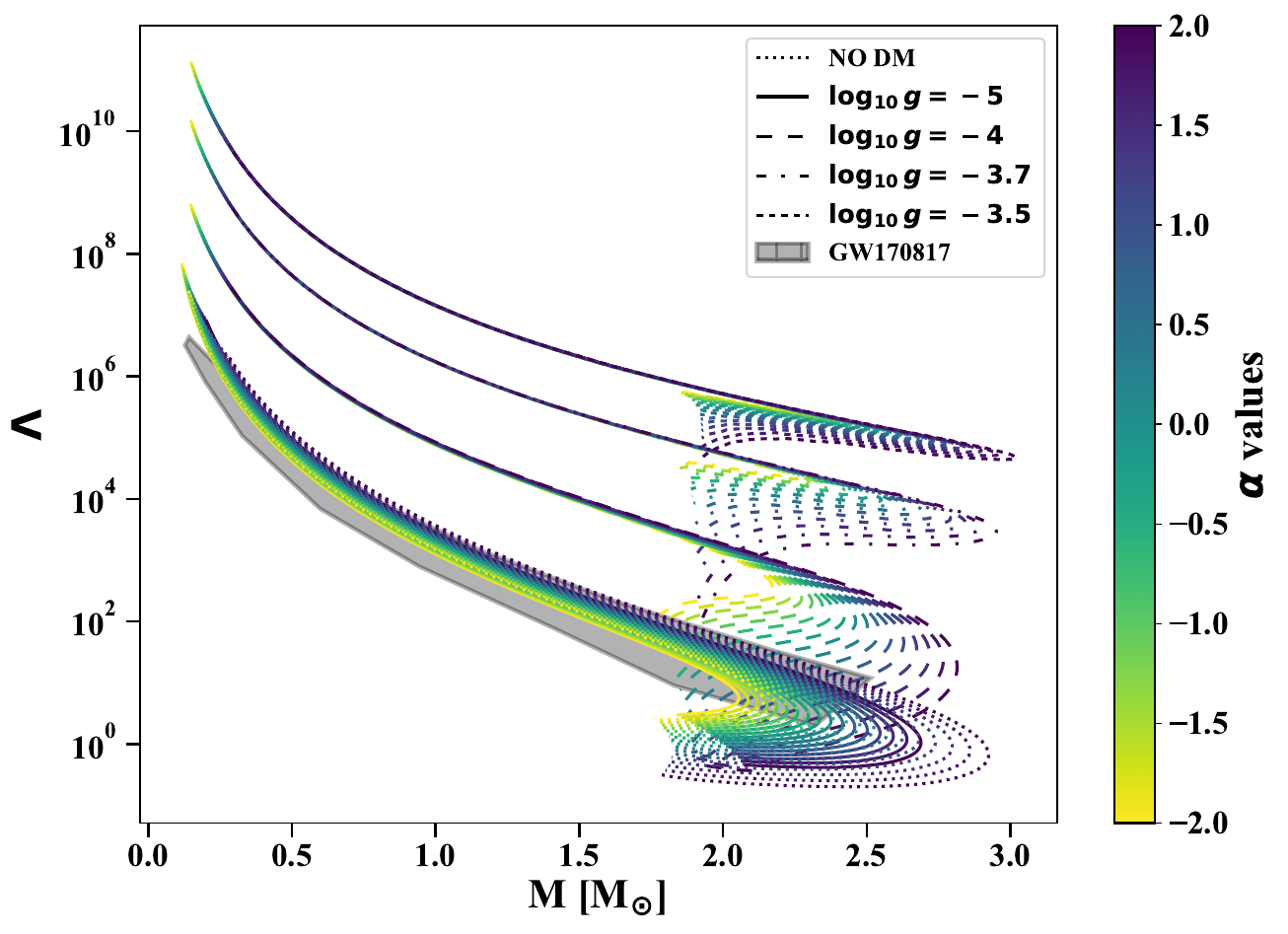}\label{fig:MPA1 g vary MvsLambda}}
\caption{\justifying Tidal deformability-mass stability profile, with the same setup as in Fig. \ref{g vary MvsR}}
\label{g vary MvsLambda}
\end{figure}

\begin{table}[]
\caption{\justifying Maximum mass $M_{\max}$ (in $M_\odot$), maximum radius $R_{\max}$ (in km), the radius at 1.4 $M_{\odot}$, i.e., $R_{1.4}$ (in km), and dimensionless tidal deformability at 1.4 $M_{\odot}$, i.e., $\Lambda_{1.4}$ for \textbf{AP3, BSk22, and MPA1} EOS as a function of anisotropic parameter $\alpha$ for different DM fractions with varying $g = 10^{-4}, 10^{-3.7}, 10^{-3.5}$, $m_\chi=1$GeV, and $m_\phi=1$KeV.}
\label{tab:g-vary}
\renewcommand{\arraystretch}{1.5}
\setlength{\tabcolsep}{2pt}
\begin{tabular}{|c|ccccccccc|}
\hline\hline
\textbf{}  & \multicolumn{3}{c}{\textbf{AP3}}                           & \multicolumn{3}{c}{\textbf{BSk22}}                         & \multicolumn{3}{c|}{\textbf{MPA1}}     \\ \hline\hline
\textbf{$\bm g$}           & \multicolumn{9}{c|}{\textbf{$\bm{10^{\bm{-4}}}$}}                                                                                                                               \\ \hline
\textbf{$\bm \alpha$}       & \textbf{-2} & \textbf{0} & \multicolumn{1}{c||}{\textbf{2}} & \textbf{-2} & \textbf{0} & \multicolumn{1}{c||}{\textbf{2}} & \textbf{-2} & \textbf{0} & \textbf{2} \\ \cline{2-10}
\textbf{M$_{\bm {max}}$}      & \multicolumn{1}{c|}{2.096847}  &  \multicolumn{1}{c|}{2.358023} & \multicolumn{1}{c||}{2.67851947}  & \multicolumn{1}{c|}{1.995432569} & \multicolumn{1}{c|}{2.266128805} & \multicolumn{1}{c||}{2.613144704} & \multicolumn{1}{c|}{2.171496105} & \multicolumn{1}{c|}{2.461700575} & \multicolumn{1}{c|}{2.814734913} \\ 
\textbf{R$_{\bm{max}}$}       & \multicolumn{1}{c|}{\pmb{22.79772}}  &  \multicolumn{1}{c|}{\pmb{20.234}}   & \multicolumn{1}{c||}{\pmb{18.06739875}} & \multicolumn{1}{c|}{\pmb{25.12823541}} & \multicolumn{1}{c|}{\pmb{22.89495037}} & \multicolumn{1}{c||}{\pmb{20.83688112}} & \multicolumn{1}{c|}{\pmb{23.39764094}} & \multicolumn{1}{c|}{\pmb{20.69282756}} & \multicolumn{1}{c|}{\pmb{18.23857036}} \\
\textbf{R$_{\bm{1.4}}$}       & \multicolumn{1}{c|}{31.72961}  &  \multicolumn{1}{c|}{32.04038} & \multicolumn{1}{c||}{32.41227465} & \multicolumn{1}{c|}{32.01891734} & \multicolumn{1}{c|}{32.38491044} & \multicolumn{1}{c||}{32.81278266} & \multicolumn{1}{c|}{31.91172787} & \multicolumn{1}{c|}{32.2032325} & \multicolumn{1}{c|}{32.53432127} \\
\textbf{$\Lambda_{\bm{1.4}}$} & \multicolumn{1}{c|}{12763.89}  &  \multicolumn{1}{c|}{13655.65} & \multicolumn{1}{c||}{14852.74291} & \multicolumn{1}{c|}{13409.35129} & \multicolumn{1}{c|}{14575.63382} & \multicolumn{1}{c||}{16306.00692} & \multicolumn{1}{c|}{13191.98364} & \multicolumn{1}{c|}{14122.89289} & \multicolumn{1}{c|}{15464.52486} \\ \hline\hline
\textbf{$\bm g$}           & \multicolumn{9}{c|}{\textbf{$\bm{10^{\bm{-3.7}}}$}}                                                                                                                             \\ \hline
\textbf{$\bm \alpha$}         &         \textbf{-2}              &         \textbf{0}               & \multicolumn{1}{c||}{\textbf{2}}  &             \textbf{-2}          &        \textbf{0}                & \multicolumn{1}{c||}{\textbf{2}}  &            \textbf{-2}           &         \textbf{0}               &             \textbf{2} \\ \cline{2-10}
\textbf{M$_{\bm {max}}$}      & \multicolumn{1}{c|}{2.15106335}  & \multicolumn{1}{c|}{2.443768316} & \multicolumn{1}{c||}{2.808332924} & \multicolumn{1}{c|}{2.039437884} & \multicolumn{1}{c|}{2.338203473} & \multicolumn{1}{c||}{2.732902374} & \multicolumn{1}{c|}{2.22634224}  & \multicolumn{1}{c|}{2.55108342}  & \multicolumn{1}{c|}{2.951210639} \\
\textbf{R$_{\bm{max}}$}       & \multicolumn{1}{c|}{\pmb{59.03971866}} & \multicolumn{1}{c|}{\pmb{55.17789549}} & \multicolumn{1}{c||}{\pmb{50.26830992}} & \multicolumn{1}{c|}{\pmb{61.16345586}} & \multicolumn{1}{c|}{\pmb{58.15801249}} & \multicolumn{1}{c||}{\pmb{53.81997682}} & \multicolumn{1}{c|}{\pmb{59.10027686}} & \multicolumn{1}{c|}{\pmb{55.14270106}} & \multicolumn{1}{c|}{\pmb{49.46734733}} \\
\textbf{R$_{\bm{1.4}}$}       & \multicolumn{1}{c|}{67.43809008} & \multicolumn{1}{c|}{67.47134441} & \multicolumn{1}{c||}{67.55036622} & \multicolumn{1}{c|}{67.41453005} & \multicolumn{1}{c|}{67.58546865} & \multicolumn{1}{c||}{67.85303692} & \multicolumn{1}{c|}{67.33798899} & \multicolumn{1}{c|}{67.36869102} & \multicolumn{1}{c|}{67.69749689} \\
\textbf{$\Lambda_{\bm{1.4}}$} & \multicolumn{1}{c|}{337536.9584} & \multicolumn{1}{c|}{338456.1153} & \multicolumn{1}{c||}{344194.0346} & \multicolumn{1}{c|}{337420.8862} & \multicolumn{1}{c|}{341012.0262} & \multicolumn{1}{c||}{350084.1383} & \multicolumn{1}{c|}{336683.2517} & \multicolumn{1}{c|}{336958.4416} & \multicolumn{1}{c|}{344876.6066} \\ \hline\hline
\textbf{$\bm g$}           & \multicolumn{9}{c|}{$\bm{10^{\bm{-3.5}}}$}                                                                                                                                      \\ \hline
\textbf{$\bm \alpha$}           &   \textbf{-2}                   &              \textbf{0}           & \multicolumn{1}{c||}{\textbf{2}}   &        \textbf{-2}                             & \textbf{0}           & \multicolumn{1}{c||}{\textbf{2}}   &           \textbf{-2}             &           \textbf{0}              &           \textbf{2} \\ \cline{2-10}
\textbf{M$_{\bm {max}}$}      & \multicolumn{1}{c|}{2.167115615}  & \multicolumn{1}{c|}{2.473355716}  & \multicolumn{1}{c||}{2.869109906}  & \multicolumn{1}{c|}{2.051072239}  & \multicolumn{1}{c|}{2.361100836}  & \multicolumn{1}{c||}{2.780046423}  & \multicolumn{1}{c|}{2.242609692}  & \multicolumn{1}{c|}{2.582685443}  &  \multicolumn{1}{c|}{3.018520585}  \\
\textbf{R$_{\bm{max}}$}      &  \multicolumn{1}{c|}{\pmb{103.1040484}}  &  \multicolumn{1}{c|}{\pmb{99.41217748}} & \multicolumn{1}{c||}{\pmb{94.24138063}}  &  \multicolumn{1}{c|}{\pmb{104.3272684}} &  \multicolumn{1}{c|}{\pmb{102.3375546}} & \multicolumn{1}{c||}{\pmb{97.66089413}}  &  \multicolumn{1}{c|}{\pmb{102.8122531}} &  \multicolumn{1}{c|}{\pmb{99.36148351}} &  \multicolumn{1}{c|}{\pmb{93.31257564}} \\
\textbf{R$_{\bm{1.4}}$}      &  \multicolumn{1}{c|}{109.6483241}  &  \multicolumn{1}{c|}{110.1221421} & \multicolumn{1}{c||}{110.1687287}  &  \multicolumn{1}{c|}{109.9484949} &  \multicolumn{1}{c|}{109.9287048} & \multicolumn{1}{c||}{110.2402982}  &  \multicolumn{1}{c|}{109.8450066} &  \multicolumn{1}{c|}{110.1162929} &  \multicolumn{1}{c|}{110.2559395} \\
\textbf{$\Lambda_{\bm{1.4}}$} & \multicolumn{1}{c|}{2863632.109}  & \multicolumn{1}{c|}{2948740.854}  & \multicolumn{1}{c||}{2953546.646}  & \multicolumn{1}{c|}{2914264.508}  & \multicolumn{1}{c|}{2914731.605}  & \multicolumn{1}{c||}{2954242.100}  & \multicolumn{1}{c|}{2912823.753}  & \multicolumn{1}{c|}{2928883.188}  & \multicolumn{1}{c|}{2972804.938}  \\ \hline\hline
\end{tabular}
\end{table}

\end{widetext}

\subsection{ Correlation between different Physical observables} \label{corelation}

\begin{widetext}

\begin{figure}[]
\centering
\subfigure[AP3]{\includegraphics[width=0.9\linewidth]{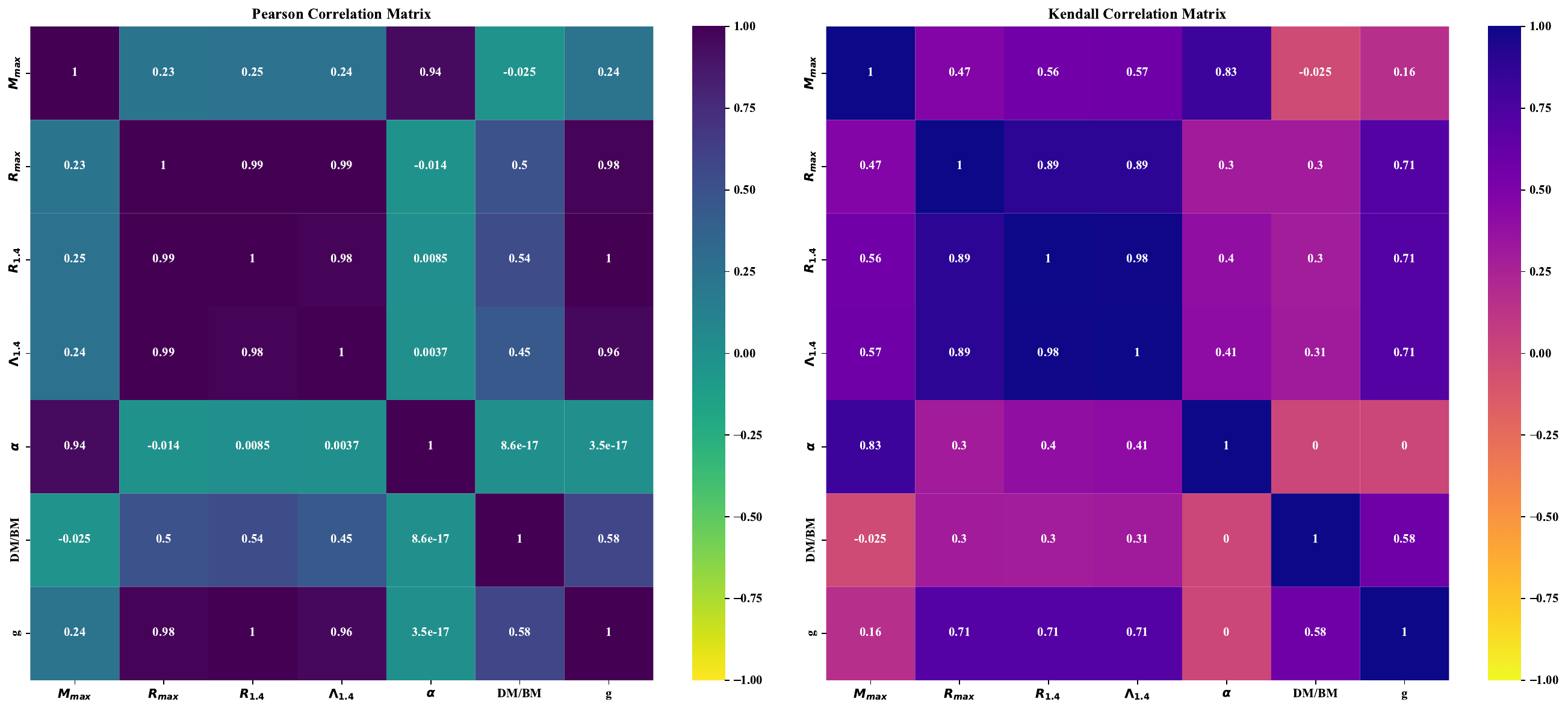}\label{fig:AP3 Correlation}}
\subfigure[BSk22]{\includegraphics[width=0.9\linewidth]{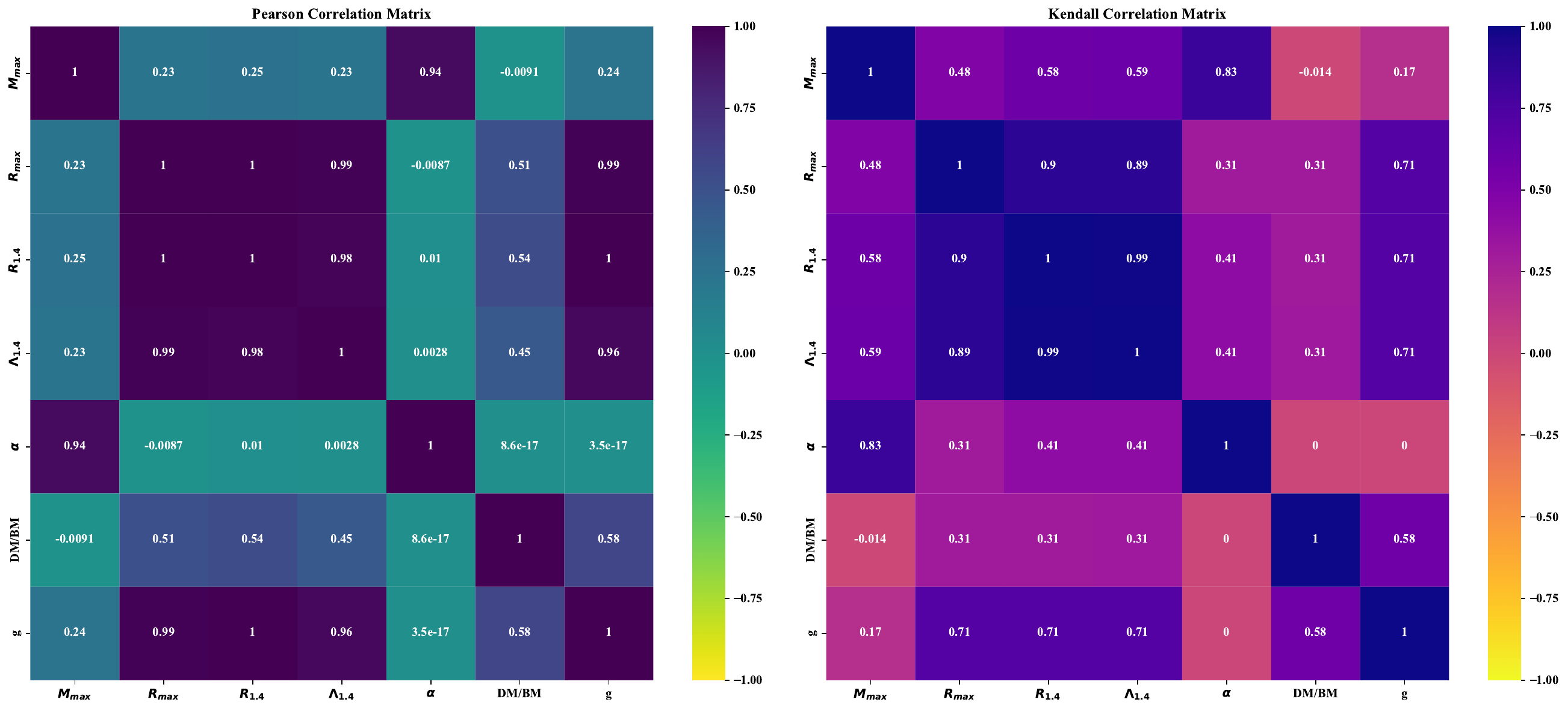}\label{fig:BSk22 Correlation}}
\subfigure[MPA1]{\includegraphics[width=0.9\linewidth]{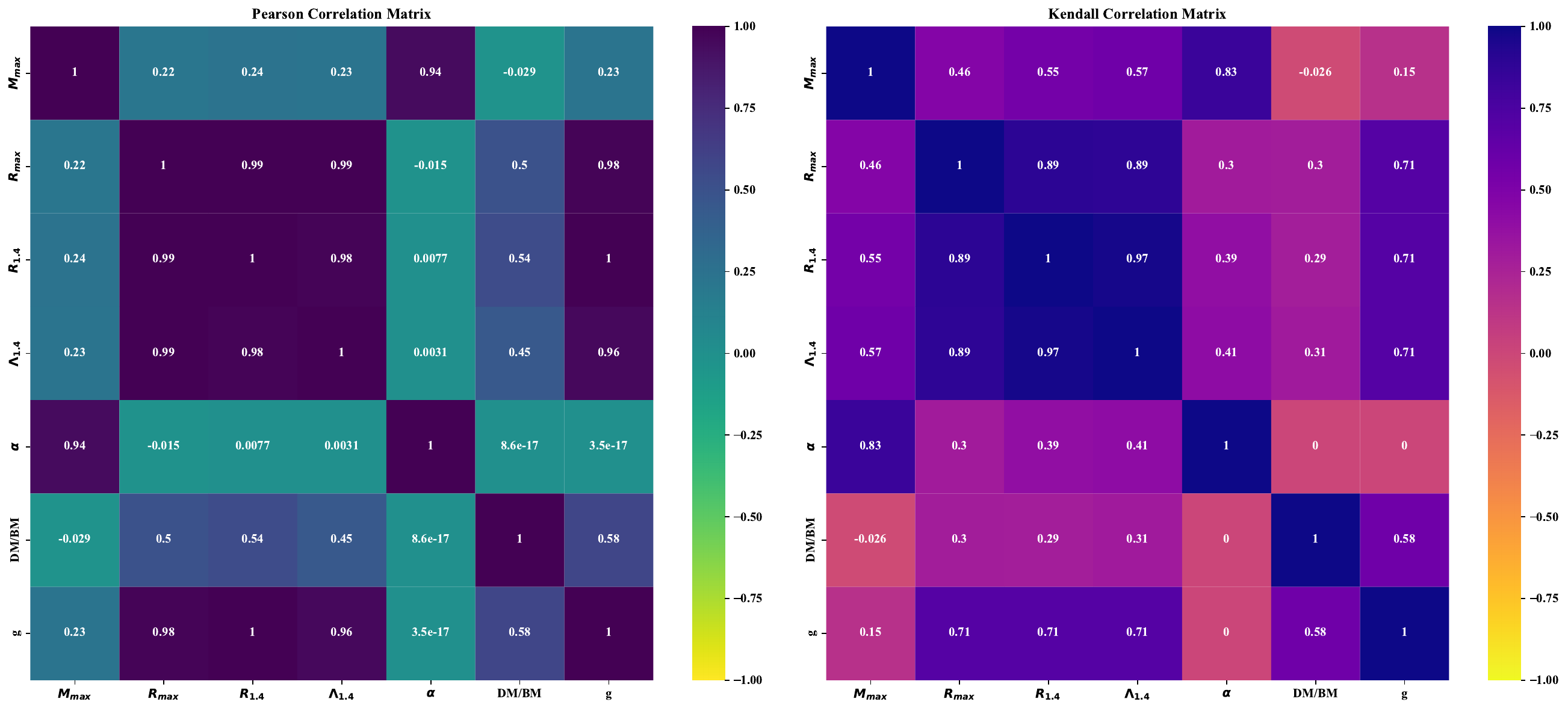}\label{fig:MPA1 Correlation}}
\caption{\justifying Correlation coefficients of different NS observables}
\label{Correlation}
\end{figure}

\end{widetext}

Fig. \ref{Correlation} presents the correlation analysis of the key physical variables involved in our admixed-NS study, using both Pearson and Kendall correlation coefficients for the considered EOS (AP3, BSk22, MPA1). The left-hand side of Fig. \ref{Correlation} shows Pearson's correlation coefficient that measures the linear relationship between two variables, with values ranging from $-1$ to $1$. Values close to $1$ indicate a strong positive linear relationship, values close to $-1$ indicate a strong negative linear relationship, and values around $0$ indicate no linear relationship. The key observations from the Pearson correlation matrix are as given.
\begin{itemize}
    \item $\alpha$ has a very high positive correlation with $M_{max}$ indicating a strong linear relationship. In contrast, it has a near-zero correlation with $R_{max}$, $R_{1.4}$, $\Lambda_{1.4}$, $DM/BM$ and $g$, indicating no significant linear relationship.
    \item $g$ shows an almost perfect correlation with $R_{max}$, $R_{1.4}$ and $\Lambda_{1.4}$, suggesting a near-perfect linear dependency among these variables. It is weakly correlated with $M_{max}$ and shows a moderate correlation with $DM/BM$.
    \item $DM/BM$ exhibits moderate correlations with $R_{max}$, $R_{1.4}$ and $\Lambda_{1.4}$ and near-zero with $M_{max}$.
    \item $R_{max}$, $R_{1.4}$ and $\Lambda_{1.4}$ indicates significant linear relationships among them. Conversely, they show a near-weakly dependence on $M_{max}$.
\end{itemize}
The right-hand side of Fig. \ref{Correlation} shows the Kendall correlation coefficient that measures the ordinal association between two variables, with values ranging from $-1$ to $1$. It is a measure of the correspondence between the rankings of the data. The key observations from the Kendall correlation matrix are as follows.
\begin{itemize}
    \item $\alpha$ has no ordinal relationships with $DM/BM$ and $g$, displays moderate correlations with $R_{max}$, $R_{1.4}$ and $\Lambda_{1.4}$, and shows a strong ordinal association with $M_{max}$.
    \item $g$ demonstrates moderate correlations with $DM/BM$, strong with $R_{max}$, $R_{1.4}$, and $\Lambda_{1.4}$, while showing negligible ordinal association with $M_{max}$.
    \item $DM/BM$ shows weak ordinal relationships with $R_{max}$, $R_{1.4}$, and $\Lambda_{1.4}$, and a negligible correlation with $M_{max}$.
    \item $R_{max}$, $R_{1.4}$ and $\Lambda_{1.4}$ exhibit a high degree of ordinal association among them, but moderate with $M_{max}$.
\end{itemize}
\par Both matrices provide insights into the relationships between the various macroscopic variables, with the Pearson correlation highlighting linear dependencies and the Kendall correlation emphasizing ordinal associations. Both matrices highlight a strong positive association between $\alpha$ and $M_{max}$, indicating $\alpha$'s significant impact on the maximum mass of the admixed-NSs. $R_{max}$, $R_{1.4}$, and $\Lambda_{1.4}$ show strong linear and ordinal correlations, reflecting their tight linkage in NS structural properties. $DM/BM$ exhibits moderate linear correlations with $R_{max}$, $R_{1.4}$ and $\Lambda_{1.4}$, but weaker ordinal relationships, indicating its lesser influence. $g$ is strongly correlated with radius and deformability parameters, underlining its crucial role in the NS equation of state. Notably, $M_{max}$ has weak linear but moderate ordinal associations with $R_{max}$, $R_{1.4}$ and $\Lambda_{1.4}$, suggesting that while linear predictability is low, rank-based relationships provide meaningful insights. These insights are crucial for understanding the underlying physics of admixed-NSs, particularly in how different EOS parameters influence key structural properties like maximum mass, radius, and tidal deformability. The strong correlations among certain parameters across both matrices suggest robust dependencies that are critical for the theoretical modeling and observational analysis of NSs.

\section{Conclusions} \label{conclusion}
This study illustrates the importance of considering both DM and anisotropy in NS modeling. The M-R curves highlight that even small amounts of DM or changes in $\alpha$ can significantly impact stability, which has profound implications for interpreting astrophysical observations and the underlying physics of NSs.  The major findings worth mentioning here are as follows.

\begin{itemize}
    \item The mass-radius stability profile reveals that increasing (decreasing) the $\alpha$ value results in stiffer (softer) NSs while increasing (decreasing) the DM subfraction softens (stiffens) the NSs, with a fixed coupling strength.
    \item It is noteworthy that dimensionless tidal deformability decreases (increases) with incrementing (decrementing) DM subfraction. Therefore, the presence of DM subfraction promotes more compact NSs that resist deformation. 
    \item It is seen that increasing the coupling parameter ($g$) between DM and its mediator initiates the formation of a core-halo structure, where a DM halo surrounds the BM. For coupling values $g = 10^{-4}$, $10^{-3.7}$, and $10^{-3.5}$, we observe a \textbf{decrease} in the maximum radius ($R_{max}$) with increasing anisotropy. This behavior suggests that stronger DM interactions lead to more compact neutron stars with significant core-halo differentiation, differing from the patterns observed at $g = 10^{-5}$ or in scenarios without DM, where such compactness is not as pronounced.
    \item It is noticeable that with coupling values $g = 10^{-4}$, $10^{-3.7}$, and $10^{-3.5}$, the Dimensionless Tidal deformability is \textbf{\textit{not observationally feasible}}. Hence, such coupling values could be discarded.
    \item Out of these 3 models taken, the color contours only for \textbf{BSk22} EOS in zero DM case with positive values of $\alpha$ are going beyond the observational bounds, but increasing DM subfraction up to 5\% gives a \textbf{\textit{unique and robust}} relation satisfying between $\Lambda_{1.4}$ and $R_{1.4}$ in the GW170817 band. The other 2 EOS don't respond much to increment of DM subfraction.
    \item The key observation from the Pearson correlation matrix is that $\alpha$ has a very high positive correlation with $M_{max}$, indicating a strong linear relationship. In contrast, it has a \textbf{near-zero correlation} with $R_{max}$, $R_{1.4}$, $\Lambda_{1.4}$, $DM/BM$ and $g$, indicating no significant linear relationship, whereas the Kendall correlation matrix suggests that $\alpha$ has \textit{\textbf{no}} ordinal relationships with $DM/BM$ and $g$, displays \textbf{moderate} correlations with $R_{max}$, $R_{1.4}$ and $\Lambda_{1.4}$, and shows a strong ordinal association with $M_{max}$.
\end{itemize}

In our study, we assumed baryonic matter (BM) to be anisotropic and dark matter (DM) to be isotropic. However, alternative configurations—where BM is isotropic, DM is anisotropic, or both are anisotropic—are also possible and warrant exploration in future research. The stability of neutron stars (NS) is influenced by the densities of BM and DM, as discussed in \cite{Hippert:2022snq}. Recent observations of high-mass pulsars, like PSR J0952+0607, challenge traditional isotropic models, which struggle to explain such extreme masses. Anisotropic models, on the other hand, provide a more realistic explanation, supporting stable configurations even at higher masses. Anisotropy also impacts the star's deformation and response to tidal forces, leading to more accurate predictions of gravitational waveforms.

In summary, our proposed study is expected to make an important and fundamental contribution to the understanding of DM-admixed NSs. DM doesn't interact electromagnetically, and gravitational wave (GW) experiments do not differentiate between various forms of matter; the resulting tidal effects are determined only by the compactness of the object. Thus, more data from LIGO and future GW experiments such as LISA \cite{LISA_2017pwj}, Einstein Telescope \cite{maggiore2020science}, and Cosmic Explorer \cite{reitze2019cosmic} will present opportunities to identify NSs that have subfractions of DM or particles of a similar type. Incorporating anisotropic pressure into DM-admixed NS models provides a more comprehensive understanding of their properties, structure, and behavior under extreme conditions. It not only aligns better with observational data but also allows for more sophisticated modeling of exotic matter and gravitational wave predictions, pushing the limits of our understanding of compact objects in the universe.

\section*{Acknowledgments}

P.M. would like to thank BITS Pilani K K Birla Goa campus for the fellowship support, Prashant Thakur for discussing \& sharing ideas on pulsars and GW170817 data, and comments on our EOS model and IUCAA, Pune, for providing a good computational facility for this project. C.S. and P.M. thank Apratim Ganguly for the rigorous discussion on the EOS models and relevant plots for this work. A.H. expresses gratitude to Michael Collier (Durham University, UK) for his assistance with the discussion on two-fluid code and to the International Centre for Theoretical Sciences (ICTS) for the opportunity to participate in the program 'Summer School on Gravitational-Wave Astronomy' (code: ICTS/GWA2024/07). This work makes use of the following Python packages: \texttt{pandas} \cite{reback2020pandas}, \texttt{scipy} \cite{2020SciPy-NMeth}, \texttt{seaborn} \cite{Waskom2021}, \texttt{matplotlib} \cite{Hunter:2007}, \texttt{numpy} \cite{harris2020array}.

\FloatBarrier
\bibliography{citations}
\bibliographystyle{unsrt}

\end{document}